\documentclass{aa}  
\usepackage{natbib}
\bibpunct{(}{)}{;}{a}{}{,}
\usepackage{lscape}
\usepackage{supertabular}
\usepackage{graphicx}
\usepackage{txfonts}
\begin{document}

   \title{{\em Herschel}-PACS\thanks{{\em Herschel} is an ESA space observatory with science instruments 
          provided by European-led Principal Investigator consortia and with important participation from NASA.} 
          photometry of faint stars for sensitivity performance assessment and establishment of faint FIR primary 
          photometric standards
          }


\titlerunning{{\em Herschel}-PACS photometry of faint stars}

   \author{U.~Klaas
          \inst{1}
          \and
           Z.~Balog
          \inst{1}
          \and
           M.~Nielbock
          \inst{1,2}
          \and
           T.G.~M\"uller
          \inst{3}
          \and
           H.~Linz
          \inst{1}
          \and
           Cs.~Kiss
          \inst{4}
          }

   \institute{Max-Planck-Institut f\"ur Astronomie (MPIA),
              K\"onigstuhl 17, 69117 Heidelberg, Germany \\
              \email{klaas@mpia.de}
         \and
              Haus der Astronomie, MPIA-Campus,
              K\"onigstuhl 17, 69117 Heidelberg, Germany
         \and
             Max-Planck-Institut f\"ur extraterrestrische Physik (MPE),
              PO Box 1312, Giessenbachstra{\ss}e, 85741 Garching, Germany
         \and
             Konkoly Observatory, Research Centre for Astronomy and Earth 
             Sciences, Hungarian Academy of Sciences, 1121 Budapest, Konkoly 
             Thege Mikl\'os \'ut 15-17, Hungary
             }

   \date{Received 23~August~2017 / Accepted 7~December~2017}

 
  \abstract
   {}
   {Our aims are to determine flux densities and their photometric accuracy 
    for a set of seventeen stars that range in flux from intermediately bright 
    ($\lesssim$2.5\,Jy) to faint ($\gtrsim$5\,mJy) in the far-infrared (FIR).
    We also aim to derive signal-to-noise dependence with flux and time, and
    compare the results with predictions from the {\em Herschel} exposure-time 
    calculation tool.
    }
   {We obtain aperture photometry from {\em Herschel}-PACS high-pass-filtered 
    scan maps and chop/nod observations of the faint stars. The issues of
    detection limits and sky confusion noise are addressed by comparison of
    the field-of-view at different wavelengths, by multi-aperture photometry,
    by special processing of the maps to preserve extended emission, and with 
    the help of large-scale absolute sky brightness maps from {\it AKARI}. 
    This photometry is compared with flux-density predictions based on 
    photospheric models for these stars. We obtain a robust noise estimate by 
    fitting the flux distribution per map pixel histogram for the area around 
    the stars, scaling it for the applied aperture size and correcting for 
    noise correlation. 
}
   {For 15 stars we obtain reliable photometry in at least one PACS filter,
    and for 11 stars we achieve this in all three PACS filters (70, 100, 
    160\,$\mu$m). Faintest fluxes, for which the photometry still has good 
    quality, are about 10 -- 20\,mJy with scan map photometry. The photometry 
    of seven stars is consistent with models or flux predictions for pure 
    photospheric emission, making them good primary standard candidates. Two 
    stars exhibit source-intrinsic far-infrared excess: $\beta$\,Gem (Pollux),
    being the host star of a confirmed Jupiter-size exoplanet, due to emission 
    of an associated dust disk, and $\eta$\,Dra due to dust emission in a 
    binary system with a K1 dwarf. The investigation of the 160\,$\mu$m sky 
    background and environment of four sources reveals significant sky 
    confusion prohibiting the determination of an accurate stellar flux at 
    this wavelength. As a good model approximation, for nine stars we obtain 
    scaling factors of the continuum flux models of four PACS fiducial 
    standards with the same or quite similar spectral type. We can verify a 
    linear dependence of signal-to-noise ratio (S/N) with flux and with square 
    root of time over significant ranges. At 160\,$\mu$m the latter relation 
    is, however, affected by confusion noise.
}
   {The PACS faint star sample has allowed a comprehensive sensitivity
    assessment of the PACS photometer. Accurate photometry allows us to 
    establish a set of five FIR primary standard candidates, namely 
    $\alpha$\,Ari, $\varepsilon$\,Lep, $\omega$\,Cap, HD\,41047 and 42\,Dra, 
    which are 2 -- 20 times fainter than the faintest PACS fiducial standard 
    ($\gamma$\,Dra) with absolute accuracy of $<$6\%. For three of these 
    primary standard candidates, essential stellar parameters are known, 
    meaning that a dedicated flux model code may be run.}


   \maketitle
%

%
%
\begin{table*}[ht!]
\caption{Faint secondary standards observed by {\em Herschel}-PACS. Source
  fluxes from \citet{gordon07} are for an effective wavelength of
  71.42\,$\mu$m and have been colour-corrected to the PACS central wavelength
  of 70\,$\mu$m by dividing by the factor 0.961 \citep[cf.\ ][]{mueller11}
  for a Rayleigh-Jeans-tail-type SED. 100 and 160\,$\mu$m fluxes for these
  sources are then extrapolated values for this adopted SED.
}             
\label{table:sources}      
\centering
    \begin{tabular}{r l c c c l l}
   \hline\hline
            \noalign{\smallskip}
        &       & \multicolumn{3}{c}{Model flux prediction (mJy)} &  Spectral type & Reference \\
 HD     &   Other name     &f$_{70}$~~~~~~~~&f$_{100}$~~~~~~~~& f$_{160}$~~~~~~~~ &             &           \\
            \noalign{\smallskip}
    \hline
             \noalign{\smallskip}
 62509  & \object{$\beta$\,Gem} & 2457~($\pm$5.73\%) &  1190~($\pm$5.73\%) & 455.9~($\pm$5.73\%) &  K0IIIb  & \citet{cohen96} \tablefootmark{a} \\
 12929  & \object{$\alpha$\,Ari}& 1707~($\pm$5.9\%)  & 831.4~($\pm$5.9\%)  & 321.0~($\pm$5.9\%)  &  K2III   & \citet{cohen96} \tablefootmark{a} \\
 32887  & \object{$\varepsilon$\,Lep} & 1182~($\pm$5.9\%)  & 576.2~($\pm$5.9\%)  & 222.7~($\pm$5.9\%)  &  K4III   & \citet{cohen96} \tablefootmark{a} \\
198542  & \object{$\omega$\,Cap}& 857.7~($\pm$6.03\%)& 418.0~($\pm$6.03\%) & 161.5~($\pm$6.03\%) &  M0III   & \citet{cohen96} \tablefootmark{a} \\
148387  & \object{$\eta$\,Dra}  & 479.5~($\pm$3.38\%)& 232.6~($\pm$3.45\%) &  89.4~($\pm$3.51\%) &  G8III   & \citet{hammersley98} \tablefootmark{b} \\
180711  & \object{$\delta$\,Dra}& 428.9~($\pm$5.7\%) & 207.7~($\pm$5.7\%)  &  79.6~($\pm$5.7\%)  &  G9III   & \citet{cohen96} \tablefootmark{a} \\
139669  & \object{$\theta$\,Umi}& 286.2~($\pm$5.67\%)& 139.5~($\pm$5.67\%) &  53.9~($\pm$5.67\%) &  K5III   & \citet{cohen96} \tablefootmark{a} \\
 41047  & \object{HR\,2131}     & 195.6~($\pm$5.96\%)&  95.4~($\pm$5.96\%) &  36.9~($\pm$5.96\%) &  K5III   & \citet{cohen96} \tablefootmark{a} \\
170693  & \object{42\,Dra}      &   153.7$\pm$4.6    &  75.3~($\pm$3.0\%)  &  29.4~($\pm$3.0\%)  & K1.5III  & \citet{gordon07} \\
138265  & \object{HR\,5755}     &   115.9$\pm$4.0    &  56.8~($\pm$3.5\%)  &  22.2~($\pm$3.5\%)  &  K5III   & \citet{gordon07} \\
159330  & \object{HR\,6540}     &    64.2$\pm$2.1    &  31.5~($\pm$3.3\%)  &  12.3~($\pm$3.3\%)  &  K2III   & \citet{gordon07} \\
152222  & \object{SAO\,17226}   &    39.4$\pm$1.9    &  19.3~($\pm$5.0\%)  &   7.5~($\pm$5.0\%)  &  K2III   & \citet{gordon07} \\
 39608  & \object{SAO\,249364}  &    30.9$\pm$1.2    &  15.1~($\pm$4.0\%)  &   5.9~($\pm$4.0\%)  &  K5III   & \citet{gordon07} \\
181597  & \object{HR\,7341}     &  28.0~($\pm$3.29\%)&  13.6~($\pm$3.34\%) &   5.2~($\pm$3.42\%) &  K1III   & \citet{hammersley98} \tablefootmark{b} \\
 15008  & \object{$\delta$\,Hyi}&    22.9$\pm$0.8    &  11.2~($\pm$3.5\%)  &   4.4~($\pm$3.5\%)  &  A1/2V   & \citet{gordon07} \\
156729  & \object{e\,Her}       &  12.0~($\pm$3.21\%)&   5.8~($\pm$3.25\%) &   2.2~($\pm$3.28\%) &   A2V    & \citet{hammersley98} \tablefootmark{b} \\
168009  & \object{HR\,6847}     &  10.0~($\pm$3.40\%)&   4.9~($\pm$3.45\%) &   1.9~($\pm$3.50\%) &   G2V    & \citet{hammersley98} \tablefootmark{b} \\
            \noalign{\smallskip}
\hline
    \end{tabular}
\tablefoot{Source flux models are from \\
\tablefoottext{a}{http://general-tools.cosmos.esa.int/iso/users/expl\_lib/ISO/wwwcal/isoprep/cohen/extraps/} \\
\tablefoottext{b}{http://general-tools.cosmos.esa.int/iso/users/expl\_lib/ISO/wwwcal/isoprep/gbpp/}
}
\end{table*}

\section{Introduction}

The photometric calibration of the PACS photometer \citep{poglitsch10}
on-board the {\em Herschel} Space Observatory \citep{pilbratt10} is based on 
celestial standard stars \citep{balog14,nielbock13}. These primary standard 
stars have well-modelled spectral energy distributions (SEDs) of their 
photospheric emission and an accurate absolute calibration in the K-band 
\citep{dehaes11}. They are still relatively bright in the far-infrared 
(in the range 1 - 10\,Jy) to achieve high signal-to-noise ratios (S/N) within 
reasonable measurement times. Besides repeated measurements of these standard 
stars, a set of fainter secondary standard stars was repeatedly measured by 
PACS as part of the calibration program during the {\em Herschel} Performance 
Verification and Routine Operations periods. This included sources down to a 
few mJy. The PACS photometer is linear over a flux range exceeding the primary 
standard fluxes, with an optimized detector set-up for the flux background 
from the telescope. Flux nonlinearity is therefore an issue for considerably 
brighter sources and has been addressed elsewhere \citep{mueller16}. However, 
including fainter sources with well known flux predictions allows to us 
address the following questions:
\begin{itemize}
  \item[1)] How does the sensitivity scale with flux and time?
  \item[2)] How does the finally achieved sensitivity compare with predictions
            by the PACS exposure-time calculator of the {\it Herschel}
            observation planning tool?
  \item[3)] What is the impact and consistency of the applied data reduction 
            scheme on the resulting source flux for fainter and fainter flux 
            contributions on top of the telescope background level?
  \item[4)] What is the impact of background confusion noise on the resulting
            fluxes and the sensitivity limit?
\end{itemize}
Ultimately, some of the faint sources may be characterized well enough to
become primary standard sources for future powerful and sensitive FIR space 
telescopes, such as SPICA~\citep[e.g.][]{sibthorpe15}, 
Millimetron~\citep[e.g.][]{smirnov12} or the Origins Space 
Telescope~\citep{meixner17}.

Most of the observations have been done in mini-scan-map mode,
but we have included also a valuable set of complementary chop/nod 
point-source photometry. We first report the scan map photometry including
the sensitivity verification. Then we present the chop/nod photometry
and compare it with the scan map results. Finally, we analyse the source
spectral energy distributions (SEDs) by comparison with model SEDs and
establish which sources are suitable as accurate celestial standards.

\section{Source selection}

In preparation of the PACS in-flight photometric calibration, secondary 
standard source lists with stars described in \citet{cohen96}, \citet{hammersley98}, 
and \citet{gordon07} were prepared by the PACS Instrument Control Centre (ICC)
team. Depending on the source visibility during the Herschel mission, a subset
of sources from these lists were observed to cover the flux range from 0.5 -
2.5\,Jy down to 2 - 10\,mJy over the three photometer wavelengths 70, 100, and
160\,$\mu$m. The finally observed 17 sources are listed in Table~\ref{table:sources}.

\section{Scan map photometry}

Fifteen out of the 17 sources were observed in the PACS mini-scan-map 
point-source observing mode. This was the recommended scientific observing 
mode for point sources after {\em Herschel's} Science Demonstration Phase 
(SDP), because it had a better sensitivity and allowed a better 
characterization of the source vicinity and larger-scale structures of the 
background than chop/nod photometry. The satellite scans were mostly done with 
the nominal 20\arcsec\,/\,s speed; a few early ones were done with the 
originally adopted speed of 10\arcsec\,/\,s. The scan map dimension parameters 
are usually 3\arcmin~leg length and ten legs with a separation of 4\arcsec 
\ with scan angles in array coordinates of 70$^{\circ}$ and 110$^{\circ}$
(along the diagonal of the bolometer arrays). Only a few early measurements 
had different parameters from these values, when still probing for the optimum
parameter set. In the case of repetition factors larger than 1, in particular 
for our faintest targets, the whole scan map was repeatedly executed according 
to the specified factor. We note that a repetition factor may have been 
optimized for the short wave filter measurement and is hence less optimal 
for the 160\,$\mu$m filter, where the star is fainter. The observations were 
usually done in high gain mode. There are a few exceptions taken for 
comparative performance checks. Selected observing parameters are listed for 
all individual scan map observations in Tables~\ref{table:scanmapphotblue6_1} 
to~\ref{table:scanmapphotred10_1}.

%
%
\begin{table*}[ht!]
\caption{Relevant scan map parameters for photometry and noise determination.
         r$_{\rm aper}^{phot}$ is the radius of the aperture used for the 
         point-source photometry, c$_{\rm aper}$ is the corresponding
         correction factor to scale the flux to its total value, 
         cc($\lambda_{\rm ref}$) is the colour-correction factor to derive the 
         source flux at the reference wavelength $\lambda_{\rm ref}$ of the 
         filter~\citep{mueller11}, HPF is the abbreviation for high pass 
         filter, pixfrac is the ratio of drop size to input pixel size 
         used for the drizzling algorithm \citep{fruchter02} within the 
         {\sf photProject()} mapper, outpix is the output pixel size in the 
         final map, N$_{\rm aper}$ is the number of output pixels inside the 
         photometry aperture with r$_{\rm aper}^{phot}$ , and f$_{\rm corr}$
         is the correlated noise correction factor depending on the 
         combination of HPF radius / pixfrac / outpix. 
}             
\label{table:scanmapparamsphotnoise}      
\centering
    \begin{tabular}{r c c c c c c c c}
   \hline\hline
            \noalign{\smallskip}
Filter  & r$_{\rm aper}^{phot}$ & c$_{\rm aper}$ & cc$_{\rm star}$($\lambda_{\rm ref}$) & HPF radius\tablefootmark{a} & pixfrac & outpix  & N$_{\rm aper}$ & f$_{\rm corr}$ \\
($\mu$m)&       (")          &              & &            &         &     (")     &              &         \\
            \noalign{\smallskip}
    \hline
             \noalign{\smallskip}
   70  &  5.6 & 1.61 & 1.016 &  15 & 1.0 & 1.1 & 81.42 & 3.13 \\
  100  &  6.8 & 1.56 & 1.033 &  20 & 1.0 & 1.4 & 74.12 & 2.76 \\
  160  & 10.7 & 1.56 & 1.074 &  35 & 1.0 & 2.1 & 81.56 & 4.12 \\
            \noalign{\smallskip}
\hline
    \end{tabular}
\tablefoot{
\tablefoottext{a}{This parameter determines the elementary section of a scan 
                  over which the high-pass filter algorithm computes a running 
                  median value. Its unit is "number of read-outs". The spatial 
                  interval between two readouts is $\alpha_{\rm ro} = 
                  \frac{v_{\rm scan}}{\nu_{\rm ro}}$. For the standard 
                  $\nu_{\rm ro}$ = 10\,Hz read-out scheme in PACS prime mode, 
                  and a scan speed $v_{\rm scan}$ = 20"/s,~the spatial
                  interval $\alpha_{\rm ro}$ between two read-outs corresponds 
                  to 2". The entire width of the HPF window (") = [(2 $\times$ HPF radius) + 1] 
          $\times~\alpha_{\rm ro}$.}
}
\end{table*}

\subsection{Data analysis and calibration}

The data reduction and calibration performed in HIPE\footnote{HIPE is a joint 
development by the Herschel Science Ground Segment Consortium, consisting of 
ESA, the NASA Herschel Science center, and the HIFI, PACS and SPIRE 
consortia.}~\citep{ott10} followed the description in \citet{balog14}, applying
the high-pass filter (HPF) algorithm to remove the $\frac{1}{f}$-noise from 
the scan data of the bolometer detectors. A few recent developments in PACS 
data reduction (gyro correction and updated pointing products, refined focal 
plane geometry calibration and more precise timing of the detector readouts) 
have been included.

The source flux is determined by aperture photometry. The relation between the 
final stellar flux at the reference wavelength of the respective filter (70, 
100 and 160\,$\mu$m), f$_{\rm star}(\lambda_{\rm ref})$, and the integrated
background subtracted map flux inside the aperture, f$_{\rm aper}$, is given by
\begin{equation}
\label{eq:apercolcorr}
f_{\rm star}(\lambda_{\rm ref}) = \frac{c_{\rm aper}(\lambda_{\rm ref}) \times
  f_{\rm aper}(\lambda_{\rm ref})}{cc_{\rm star}(\lambda_{\rm ref})} 
,\end{equation}
where c$_{\rm aper}$ is the aperture correction factor to get the total 
non-colour-corrected source flux, f$_{\rm tot}$. Since the PACS calibration 
scheme yields a flux related to a SED $\nu \times f = const.$ the 
colour-correction factor cc$_{\rm star}$($\lambda_{\rm ref}$) provides the 
appropriate correction for the stellar SED (5000\,K blackbody). The aperture 
and colour-correction factors are listed in 
Table~\ref{table:scanmapparamsphotnoise}.

For the investigation of background contamination we also used the JScanam 
algorithm~\citep{gracia15}, which better preserves extended emission. For the 
final projection of all data, the HIPE algorithm {\sf photProject()} was 
applied; the selected mapping parameters pixfrac and output pixel size are 
listed in Table~\ref{table:scanmapparamsphotnoise}.

\subsection{Optimum aperture size for faint star photometry}

For the faint star photometry we have selected smaller apertures
(cf.\ Table~\ref{table:scanmapparamsphotnoise}) than were used for the 
fiducial star photometry in \citet{balog14} (12\arcsec, 12\arcsec, 22\arcsec, 
respectively). These are the same aperture sizes as for chop-nod photometry.

These smaller apertures, which are adapted to the PSF FWHM in the respective 
filter, result in a much higher flux reproducibility among the individual 
measurements and hence a smaller standard deviation of the mean source flux,
as well as more reliable and consistent (with regard to the relative spectral 
shape) source flux measurements for the faintest sources. This is shown in 
Table~\ref{table:photminiscanapercom} in Appendix~\ref{sect:compapersize}, 
where photometry with the large standard apertures is compared with the 
photometry applying the smaller apertures. For the cases with $\ge$4 
individual measurements the improvement in reproducibility can be up to a 
factor of 2 -- 3. The finally achieved average reproducibility for sources 
with eight individual measurements in each filter is listed in 
Table~\ref{table:photreproducibility}. 

%
%

\begin{table}
\caption{Average photometric reproducibility and its standard deviation 
for the six brightest stars with at least eight individual measurements per 
filter.
}             
\label{table:photreproducibility}      
\centering
    \begin{tabular}{r c}
   \hline\hline
            \noalign{\smallskip}
Filter  & Photometric reproducibility \\
($\mu$m)&      (\%)        \\
            \noalign{\smallskip}
    \hline
             \noalign{\smallskip}
   70  &  0.23 $\pm$ 0.15 \\
  100  &  0.57 $\pm$ 0.63 \\
  160  &  1.85 $\pm$ 1.80 \\
            \noalign{\smallskip}
\hline
    \end{tabular}
\end{table}

From a statistical analysis of the signals of the approximately 21,000 PACS 
photometer internal calibration source measurements, \citet{moor14} derived 
a stability of the PACS bolometer response of about 0.2\% standard deviation 
or 2\%, 3\%, and 5\% peak-to-peak at 70, 100, and 160\,$\mu$m, respectively, 
after correction for evaporator temperature effects and initial signal drifts 
after cooler recycling and photometer switch-on. Our photometry includes the 
evaporator temperature correction and practically all measurements are outside 
phases with noticeable initial signal drifts. The mean reproducibility of the 
70\,$\mu$m stellar fluxes comes close to the standard deviation of the 
bolometer response. At 100 and 160\,$\mu$m the mean reproducibility is less 
good and shows a larger scatter, firstly because the sources are weaker and 
secondly because the uncertainties in background subtraction are higher.

\subsection{Noise and S/N determination}
\label{sect:noise_sn_determination}

\begin{figure}[h!]
   \centering
   \includegraphics[width=0.48\textwidth]{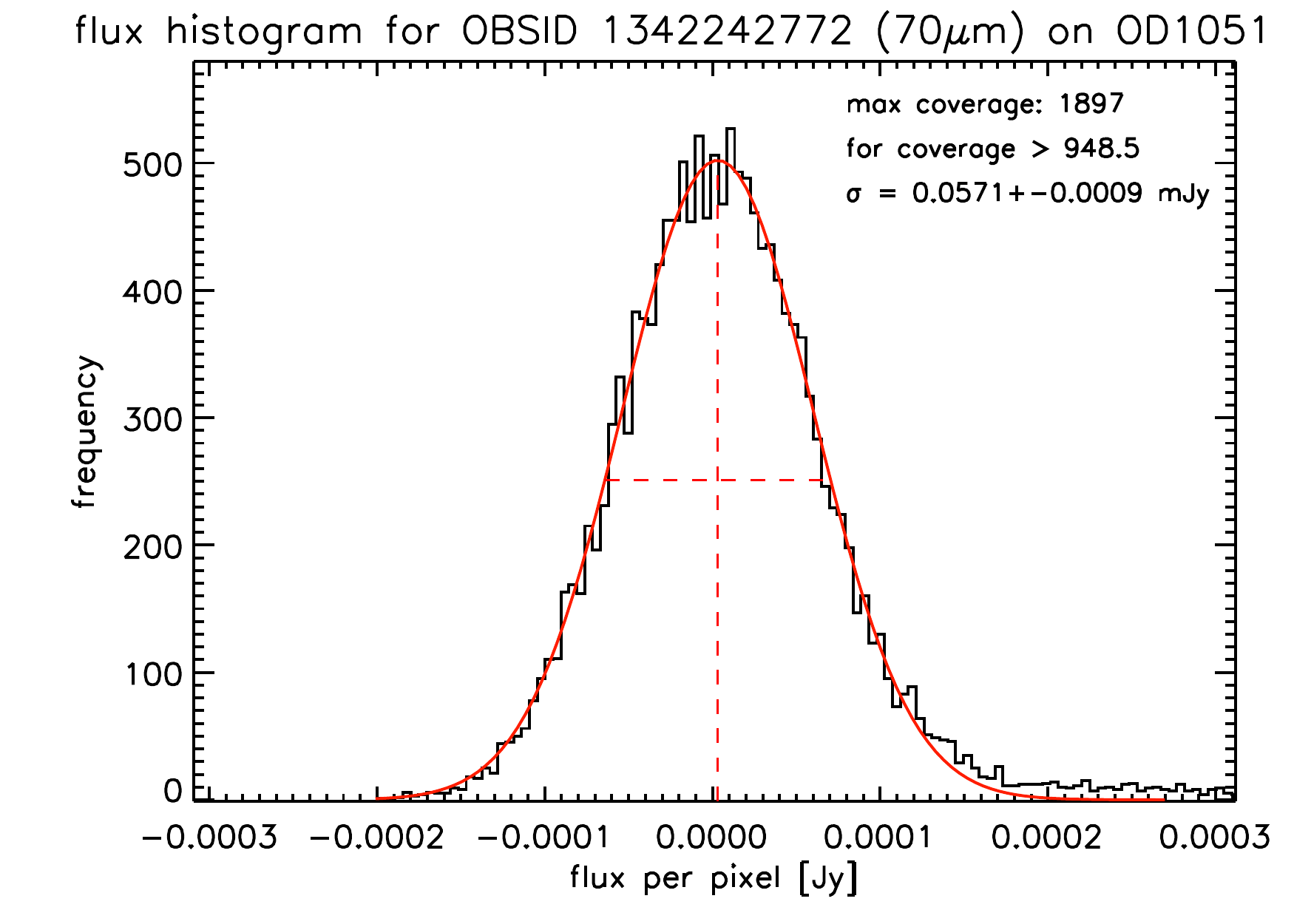}
      \caption{Illustration of the histogram method to determine
               the background noise. The example shows the number of pixels per
               flux bin of the 70\,$\mu$m map of OBSID 1342242772 
               ($\beta$\,Gem on OD\,1051) for all pixels with a coverage value 
               $>\,\frac{1}{2}~cover_{\rm max}$ ($>$948.5). The displayed flux 
               distribution is cut off towards higher fluxes. The red curve is 
               the Gaussian fit to this histogram. For this fit we took all 
               bins left of the distribution maximum into account, but limited 
               the right fitting range to ten bins beyond the distribution 
               maximum in order to avoid a bias of the fitted width by true 
               source flux. The vertical and horizontal red dashed lines 
               indicate the mean background level and the $FWHM =
               \sqrt{2\,log(2)}~\sigma_{\rm pix}$, respectively. For fluxes
               per pixel above $\approx$0.12\,mJy the contribution by true 
               sources becomes noticeable.  
              }
         \label{fig:noise_det_histo}
   \end{figure}

A flux histogram has been constructed for all output pixels of the image map, 
where the corresponding coverage map\footnotemark[2]\footnotetext[2]{The 
coverage map gives the sum of all complete (=~1.0) or partial ($<$~1.0) 
coverage occurrences of each map output pixel by any physical array pixel, 
reduced to the specified drop size, from all unmasked read-out frames 
along the scan time-line.} indicates that $cover_{\rm pix} \gtrsim 
\frac{1}{2}~cover_{\rm max}$. This is justified, since the stars are located
in the central part of the map around the highest coverage. A Gauss fit has 
been performed to the histogram but restricted to the part with fluxes below 
the bin associated with the maximum number, representing in first approximation
the background level, and hence avoiding contamination of the derived noise 
per pixel, $\sigma_{\rm pix}$, by flux of faint sources (to optimize the
quality of the fit, actually about 10 bins above the bin with the maximum 
number are included in the fit). An example of this procedure is shown in 
Fig.~\ref{fig:noise_det_histo}. This method provides very reliable and 
homogeneous noise figures.

%
%

For our photometric measurements, the noise inside the measurement aperture 
must be determined from the noise per pixel $\sigma_{\rm pix}$. This is given by 
\begin{equation}
\label{eq:sigaper}
\sigma_{\rm aper} = \sqrt{N_{\rm aper}} \times \sigma_{\rm pix}
,\end{equation}
with N$_{\rm aper}$ being the number of map output pixels inside the
measurement aperture. The respective numbers of N$_{\rm aper}$ are listed in 
Table~\ref{table:scanmapparamsphotnoise}.

The high pass filtering and map projection lead to correlated noise which must 
be corrected to reconstruct the real detector noise~\citep{popesso12}. This is 
achieved by applying the correlated noise correction factor 
f$_{\rm corr}$\footnotetext[3]{$c_{ijk}$ is related to the 20 parameters P(0) 
$\ldots$ P(19) in Table 9 of \citet{popesso12} by running three nested DO-loops with (from outer to inner) k = 0, n; j = 0, (n-k); and i = 0, (n-k-j).}
\begin{equation}
f_{\rm corr} = \sum_{0 \le i+j+k \le n}^{n=3}\,c_{ijk}\,hpf^i\,outpix^j\,pixfrac^k \\
\end{equation}
\begin{equation*}
k = 0,n;~j = 0,(n-k);~i = 0,(n-k-j)~\footnotemark[3].
\end{equation*}

Hence, the noise corrected for correlated noise inside the measurement 
aperture is
\begin{equation}
\label{eq:sigapercorr}
\sigma_{\rm aper,corr} = \sqrt{N_{\rm aper}} \times f_{\rm corr} \times \sigma_{\rm pix}
.\end{equation}

The S/N of the measurement is then determined as
\begin{equation}
\label{eq:sn_meas}
\frac{S}{N} = \frac{f_{\rm aper}}{\sigma_{\rm aper,corr}}
,\end{equation}
where f$_{\rm aper}$ is the part of the source flux measured inside the aperture.

\subsection{The dependence of S/N on exposure time and flux}

The measured S/Ns are compared with the S/N predictions of the 
exposure time calculation tool in the {\it Herschel} Observatory Planning 
Tool HSpot \citep{hspot13}. HSpot calculates the S/Ns based on an 
rms noise due to telescope thermal noise emission and the electrical noise
of the read-out electronics, cf.~Table~\ref{table:HSpotrmsnoise}: 

\begin{equation}
\label{eq:sn_hspot}
{\frac{S}{N}}_{\rm HSpot} = \frac{f_{\rm star}}{f_{1\,\sigma,1\,s}} \sqrt{n_{\rm rep}~t_{\rm obscover}},
\end{equation}
with f$_{\rm star}$ using the colour- and aperture-corrected total stellar flux. 

This S/N scales with the square root of the coverage time of the source
during one scan map, t$_{\rm obscover}$ and the number of scan map repetitions, 
n$_{\rm rep}$. For mini-scan-maps, t$_{\rm obscover}$ is maximum at the map centre 
and decreases towards the boarders. In analogy to the noise determination in 
the final maps, as described in Sect.~\ref{sect:noise_sn_determination}, we 
use t$_{\rm obscover}$ = $\frac{1}{2}$~t$_{\rm obscovercent}$.The value of 
t$_{\rm obscover}$ depends on the scan map parameters (scan leg length, scan leg 
separation and number of scan legs) and is listed in 
Table~\ref{table:HSpotcovercentre} for all scan map parameter combinations used 
for our observations.

%
%
\begin{table}
\caption{RMS noise values f$_{1\,\sigma,1\,s}$, the 1\,$\sigma$ ($\frac{S}{N}$ 
= 1) flux level being achievable with an integration time of 1\,s, used in 
HSpot for S/N calculation.
}             
\label{table:HSpotrmsnoise}      
\centering
    \begin{tabular}{r c}
   \hline\hline
            \noalign{\smallskip}
Filter  & f$_{1\,\sigma,1\,s}$ \\
($\mu$m)&      (mJy)        \\
            \noalign{\smallskip}
    \hline
             \noalign{\smallskip}
   70  &  30.6 \\
  100  &  36.0 \\
  160  &  68.5 \\
            \noalign{\smallskip}
\hline
    \end{tabular}
\end{table}

%
%
\begin{table}
\caption{Central coverage time of a source during a scan map
execution depending on the scan map parameters (scan leg length, 
scan leg separation and number of scan legs) as calculated by HSpot. 
The combination in bold face is the default combination used for the 
majority of the measurements.
}             
\label{table:HSpotcovercentre}      
\centering
    \begin{tabular}{c c r}
   \hline\hline
            \noalign{\smallskip}
Scan map parameters & Map angle & t$_{\rm obscovercent}$ \\
   ("/"/\#)         &   (deg)   &         (s)        \\
            \noalign{\smallskip}
    \hline
             \noalign{\smallskip}
\multicolumn{3}{c}{scan speed 20"/s} \\
\hline
 150/4/10  & 70/110 &  80 \\
{\bf 180/4/10}& 70/110 &  90 \\
 210/4/20  &   90   & 220 \\
 210/4/25  &   90   & 275 \\
 240/4/8   & 63/117 &  96 \\
  90/5/9   & 70/110 &  45 \\
 210/15/4  &   90   &  44 \\
\hline
\multicolumn{3}{c}{scan speed 10"/s} \\
\hline
 120/3/21  & 80/100 & 252 \\
 150/4/9   & 85/95  & 144 \\
 210/4/20  &   90   & 440 \\
 210/4/25  &   90   & 550 \\
  90/5/9   & 60/120 &  90 \\
 120/5/9   & 80/100 & 108 \\
 210/15/4  &   90   &  88 \\
            \noalign{\smallskip}
\hline
    \end{tabular}
\end{table}

Figure~\ref{fig:scanmap_sn_time1} shows the dependence of the achieved S/Ns on 
time, represented as number of scan repetitions, and the comparison with the 
HSpot prediction. This includes combined maps of scan and cross-scans, which 
have the sum of the scan repetitions of the individual maps. 

For 70 and 100\,$\mu$m measurements we find S/N $\propto~\sqrt{n_{\rm rep}}$. 
There are deviations from this relation in that respect that the S/N of the 
combined maps is higher than the expected factor of $\sqrt{2}$ by a few 
percent. The ratio of the average measured S/N to the HSpot prediction is 
1.14--1.22 at 70\,$\mu$m and 1.03--1.09 at 100\,$\mu$m, respectively. Given 
the fact that the HSpot prediction is for half maximum coverage and the noise 
determination in the maps is above a threshold of half maximum coverage, the 
measured S/N can be considered as consistent with the HSpot prediction.  

For the 160\,$\mu$m measurements we find for small repetition numbers
($n_{\rm rep} \le$ 12) that the S/N increases with the $\sqrt{n_{\rm rep}}$ for 
single and combined maps. For higher repetition numbers it is obvious that the 
increase of the measured S/Ns is flatter. This flattening is caused by 
confusion noise, which will be discussed in the following Section. The ratio 
of the average measured S/N to the HSpot prediction is around 0.80. We note, 
however, that there is some margin in achievable S/N depending on the selection
of the high-pass filter (HPF) radius. We calculate a decrease of the resulting 
noise by $\approx$23\% between HPF radius = 40 read-outs and HPF radius = 15 
read-outs for pixfrac = 1.0 and output pixel size of 2\farcs0 according to the 
formalism in~\citet{popesso12}. The latter harsh filter width would only be 
applied for maps with very weak sources covering only a few pixels. The HSpot 
values were derived from cosmological fields, where harsh HPF filter widths 
could be applied, and that would explain the somewhat worse performance for 
our milder HPF radius of 35\,read-outs.   
 
After OD\,1375, half of the red photometer array was lost. We have one good
case, namely the 160\,$\mu$m observations of $\varepsilon$\,Lep from OD\,1377, 
to check the performance relative to full array observations. With regard to
comparable observations from ODs\,502, 833, and 1034 
(cf.\ Table~\ref{table:scanmapphotred10_1}), we find the following:
The coverage is 0.51 of the full array map, but the noise is increased
by only a factor of 1.21.

In the case of $\eta$\,Dra the performance of scan speeds 10"/s and 20"/s
can be inter-compared. While the coverage time of the 10"/s scan speed 
maps is always greater than or equal to twice the coverage time of the 20"/s
scan speed (cf.\ configurations in Tables~\ref{table:scanmapphotblue6_1} 
to~\ref{table:scanmapphotred10_1} and corresponding coverage times in 
Table~\ref{table:HSpotcovercentre}),   the measured S/N of the 20"/s 
scan speed is above the HSpot prediction at 70 and 100\,$\mu$m, while the 
measured S/Ns of all 10"/s scan speed combinations are below the HSpot 
prediction. This is a clear demonstration that the 20"/s scan speed maps are 
relatively more sensitive than their 10"/s scan speed counterparts. At 
160\,$\mu$m this is even more pronounced; the S/N of the 20"/s scan speed map 
with half of the coverage time is better than that of the 10"/s scan speed map 
with otherwise identical map parameters.    

Figure~\ref{fig:scanmap_sn_flux} shows the dependence of the achieved S/Ns on 
flux. For the 70 and 100\,$\mu$m filters we can verify that the S/N scales 
linearly with flux over two decades of flux and at least down to total 
(aperture corrected) source fluxes of 30\,mJy and 18\,mJy, respectively. For 
the 160\,$\mu$m filter the linearity with flux can be verified over about one 
decade in flux down to a total (aperture corrected) source flux of 85\,mJy for 
repetition factors 1 -- 12. For fainter fluxes measurements with higher 
($\ge$\,20) repetition factors are necessary to achieve a S/N which is 
sufficiently above values close to the detection limit (S/N $\lesssim$ 1.5, 
cf.\ Sect.~\ref{sect:detlimits}). These high repetition factors give an 
increase in S/N that is smaller than expected, which we explain in the 
following Section as being due to a confusion noise contribution. Since the 
confusion noise contribution is not the same in the different source fields, 
the linearity of the S/N with flux cannot be verified any more 
straightforwardly in the 160\,$\mu$m flux range below 85\,mJy.

\subsection{\bf Impact on S/N by background confusion noise}

In particular at 160\,$\mu$m, there may be another relevant noise source, 
which is FIR background confusion noise. This is composed of a cosmic infrared 
background component and a galactic cirrus component. Examples of background 
confusion, which also affects the source photometry, are shown in 
Figs.~\ref{fig:hd159330maps} and~\ref{fig:hd148387cirrusmaps}. The confusion 
noise is independent of on-source observation time, that is, in the case of 
approaching the confusion noise limit, the S/N does not improve anymore with 
on-source observation time.

\begin{equation}
\label{eq:sn_hspot_conf}
{\frac{S}{N}}_{\rm HSpot,conf} = \frac{f_{\rm star}}{\sqrt{\frac{f_{1\,\sigma,1\,s}^{2}}{n_{\rm rep}~t_{\rm obscover}} + f_{\rm conf noise}^{2}}}
.\end{equation}

This leads to the effect that the S/N curve with time flattens out as observed 
for the 160\,$\mu$m S/Ns in Fig.~\ref{fig:scanmap_sn_time1}. For our source 
fields, HSpot returns a 160\,$\mu$m point source equivalent confusion noise 
estimate f$_{\rm conf noise}$ between 1.3 and 1.5\,mJy. The typical on-source 
observation time for repetition factor 1 is 45\,s, resulting in a 160\,$\mu$m 
noise level of 10\,mJy. This is about a factor of 7 higher than the estimated 
confusion noise and only for large scan map repetition factors ($>$10), does 
the confusion noise become significant.

%
   \begin{figure*}[ht!]
   \centering
   \includegraphics[width=0.94\textwidth]{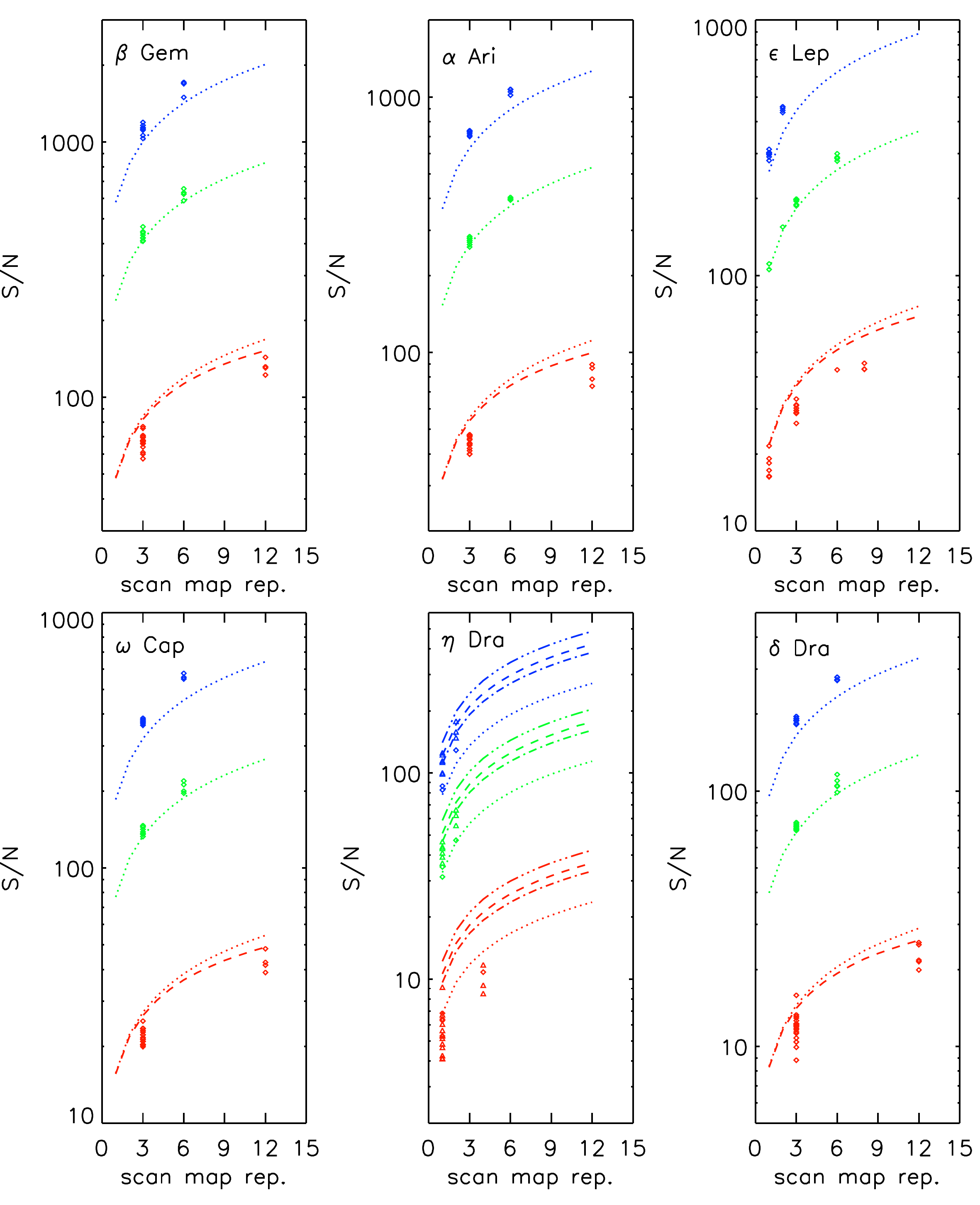}
      \caption{Measured S/Ns for mini-scan-map photometry depending on the 
               number of repetitions. Blue, green, and red symbols represent 
               measurements in the three filters, 70, 100, and 160\,$\mu$m, 
               respectively. Diamond symbols indicate a scan speed of 20"/s, 
               triangles a scan speed of 10"/s. The dotted lines in the 
               respective colours show the S/N prediction by the PACS exposure 
               time calculator of the {\em Herschel} observation planning tool 
               HSpot for the measured colour corrected stellar flux. Long 
               dashed red lines indicate the S/N prediction including confusion
               noise. An exception is the panel of $\eta$\,Dra, where the sets 
               of four dotted, dashed, and dashed-dotted lines represent the 
               sensitivity predictions for four different map parameter 
               combinations; the upper three are with 10"/s scan speed, the 
               lowest one is with 20"/s scan speed. For more details, see text.
               }
         \label{fig:scanmap_sn_time1}
   \end{figure*}

\addtocounter{figure}{-1}

%
   \begin{figure*}[ht!]
   \centering
   \includegraphics[width=0.94\textwidth]{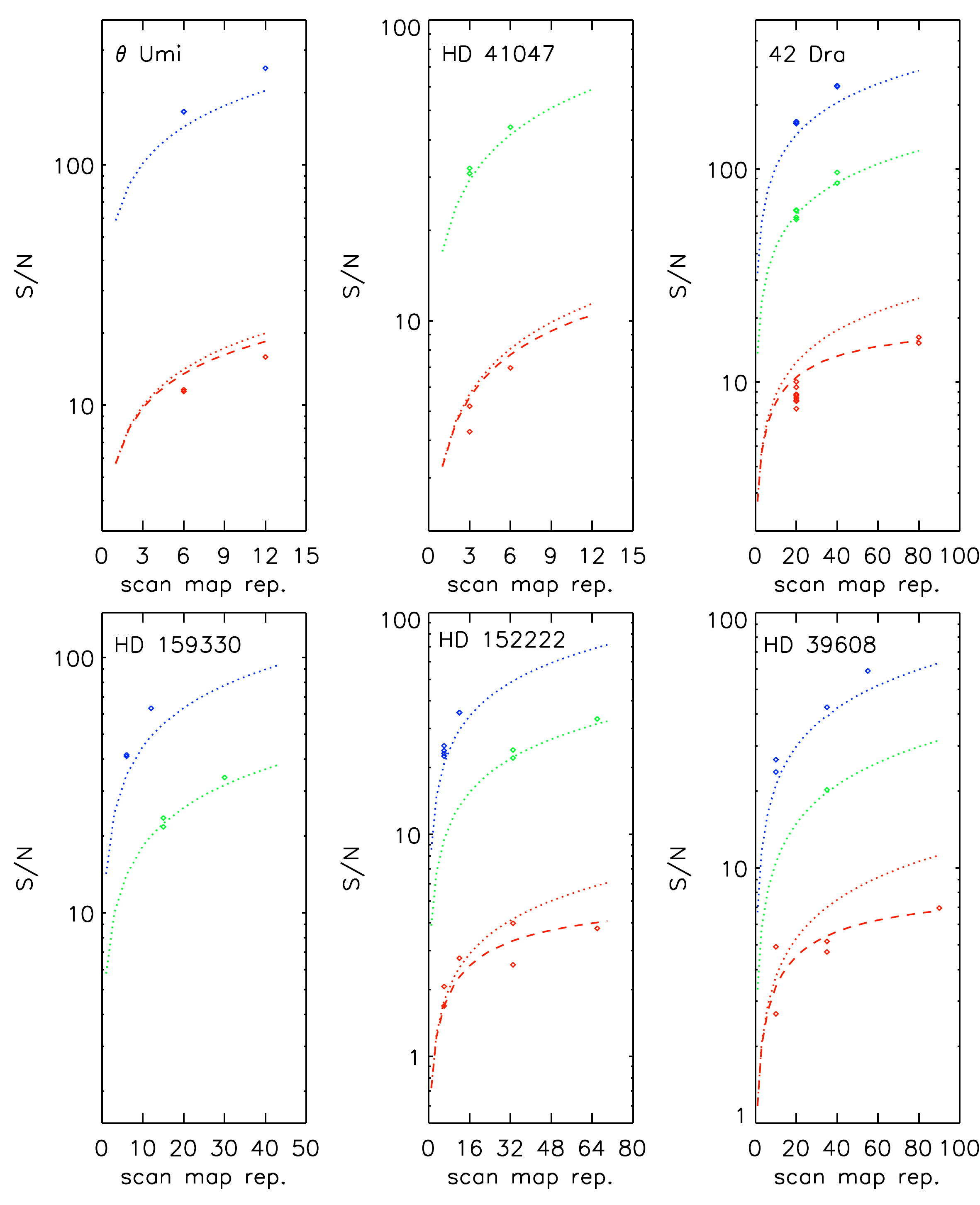}
      \caption{continued. Measured S/Ns for mini scan map photometry
               depending on the number of repetitions.
              }
         \label{scanmap_sn_time2}
      \vspace{2.0cm}
   \end{figure*}

%
   \begin{figure*}[ht!]
   \centering
   \includegraphics[width=0.50\textwidth]{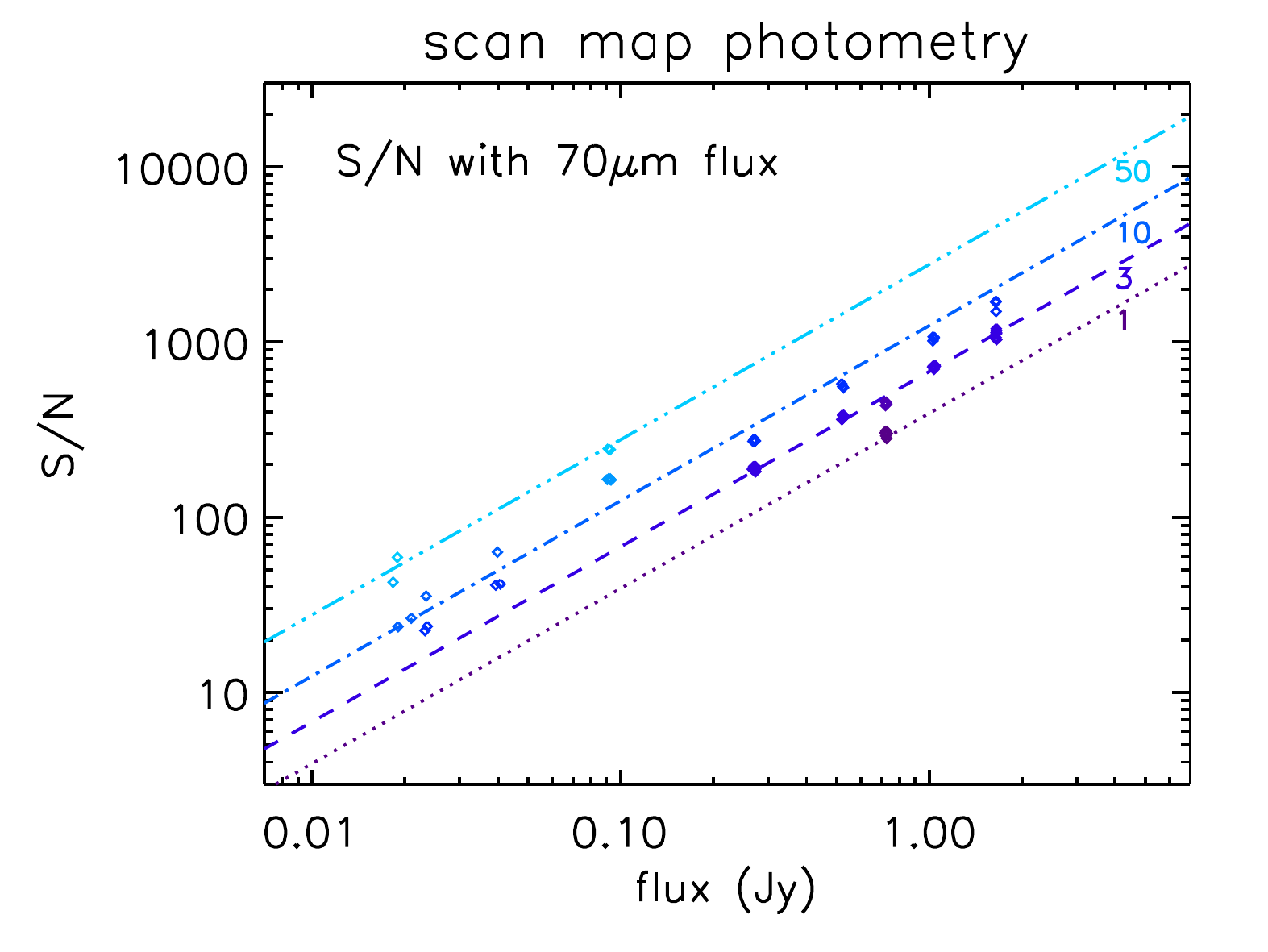}
   \includegraphics[width=0.50\textwidth]{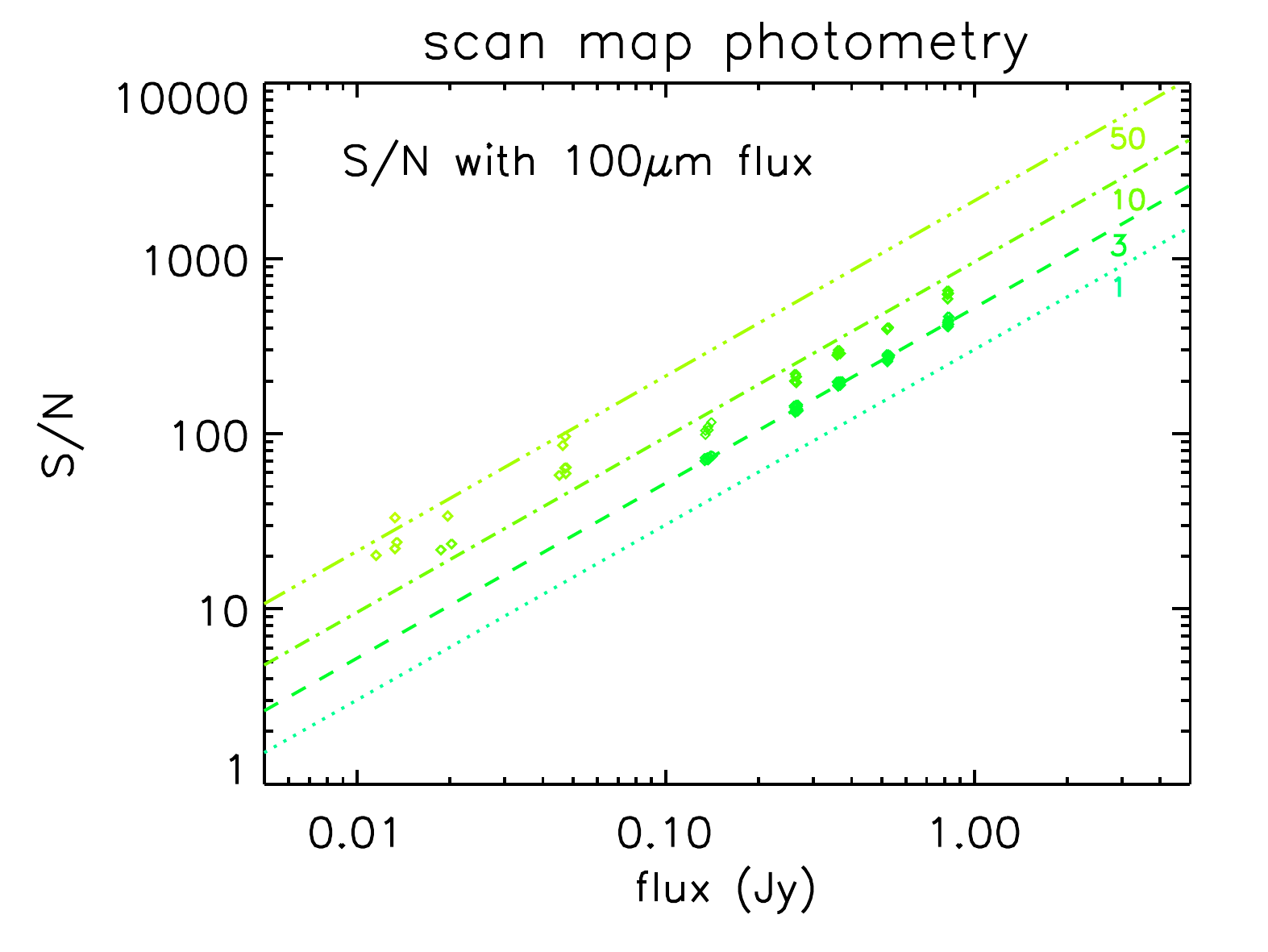}
   \includegraphics[width=0.50\textwidth]{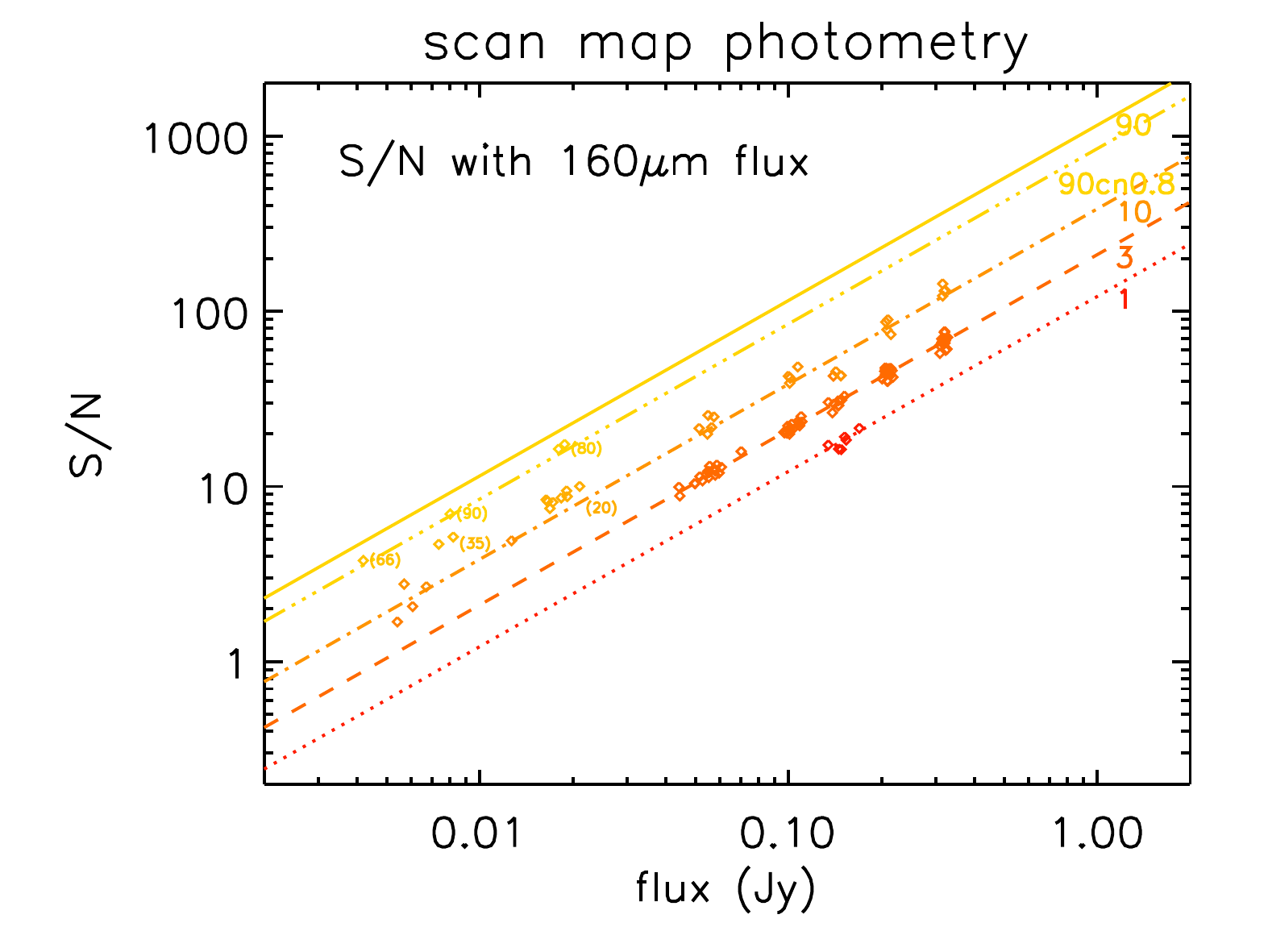}
      \caption{Measured S/Ns for mini-scan-map photometry depending on the 
               source flux (Note: fluxes measured inside the aperture are used 
               here). For better comparability only measurements with an
               observational set-up identical with the final mini-map set-up 
               (ten 180\arcsec scan legs with 4\arcsec separation and scan 
               speed 20\arcsec\,/\,s) are considered. Lighter colour tones are 
               measurements with higher scan map repetition factors. We note 
               that here the dotted, dashed, and dashed dotted lines in 
               different colour tones do not represent the S/N prediction by 
               the PACS exposure time calculator of the {\em Herschel} 
               observation planning tool HSpot, but are empirical adjustments 
               to the average measured S/N for the respective scan map 
               repetitions. In the 160\,$\mu$m panel, numbers in parentheses 
               mark measurements with high repetition factors whose S/N is 
               degraded by confusion noise. This is also indicated by two S/N 
               with flux lines for repetition factor 90, where the lower one 
               includes additional confusion noise (cn) of 0.8\,mJy.
              }
         \label{fig:scanmap_sn_flux}
   \end{figure*}

%
   \begin{figure*}[ht!]
   \centering
   \includegraphics[width=0.4\textwidth]{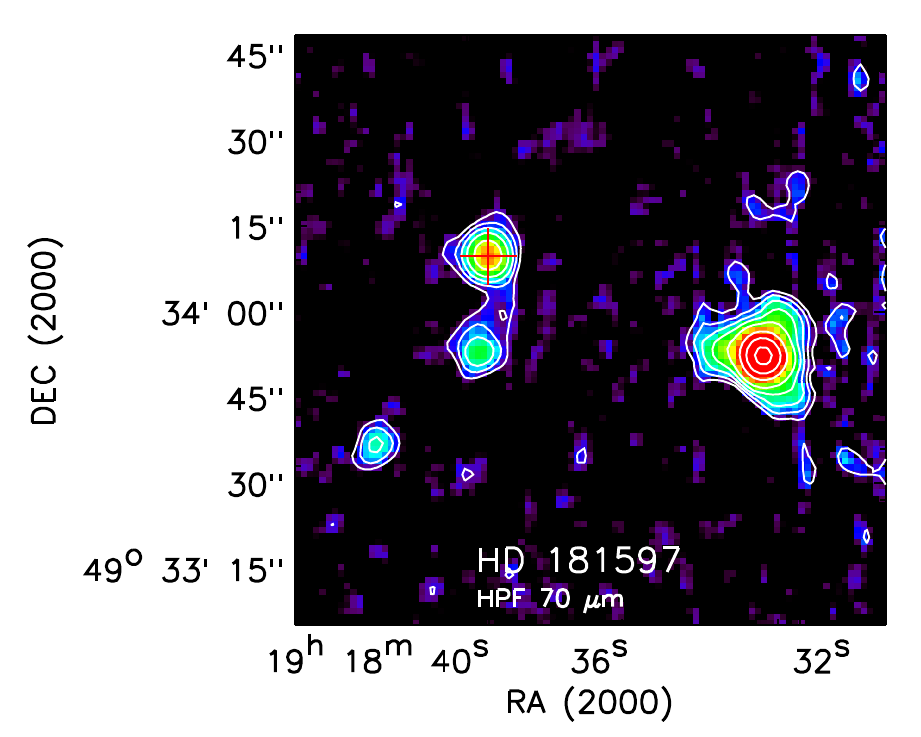}
   \includegraphics[width=0.4\textwidth]{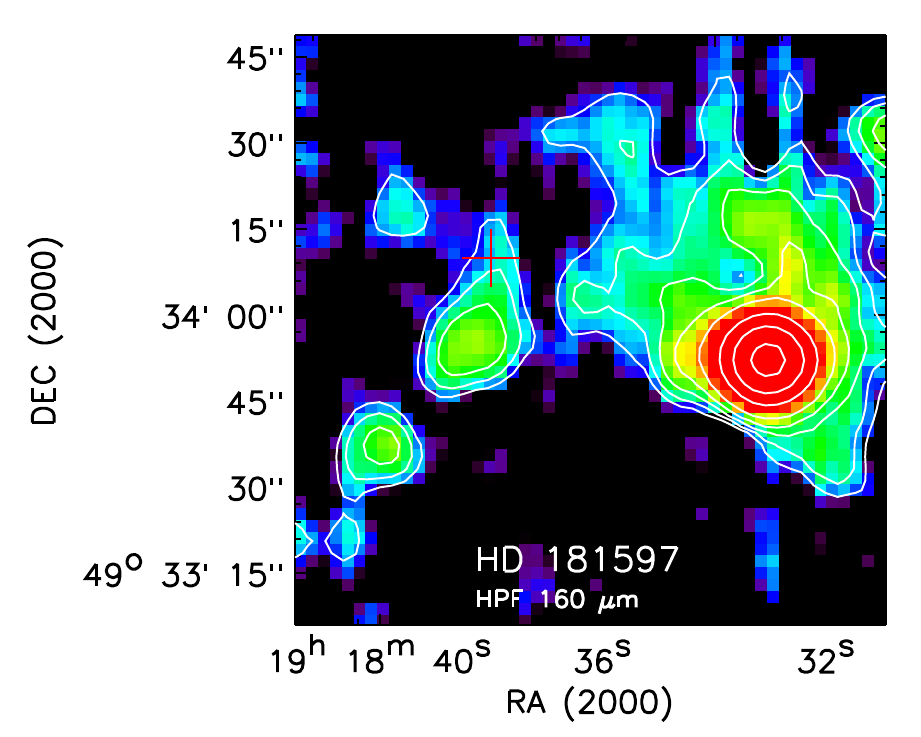}
   \includegraphics[width=0.4\textwidth]{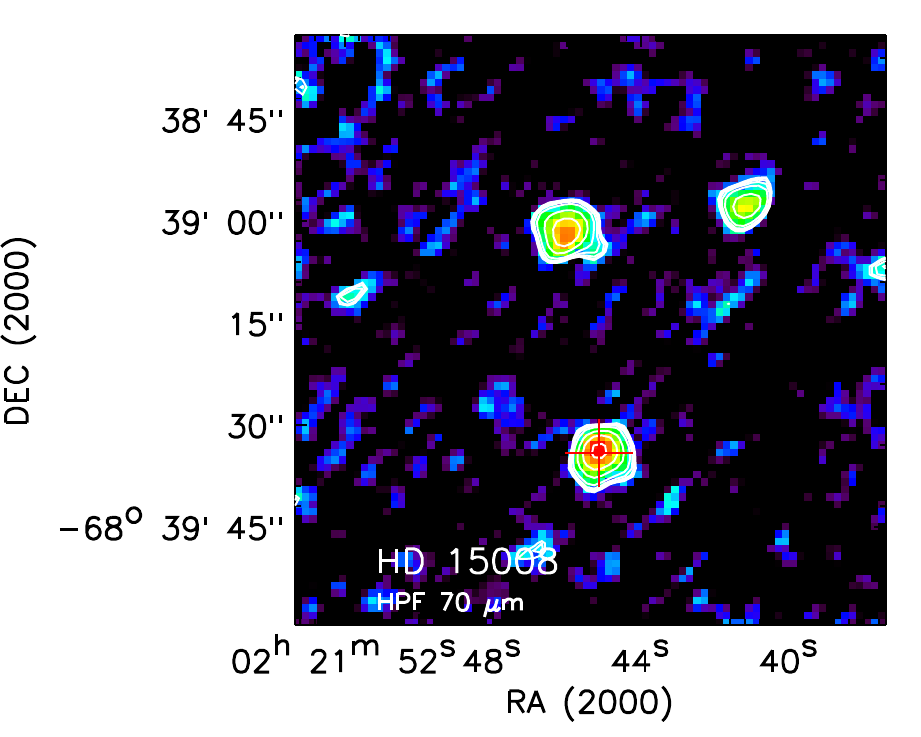}
   \includegraphics[width=0.4\textwidth]{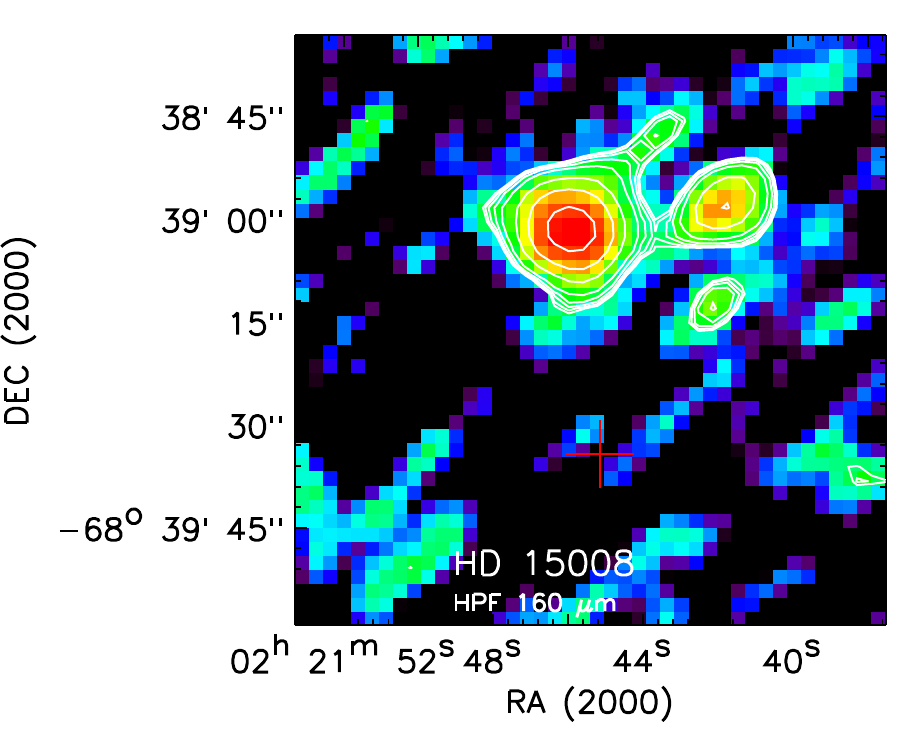}
      \caption{Examples of detection at 70\,$\mu$m and non-detection at 
               160\,$\mu$m. The red cross (arm length equal to 5\arcsec) 
               indicates the common {\em Herschel} position. We note that for 
               both sources there exists only one scan map orientation which 
               leads to some residual scan artefacts, in vertical and diagonal 
               direction, respectively; see 
               Table~\ref{table:scanmapphotblue6_3} for details.
               Top: HD\,181597, OBSID 1342185451 on OD\,146. The bright point 
               source at the right is KIC\,11555225. Bottom: HD\,15008 
               ($\delta$\,Hyi), OBSID 1342189130 on OD\,241.
              }
         \label{fig:detlimit160}
   \end{figure*}

\subsection{Detection limits}
\label{sect:detlimits}

HD\,181597 and HD\,15008 ($\delta$\,Hyi) are good examples for non-detections 
at 160\,$\mu$m, because the expected source flux is below the detection limit. 
Both sources have a clear detection at 70\,$\mu$m (S/N = 30 and 15, 
respectively), which allows to identify the expected source position on the 
160\,$\mu$m maps (see Fig.~\ref{fig:detlimit160}). Table~\ref{table:detlimits} 
lists the determined S/Ns, which are $<$1.5 (we note, that the S/N measurement 
in the map is actually higher by the factor f$_{\rm corr}$ = 4.12, which 
corresponds to the classic S/N detection limit of 5 -- 6). This is in 
accordance that no source can be detected at the location of the star in the 
160\,$\mu$m map.

%
%
\begin{table}[h!]
\caption{S/N determination at 160\,$\mu$m for HD\,181597 and HD\,15008,
which are below the detection limit (S/N $\lesssim$ 1.5) at this
wavelength.
}             
\label{table:detlimits}      
\centering
    \begin{tabular}{r c c c c}
   \hline\hline
            \noalign{\smallskip}
    HD  & $f_{\rm predict}$ &  $f_{\rm aper}$ & $\sigma_{\rm aper,corr}$ & $\frac{S}{N}$ \\
($\mu$m)&      (mJy)     &      (mJy)     &        (mJy)           &               \\
            \noalign{\smallskip}
    \hline
             \noalign{\smallskip}
181597  &       5.2      &      3.3       &         2.6            &      1.3      \\
 15008  &       4.4      &      2.8       &         3.2            &      0.9      \\
            \noalign{\smallskip}
\hline
    \end{tabular}
\end{table}

\subsection{Confusion by neighboring sources and cirrus}
\label{sect:bgconfusion}

The fainter the star, the higher the probability, in particular at 100 and 
160\,$\mu$m, that nearby sources confuse the source flux inside the measurement
aperture. 

\begin{figure*}[ht!]
   \centering
   \includegraphics[width=0.4\textwidth]{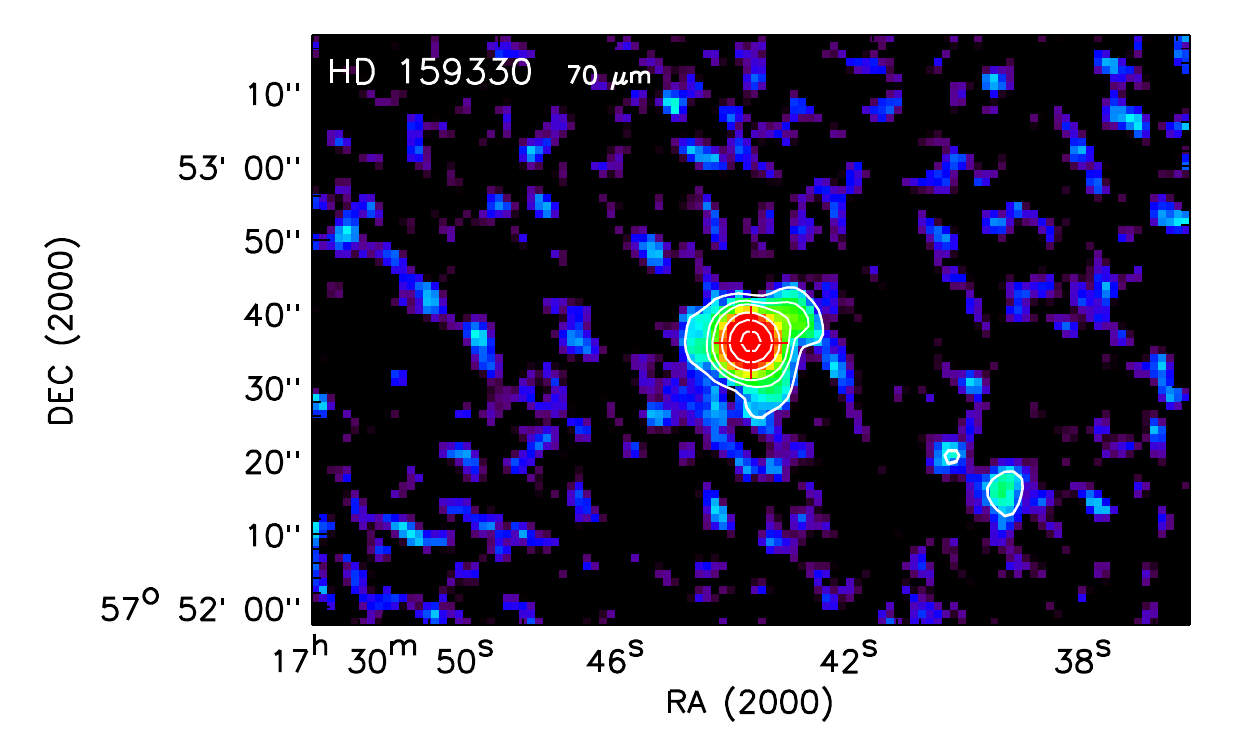}
   \includegraphics[width=0.4\textwidth]{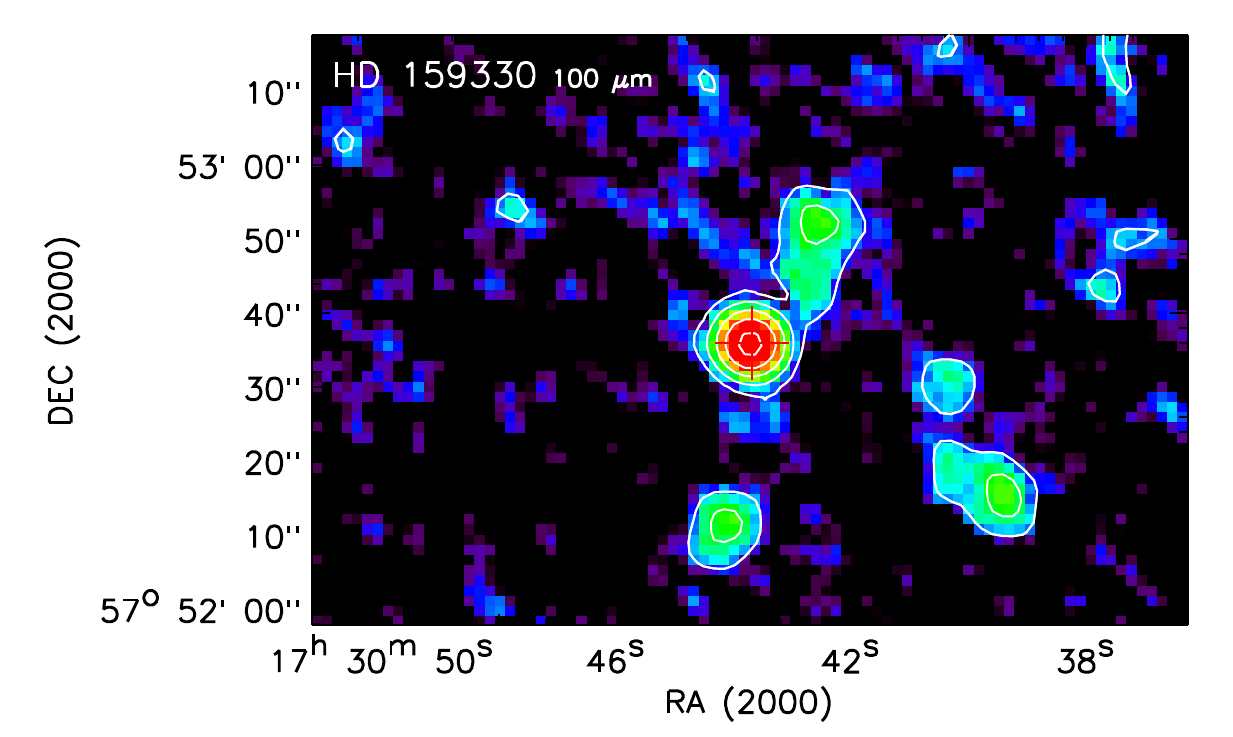}
   \includegraphics[width=0.4\textwidth]{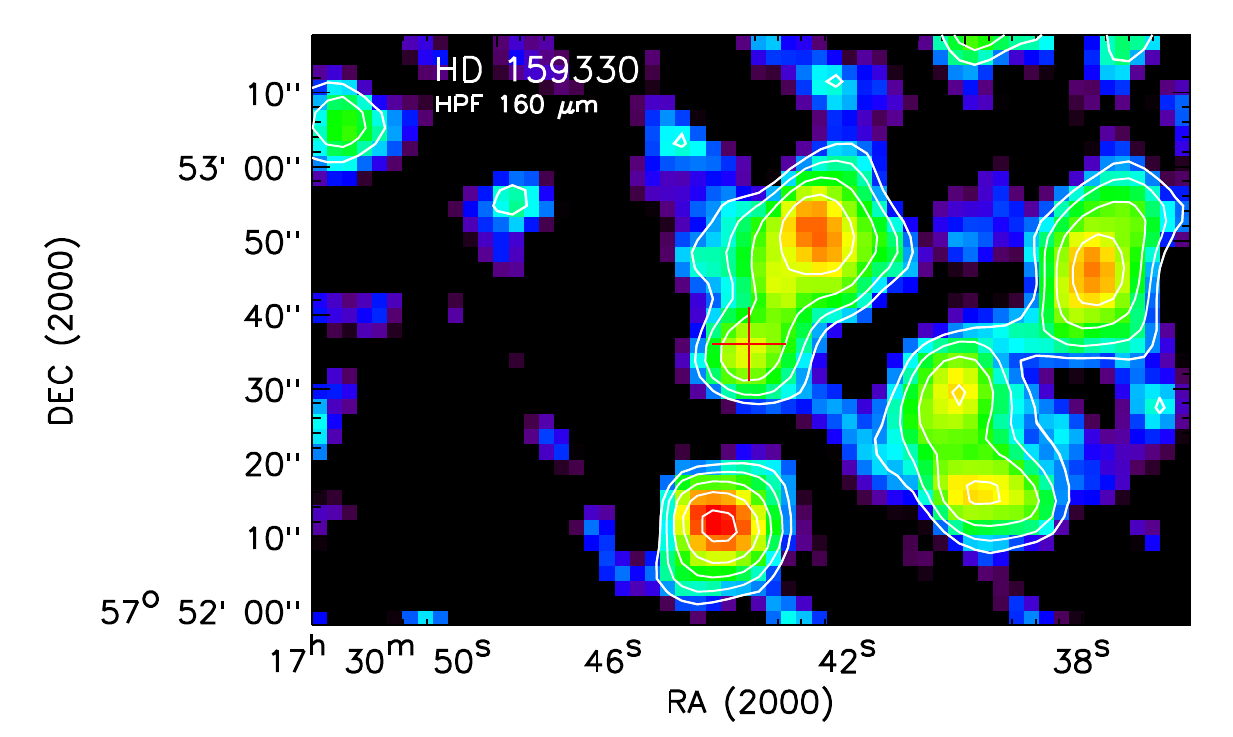}
   \includegraphics[width=0.4\textwidth]{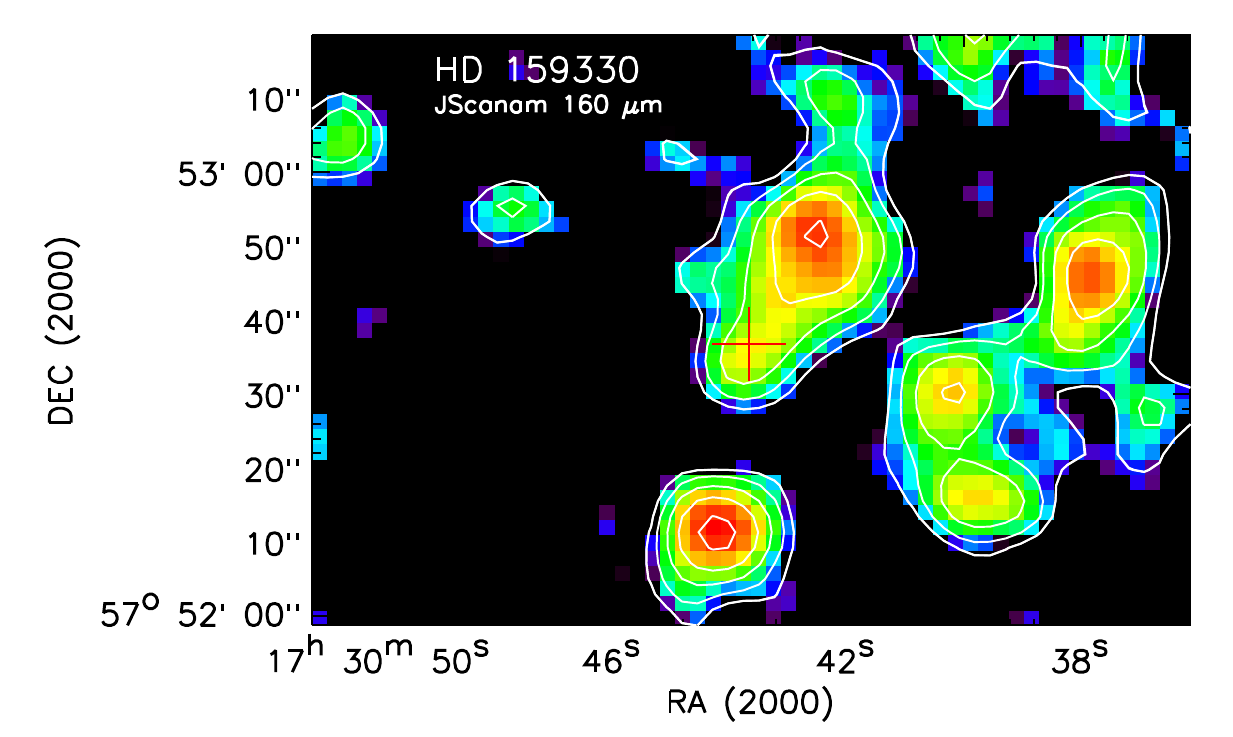}
      \caption{Example of background confusion noise around the star
               HD\,159330 (OBSIDs 1342213583-86 on OD\,628, see 
               Tables~\ref{table:scanmapphotblue6_3} 
               and~\ref{table:scanmapphotgreen6_3} for details) by comparing 
               maps in the three filters 70, 100 and 160\,$\mu$m. At 
               160\,$\mu$m both the high-pass filtered (HPF) and the JScanam 
               maps are shown to explore the nature of the background sources. 
               The red cross (arm length equal to 5\arcsec) indicates the best 
               common {\em Herschel} position of the star after frame centering
               at RA = 17:30:43.69 and DEC = $+$57:52:36.0. 
              }
         \label{fig:hd159330maps}
   \end{figure*}

A clear case of confusion by neighboring sources is shown in 
Fig.~\ref{fig:hd159330maps} for the star HD\,159330. While at 70\,$\mu$m the 
star is more or less the only visible source inside the field, at 100\,$\mu$m 
a small cluster of sources around the star starts to pop up, but the star is 
still the dominant source inside the field. At 160\,$\mu$m, all surrounding 
sources are brighter than the star, which appears only as an appendix of the 
source located north-west of it. Also the local brightness maximum is not as 
well located on the cross as is the case for the stellar images at 70 and 
100\,$\mu$m. It is therefore not possible to obtain reliable photometry for 
HD\,159330 at 160\,$\mu$m. The compactness of the surrounding sources both in 
the HPF and the JScanam image points to an extragalactic nature of the 
confusing sources. This is difficult to verify in the optical, since 
HD\,159330 is a 6.2\,mag (V band) bright star.

%

%
   \begin{figure*}[ht!]
   \centering
   \includegraphics[width=0.33\textwidth]{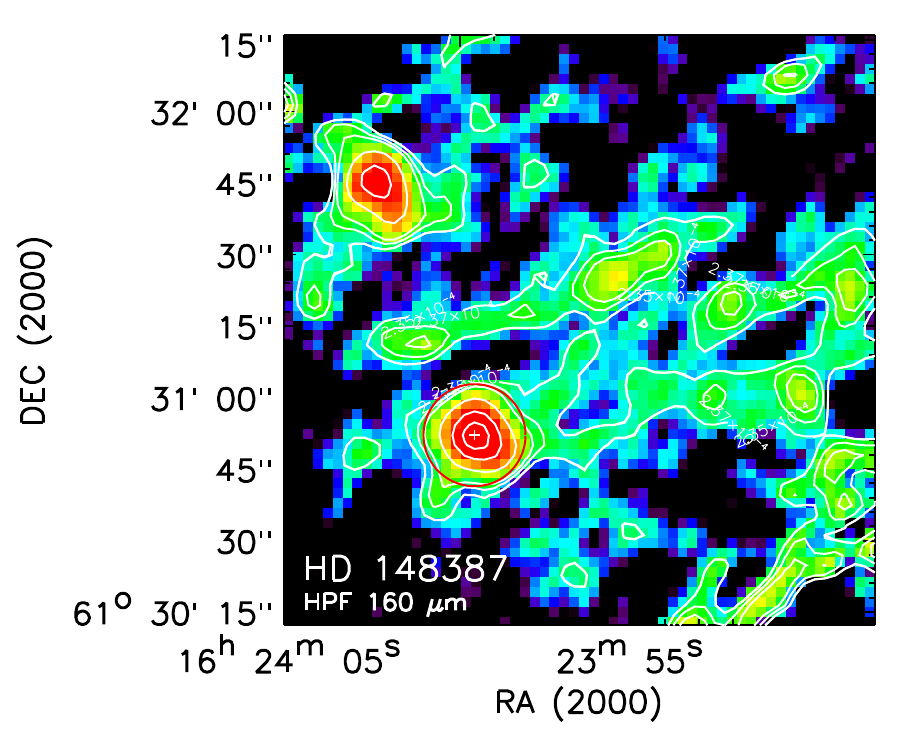}
   \includegraphics[width=0.33\textwidth]{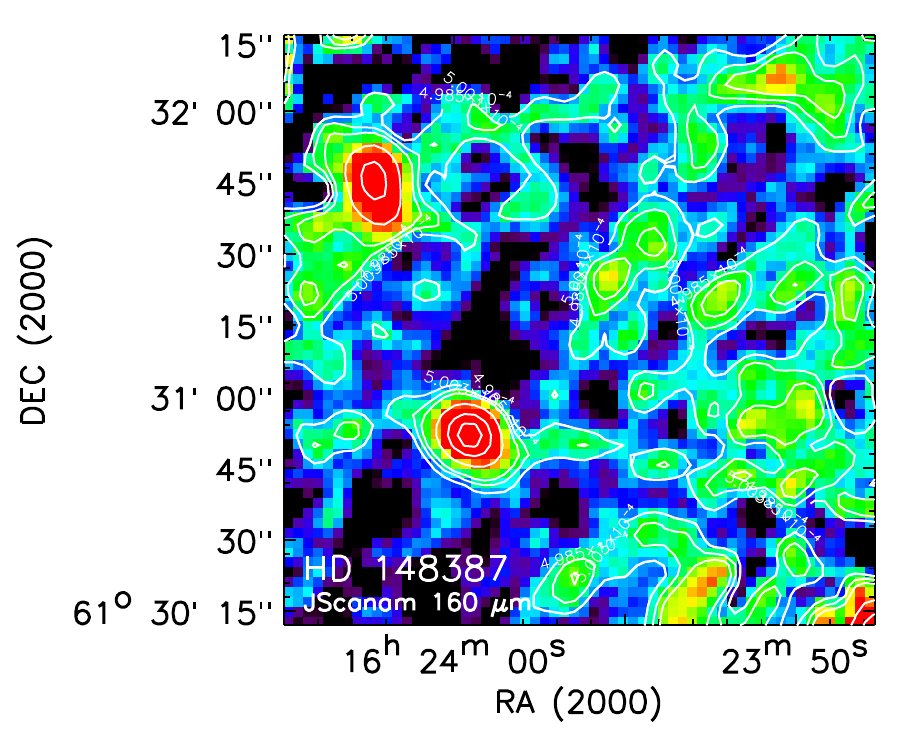}
   \includegraphics[width=0.31\textwidth]{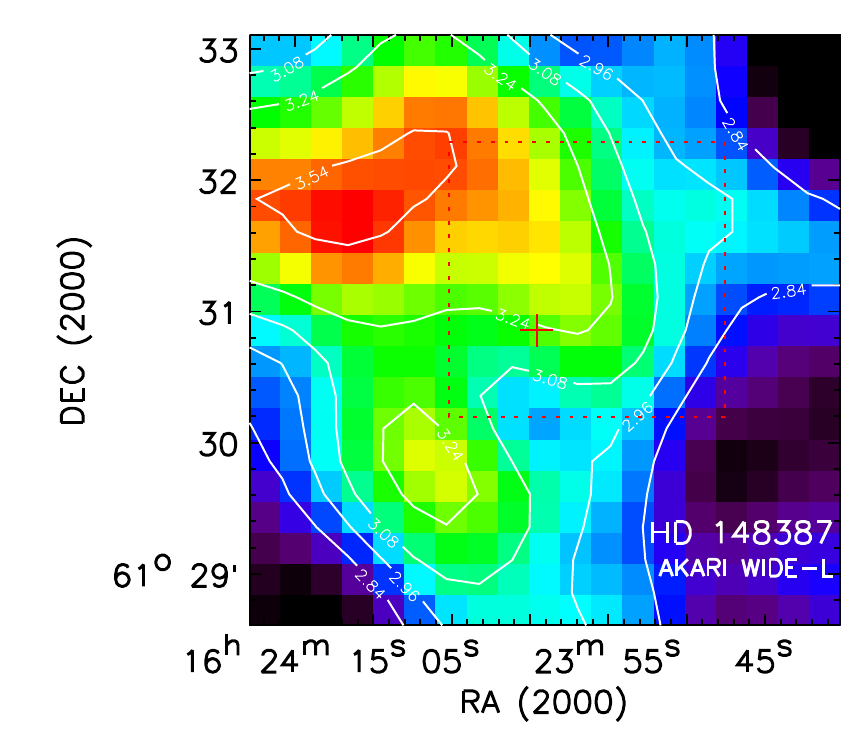}
      \caption{Example of cirrus confusion noise around the star HD\,148387 
               ($\eta$\,Dra, OBSIDs 1342186146, ..47, ..55, ..56 from OD\,160).
               The left panel shows the high-pass filter processed map used 
               for the photometry, the photometric aperture with 10\farcs7 
               radius is indicated by the red circle and a small white cross 
               at its centre. The middle panel shows the Jscanam processed map 
               which reproduces extended emission more reliably. The right panel
               shows the {\em AKARI} WIDE-L (140\,$\mu$m) background emission 
               around the source (red cross), the {\em AKARI} map area is 
               about four times as large as the PACS map area, which is 
               indicated by the red dashed square. 
              }
         \label{fig:hd148387cirrusmaps}
   \end{figure*}

An example of likely cirrus confusion is shown in 
Fig.~\ref{fig:hd148387cirrusmaps} around the star $\eta$\,Dra (HD\,148387). 
There is relatively significant similarity between the HPF and JScanam 
processed map concerning the brighter spots and features, while on the low 
level there are differences, because the HPF algorithm is not designed to 
preserve faint extended emission. Nevertheless both maps indicate a 
filamentary emission around the star. Indeed, $\eta$\,Dra, with l = 92.6$^{o}$ 
and b = $+$41.0$^{o}$ , is located at the edge of the Draco nebula~\citep[cf.\ 
e.g. Fig.~3 in][w.r.t.\ its location]{herbstmeier98}, a pronounced cirrus 
cloud. The extract of the {\it AKARI} Wide-L (140\,$\mu$m) all sky 
map\footnotemark[4]\footnotetext[4]{{\it AKARI} Far-infrared All-Sky Survey 
maps query service \\
http://www.ir.isas.jaxa.jp/AKARI/Archive/Images/FIS\_AllSkyMap/\\search/\\We
use the {\it AKARI} WIDE-L (140\,$\mu$m) maps instead of the 
N160 (160\,$\mu$m) maps, because the latter ones do not have sufficient
S/N over the whole field for illustration of the background structure.}~\citep{doi15} reveals that $\eta$\,Dra is located at the southern edge of a cirrus
knot with an extension of it passing north-west into the PACS map area.  
The cirrus confusion affects the derived 160\,$\mu$m flux noticeably, 
as is discussed in Sect.~\ref{sect:starsfirexcess}. Other cases of suspected 
cirrus confusion are also discussed there. 

%
%
\begin{table*}[ht!]
\caption{Combined mini-scan-map photometry. Model fluxes are listed in 
Table~\ref{table:sources}. The quoted uncertainties of the measurements 
include the absolute calibration uncertainty of 5\%. In the case of only one 
observation for a specific source, the statistical error of this flux 
measurement is used in the uncertainty determination, and in the case of more 
than one observation for a source, as given in columns n$_{70}$, n$_{100}$, and 
n$_{160}$, the standard deviation of the weighted mean as given in 
Table~\ref{table:photminiscanapercom}, column phot\_s, is used in the 
uncertainty calculation.
}             
\label{table:photminiscan}      
\begin{center}                          
\begin{tabular}{r r r c c r c c r c c}        
\hline\hline                 
            \noalign{\smallskip}
  HD   &    Name          &n$_{70}$&f$_{\rm star}$(70\,$\mu$m)  &$\frac{f_{\rm star}(70\,{\mu}m)}{f_{\rm model}}$ &n$_{100}$&  f$_{\rm star}$(100\,$\mu$m) &$\frac{f_{\rm star}(100\,{\mu}m)}{f_{\rm model}}$ &n$_{160}$&  f$_{\rm star}$(160\,$\mu$m) & $\frac{f_{\rm star}(160\,{\mu}m)}{f_{\rm model}}$\\    
       &                  &       &  (mJy)  &                            &        &     (mJy)    &                             &        &   (mJy)    &                             \\  
            \noalign{\smallskip}
\hline                        
            \noalign{\smallskip}
 62509 &  $\beta$\,Gem    &  8 & 2649$\pm$132 & 1.08$\pm$0.08 &  8 & 1284$\pm$64  & 1.08$\pm$0.08 & 16 & 497$\pm$25 & 1.09$\pm$0.08 \\
 12929 & $\alpha$\,Ari    &  8 & 1664$\pm$83  & 0.97$\pm$0.08 &  8 &  820$\pm$41  & 0.99$\pm$0.08 & 16 & 328$\pm$17 & 1.02$\pm$0.08 \\
 32887 &$\varepsilon$\,Lep&  8 & 1166$\pm$58  & 0.99$\pm$0.08 & 12 &  568$\pm$28  & 0.99$\pm$0.08 & 18\tablefootmark{a} & 224$\pm$11 & 1.00$\pm$0.08 \\
198542 &   $\omega$\,Cap  &  8 &  845$\pm$42  & 0.99$\pm$0.08 &  8 &  413$\pm$21  & 0.99$\pm$0.08 & 16 & 161$\pm$8.2 & 0.99$\pm$0.08 \\
148387 &   $\eta$\,Dra    &  8 &  506$\pm$25  & 1.06$\pm$0.06 &  8 &  250$\pm$13  & 1.07$\pm$0.06 & 16 &  98$\pm$7.1 & 1.10$\pm$0.08 \\
180711 & $\delta$\,Dra    & 12 &  436$\pm$22  & 1.02$\pm$0.08 & 10 &  214$\pm$11  & 1.03$\pm$0.08 & 22 &  85$\pm$4.8 & 1.07$\pm$0.08 \\
139669 &   $\theta$\,Umi  &  4 &  284$\pm$14  & 0.99$\pm$0.08 &  2 &  144$\pm$8.8 & 1.03$\pm$0.08 &  2\tablefootmark{b} &  62$\pm$4.0\tablefootmark{f} & 1.16$\pm$0.10 \\ 
 41047 &   HR\,2131       &    &\multicolumn{2}{c}{no measurement} &  2 &   97$\pm$4.9 & 1.01$\pm$0.08 &  2 &  36$\pm$5.9 & 0.97$\pm$0.17 \\
170693 &   42\,Dra        &  4 &  148$\pm$7.4 & 0.96$\pm$0.06 &  4 &   73$\pm$3.8 & 0.97$\pm$0.06 &  8 &  28$\pm$1.7 & 0.96$\pm$0.07 \\
138265 &   HR\,5755       &  4 &  113$\pm$5.7 & 0.97$\pm$0.06 &  6 &   57$\pm$2.9 & 0.99$\pm$0.06 &5\tablefootmark{c}& 31$\pm$1.8 & 1.40$\pm$0.10 \\
159330 &   HR\,6540       &  4 &   65$\pm$3.4 & 1.01$\pm$0.06 &  6 &   31$\pm$1.9 & 0.99$\pm$0.07 & 10 &\multicolumn{2}{c}{source confusion}\\
152222 &                  &  4 &   39$\pm$2.1 & 0.99$\pm$0.07 &  2 &   21$\pm$1.0 & 1.08$\pm$0.07 &4\tablefootmark{d}& 7.4$\pm$0.9 & 0.99$\pm$0.13 \\
 39608 &                  &  3 &   31$\pm$1.9 & 0.99$\pm$0.07 &  1 &   18$\pm$1.3 & 1.19$\pm$0.08 &2\tablefootmark{e}& 12$\pm$0.9 & 2.05$\pm$0.17 \\
181597 &   HR\,7341       &  2 &   29$\pm$2.5 & 1.04$\pm$0.09 &    &\multicolumn{2}{c}{no measurement}&  2 &\multicolumn{2}{c}{below detection limit}\\
 15008 &   $\delta$\,Hyi  &  1 &   22$\pm$1.6 & 0.96$\pm$0.08 &    &\multicolumn{2}{c}{no measurement}&  1 &\multicolumn{2}{c}{below detection limit}\\ 
          \noalign{\smallskip}
\hline
    \end{tabular} \\
\end{center}
$\tablefoottext{a}$~OBSIDs 1342205202~\&~1342263904~excluded \\
$\tablefoottext{b}$~OBSIDs 1342184574, 1342184575, 1342184585, 1342184586~excluded \\
$\tablefoottext{c}$~OBSIDs 1342185446, 1342185448, 1342185447, 1342185449, 1342191986~excluded \\
$\tablefoottext{d}$OBSIDs 1342240702, 1342240703, 1342227973~\&~1342227974 \\
$\tablefoottext{e}$Only OBSIDs 1342198537~\&~1342198538 \\
$\tablefoottext{f}$Due to cirrus confusion, a background subtraction uncertainty of 10\% must be added: 62$\pm$7.1\,mJy, see discussion in Sect.~\ref{sect:starsfirexcess} 
\end{table*}

%
%
\begin{table*}[ht!]
\caption{Combined chop-nod photometry. Model fluxes are listed in 
Table~\ref{table:sources}. Values in italics are uncertain. The quoted 
uncertainties of the measurements include the absolute calibration uncertainty 
of 5\%. In the case of only one observation for a specific source, the 
statistical error of this flux measurement is used in the uncertainty 
determination, in case of more than one observation for a source, as given in 
columns n$_{70}$, n$_{100}$, and n$_{160}$, the standard deviation of the 
weighted mean from the individual chop/nod fluxes listed in 
Tables~\ref{table:chopnodphotblue} to~\ref{table:chopnodphotred1} is used in 
the uncertainty calculation.
}             
\label{table:photchopnod}      
\begin{center}                          
\begin{tabular}{r r r c c r c c r c c}        
\hline\hline                 
            \noalign{\smallskip}
   HD  &    Name          &n$_{70}$&f$_{\rm star}$(70\,$\mu$m) &$\frac{f_{\rm star}(70\,{\mu}m)}{f_{\rm model}}$&n$_{100}$&f$_{\rm star}$(100\,$\mu$m) &$\frac{f_{\rm star}(100\,{\mu}m)}{f_{\rm model}}$&n$_{160}$&f$_{\rm star}$(160\,$\mu$m) &$\frac{f_{\rm star}(160\,{\mu}m)}{f_{\rm model}}$\\    
       &                  &       &  (mJy) &                            &        &  (mJy)  &                            &        &  (mJy)  &                            \\     
            \noalign{\smallskip}
\hline                        
            \noalign{\smallskip}
 62509 &  $\beta$\,Gem    &  1 & 2570$\pm$129& 1.05$\pm$0.08 &  1 & 1267$\pm$65 & 1.07$\pm$0.08 &  2 & 497$\pm$25 & 1.09$\pm$0.08 \\
 32887 &$\varepsilon$\,Lep&  2 & 1181$\pm$59 & 1.00$\pm$0.08 &  2 &  558$\pm$31 & 0.97$\pm$0.08 &  4 & 205$\pm$12 & 0.92$\pm$0.08 \\
148387 &   $\eta$\,Dra    &  1 &  509$\pm$31 & 1.06$\pm$0.07 &  1 &  236$\pm$19 & 1.01$\pm$0.10 &  2 & 101$\pm$10 & 1.13$\pm$0.10 \\
180711 &   $\delta$\,Dra  &  5 &  440$\pm$22 & 1.03$\pm$0.08 &  3 &  214$\pm$12 & 1.03$\pm$0.08 &  8 &  83$\pm$5.0& 1.05$\pm$0.08 \\
139669 &   $\theta$\,Umi  & 10 &  282$\pm$14 & 0.98$\pm$0.08 &  9 &  142$\pm$7.9& 1.02$\pm$0.08 & 16\tablefootmark{a}&  55$\pm$3.8 & 1.02$\pm$0.09 \\
 41047 &   HR\,2131       &    &             &               &  1 &   88$\pm$5.6& 0.92$\pm$0.09 &  1 &{\it 29$\pm$15}&{\it 0.77$\pm$0.54} \\
138265 &   HR\,5755       &  9 &  113$\pm$5.7& 0.97$\pm$0.06 &  4 &   59$\pm$3.0& 1.04$\pm$0.06 & 11\tablefootmark{b}&  31$\pm$1.9  & 1.39$\pm$0.07 \\
159330 &   HR\,6540       &  1 &   60$\pm$7.6& 0.94$\pm$0.13 &  3 &   31$\pm$2.2& 0.98$\pm$0.08 &  4 &\multicolumn{2}{c}{source confusion} \\
152222 &                  &  1 &   39$\pm$4.9& 0.98$\pm$0.14 &    &\multicolumn{2}{c}{no measurement}&  1 &\multicolumn{2}{c}{below detection limit}\\
181597 &   HR\,7341       &  1 &\multicolumn{2}{c}{below detection limit}&    &\multicolumn{2}{c}{no measurement}&  1 &\multicolumn{2}{c}{below detection limit}\\        
 15008 &   $\delta$\,Hyi  &  4 &   20$\pm$1.4& 0.87$\pm$0.08 &  2 &\multicolumn{2}{c}{below detection limit}&  6 &\multicolumn{2}{c}{below detection limit}\\
156729 &   e\,Her         &    &\multicolumn{2}{c}{no measurement}&  1 &\multicolumn{2}{c}{below detection limit}&  1 &\multicolumn{2}{c}{below detection limit}\\
168009 &   HR\,6847       &    &\multicolumn{2}{c}{no measurement}&  1 &\multicolumn{2}{c}{below detection limit}&  1 &\multicolumn{2}{c}{below detection limit}\\
            \noalign{\smallskip}
\hline                                   
\end{tabular} \\
\end{center}
$\tablefoottext{a}$~1342184583, 1342184584 \& 1342184595 excluded \\
$\tablefoottext{b}$~OBSIDs 1342185441 \& 1342185442 excluded
\end{table*}

\subsection{Photometry results}

Results of individual photometric measurements are given in 
Appendix~\ref{sect:appa}, Sect.~\ref{sect:individualscanphot} in 
Tables~\ref{table:scanmapphotblue6_1} to~\ref{table:scanmapphotred10_1}.
Here we report the combined aperture and colour-corrected photometry of all 
measurements in Table~\ref{table:photminiscan}. This is identical with the 
phot\_s photometry in Table~\ref{table:photminiscanapercom}. The quoted 
uncertainties of the measurements in Table~\ref{table:photminiscan} include the 
absolute calibration uncertainty of 5\%, due to the fiducial star model 
uncertainties, which is quadratically added to the rms of the mean flux value 
as quoted in Table~\ref{table:photminiscanapercom}.

For 11 stars we obtain reliable photometry in all three PACS bands. There is 
no 70\,$\mu$m flux for HD\,41047, since there are only measurements in the 
100 and 160\,$\mu$m filters. There is no 160\,$\mu$m detection for HD\,159330 
because of confusion noise. For HD\,181597 and HD\,15008 we obtain reliable 
fluxes only at 70\,$\mu$m, since there are no 100\,$\mu$m scan map measurements
and at 160\,$\mu$m the stars are too faint for the applied repetition numbers. 
Faintest fluxes, for which the photometry has still good quality (accuracy 
$\le$ 15\%), are about 10 -- 20\,mJy.

In Appendix Section~\ref{sect:mappersoftware}, we conduct a qualitative 
inter-comparison of the high-pass filter (HPF) photometry with three other 
{\em Herschel} mapper softwares for HD\,152222, the faintest star at 
160\,$\mu$m with reliable photometry in all three filters. Aspects like noise 
behaviour and shape of the intensity profiles are investigated and discussed.
The main conclusion is that for the other mappers adapted aperture correction 
factors need to be established which will be determined on the basis of the 
high S/N fiducial standard star observational database in a forthcoming paper 
(Balog et al., 2018, in preparation).

The evaluation of the correspondence with the models is done in 
Sect.~\ref{sect:comp_meas_model}.

%
   \begin{figure*}[ht!]
   \centering
   \includegraphics[width=0.95\textwidth]{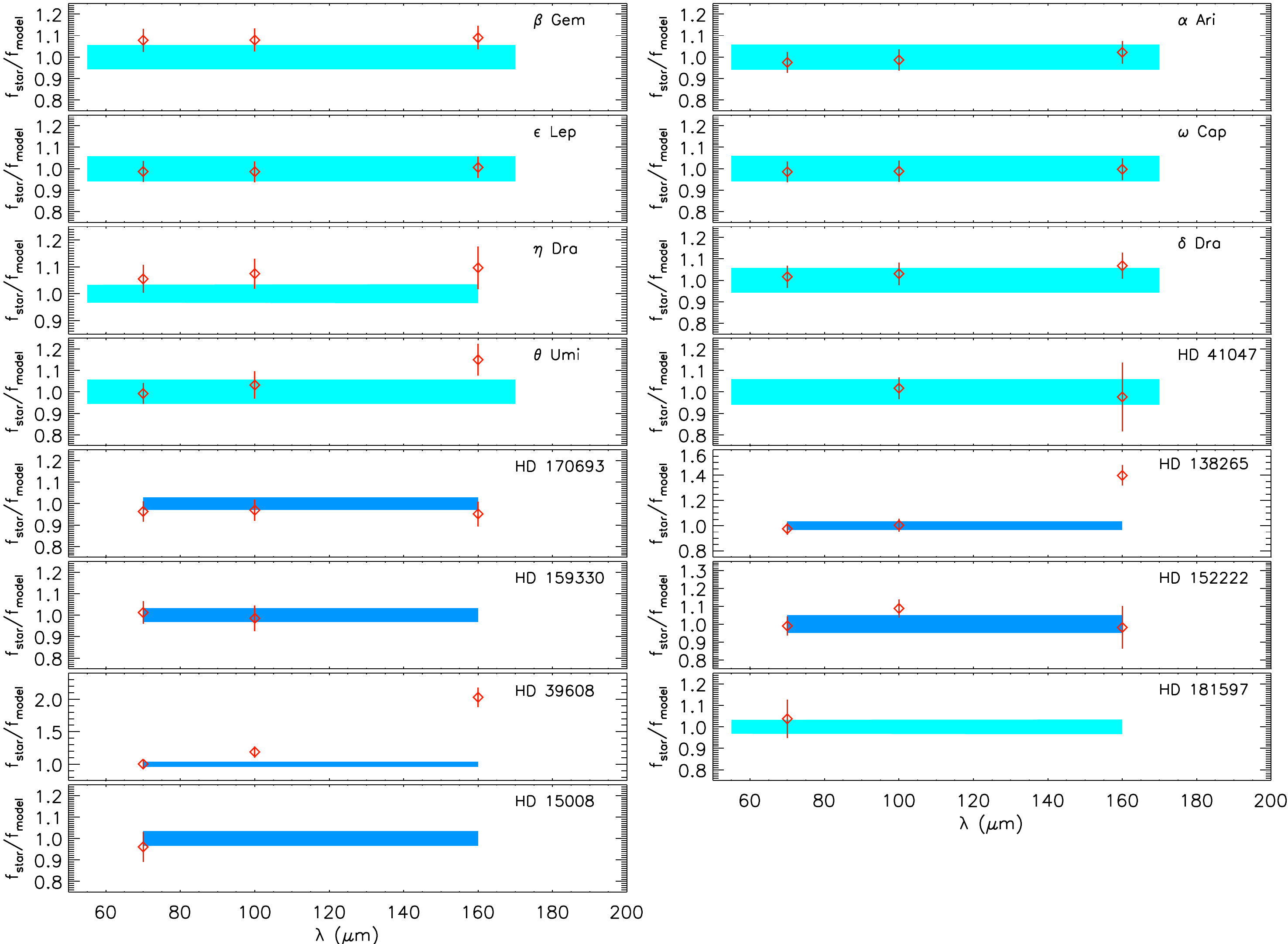}
      \caption{Ratio of the observed and colour-corrected (cc) scan map 
               photometry fluxes with either the photometric standard model 
               flux or the \citet{gordon07} 70\,$\mu$m flux prediction and 
               Rayleigh-Jeans extrapolation (red diamonds). The uncertainty 
               of the models is shown by the light blue range, the uncertainty 
               of the flux prediction and the Rayleigh-Jeans extrapolation by 
               the dark blue range. The uncertainty of the flux ratio includes 
               the absolute photometric error of the measurement.
              }
         \label{scanmap_comp_model}
   \end{figure*}
%

\section{Chop-nod photometry}

\subsection{Observations}

Thirteen out of the 17 sources were observed in the PACS chop-nod point source 
observing mode. This was the originally recommended PACS photometer observing 
mode for point and compact sources. This mode used the PACS chopper to move
the source on-array by about 50\arcsec, corresponding to the size of about one 
blue/green bolometer matrix (16 pixels) or the size of about half a red matrix 
(8 pixels), with a chopper frequency of 1.25\,Hz. The nodding was performed by 
a satellite movement of the same amplitude but perpendicular to the chopping 
direction. On each nod position the chopper executed 3$\times$25 chopper 
cycles. The three sets of chopper patterns were either on the same array 
position (no dithering) or on three different array positions (dither option). 
In the latter case the chopper pattern was displaced parallel to the chopper 
deflection by 8\farcs5 (2$\rm\frac{2}{3}$ blue pixels or 1$\rm\frac{1}{3}$ red 
pixels). Most of the faint star observations were performed with the dither 
option on; Tables~\ref{table:chopnodphotblue} to~\ref{table:chopnodphotred1}
indicate for each observation the selection of the respective dither/no-dither 
mode. Each chopper plateau lasted for 0.4\,s (16 readouts on-board) producing 
four frames per plateau in the telemetry down-link. The full 3$\times$25 
chopper cycles per nod position were completed in less than 1 minute. In the 
case of repetition factors larger than 1, in particular for our faintest 
targets, the nod-cycles were repeated in the following way (example for 4 
repetitions): {\it nodA-nodB-nodB-nodA-nodA-nodB-nodB-nodA} to minimize
satellite slew times. Selected repetition factors are given in 
Tables~\ref{table:chopnodphotblue} to~\ref{table:chopnodphotred1}. We note 
that a repetition factor may have been optimized for the short-wave filter
measurement and is hence less optimal for the 160\,$\mu$m filter, where the 
star is fainter. The observations were usually done in high gain mode, but 
there were a few exceptions taken for comparative performance checks.

%
%
\begin{table*}[ht!]
\caption{Evaluation of correspondence of PACS photometry with models or
flux predictions (C = Cohen, H = Hammersley, G = Gordon). Column "mode" 
specifies the PACS observing mode (s = scan map, c = chop/nod). Column 
"Reason for deviation" gives a short summary of the discussion in the 
reference sections. 
}             
\label{table:modelcorrespondence}      
\centering
    \begin{tabular}{r l c c l l l}
   \hline\hline
            \noalign{\smallskip}
 HD     &   Other name       & Model & Mode & Correspondence with model  & Reason for deviation & Sect. \\
            \noalign{\smallskip}
    \hline
             \noalign{\smallskip}
 62509  & $\beta$\,Gem       & C & s,c  & no, excess of $\approx\,+$8\% & dust disk in planetary system & \ref{sect:intrinsicfirexcess}\\
 12929  & $\alpha$\,Ari      & C & s    & yes, better than $\pm$3\%     &                               & \\
 32887  & $\varepsilon$\,Lep & C & s,c  & yes, better than $\pm$1\%     &                               & \\
198542  & $\omega$\,Cap      & C & s    & yes, better than $\pm$1\%     &                               & \\
148387  & $\eta$\,Dra        & H & s,c  & no, excess of $\approx\,+$8\% & dust in binary system         & \ref{sect:intrinsicfirexcess} \\
180711  & $\delta$\,Dra      & C & s,c  & partially, $\approx\,+$3\% offset for $\lambda \le$100\,$\mu$m& & \\
        &                    &   &      & $+$7\% @160\,$\mu$m           & cirrus confusion              & \ref{sect:firbackconf}\\ 
139669  & $\theta$\,Umi      & C & s,c  & partially, $\approx\,+$3\% offset for $\lambda \le$100\,$\mu$m& & \\
        &                    &   &      & $+$16\% @160\,$\mu$m          & cirrus confusion              & \ref{sect:firbackconf}\\
 41047  & HR\,2131           & C & s,c  & yes, better than $\pm$3\%     &                               & \\
170693  & 42\,Dra            & G & s    & yes, $\approx\,-$4\% offset   &                               & \\
138265  & HR\,5755           & G & s,c  & partially, $\approx\,-$2\% offset for $\lambda \le$100\,$\mu$m& & \\
        &                    &   &      & $+$40\% @160\,$\mu$m          & cirrus confusion              & \ref{sect:firbackconf}\\
159330  & HR\,6540           & G & s,c  & partially, better than $\pm$1\% for $\lambda \le$100\,$\mu$m  & & \\
        &                    &   &      & no flux determination @160\,$\mu$m& source confusion              & \ref{sect:bgconfusion}\\
152222  & SAO\,17226         & G & s,c  & yes, better than $\pm$5\%     &                               & \\
 39608  & SAO\,249364        & G & s    & yes, better than 1\% @70\,$\mu$m&                             & \\
        &                    &   &      & $+$19\% @100\,$\mu$m, $+$105\% @160\,$\mu$m & background confusion& \ref{sect:firbackconf}\\
181597  & HR\,7341           & H & s,c  & yes, better than 4\% @70\,$\mu$m&                             & \\
        &                    &   &      & no flux determination @160\,$\mu$m&  below detection limit    & \ref{sect:detlimits}\\
 15008  & $\delta$\,Hyi      & G & s,c  & yes, better than 4\% @70\,$\mu$m&                             & \\
        &                    &   &      & no flux determination @160\,$\mu$m&  below detection limit    & \ref{sect:detlimits}\\
156729  & e\,Her             & H & c    & no                            & below detection limit         & \\
168009  & \object{HR\,6847}  & H & c   & no                             & below detection limit         & \\
            \noalign{\smallskip}
\hline
    \end{tabular}
\end{table*}

\subsection{Data analysis and calibration}

The data reduction and calibration performed in HIPE~\citep{ott10} followed 
the description in \citet{nielbock13}. For the reduction of our faint star 
targets we adjusted a few aspects and used very recent software developments 
for PACS photometer observations (gyro correction and updated pointing products
and refined focal plane geometry calibration). These new corrections are 
meanwhile part of the Standard Product Generation (SPG) pipelines version 13.0 
and higher. For the stellar flux derivation the same procedures and parameters 
as for scan map photometry, and as summarized in Eq.~\ref{eq:apercolcorr}
and Table~\ref{table:scanmapparamsphotnoise}, are applied.

The photometric uncertainty was estimated with the histogram method with 
a coverage threshold as described in detail in 
Sect.~\ref{sect:noise_sn_determination} for the scan maps. Correlated noise is 
corrected for via an empirical function to obtain a conservative upper limit 
for the measurement uncertainties. The applied correction factors f$_{\rm corr}$
are 6.33, 4.22, and 7.81 for the 70, 100, and 160\,$\mu$m filters,
respectively.

\subsection{Photometry results}

Results of individual measurements are given in Appendix~\ref{sect:appb} 
in Tables~\ref{table:chopnodphotblue} to~\ref{table:chopnodphotred1}.
We note that there are observations of HD\,138265 on OD\,146 for which the 
noise does not seem to scale with the number of repetitions. The reason is 
that for these measurements the basic length of the nod period was varied and 
compensated by the corresponding number of nod cycle repetitions.

Here we report the combined photometry of all measurements in 
Table~\ref{table:photchopnod}. The quoted uncertainties of the measurements 
include the absolute calibration uncertainty of 5\%, due to the fiducial star 
model uncertainties, which is quadratically added to the the rms of the mean 
flux value.

For six stars we obtain reliable photometry in all three PACS bands. There is 
no 70\,$\mu$m flux for HD\,41047, since there are only measurements in the 100 
and 160\,$\mu$m filters. There is no 160\,$\mu$m detection for HD\,159330 
because of confusion noise. For HD\,15008 we only obtain a reliable flux at 
70\,$\mu$m, since at 100 and 160\,$\mu$m the star is too faint for the applied 
repetition numbers. HD\,181597, HD\,156729 and HD\,168009 have non-detections
despite a high repetition factor of 50. The non-detection is likely due to a 
not-yet-perfect knowledge of the optimum observing strategy early in the 
mission (the observations were Astronomical Observation Template (AOT) test 
cases). Faintest fluxes, for which the photometry has still good quality 
(accuracy $\le$ 15\%), are approximately 30\,mJy.

In Appendix \ref{sect:appc} we conduct a detailed comparison between the 
chop/nod and scan map stellar photometry. In summary the results are very 
consistent and confirm each other. A few cases with larger discrepancy are due 
to only a small number of measurements or low S/N in chop/nod mode.

\section{Comparison with model fluxes or flux predictions}
\label{sect:comp_meas_model}

\subsection{Overview of findings}

Since all detected stars are observed in scan map mode and we have more
and better S/N measurements in this mode, we restrict the following 
inter-comparison with the models to scan map photometry. For each star a 
quantitative comparison per filter is given in Table~\ref{table:photminiscan}. 
Figure~\ref{scanmap_comp_model} shows a graphical comparison of the measured 
fluxes with the model and Table~\ref{table:modelcorrespondence} provides a 
summary of the correspondence evaluation. 

We find an excellent correspondence with the model or flux prediction 
over the full PACS wavelength range for $\alpha$\,Ari, $\varepsilon$\,Lep, 
$\omega$\,Cap, HD\,41047, 42\,Dra and HD\,152222. We find a partial 
correspondence up to 100\,$\mu$m for $\delta$\,Dra, $\theta$\,Umi, HD\,138265 
and HD\,159330, while the 160\,$\mu$m flux exceeds the model flux or, as in 
the latter case, cannot be determined due to confusion by nearby sources. For 
HD\,39608, the 70\,$\mu$m flux still corresponds excellently with the flux 
prediction, but at 100 and 160\,$\mu$m a noticeable flux excess is discovered. 
$\beta$\,Gem and $\eta$\,Dra both exhibit a significant offset above the model 
for all wavelengths. For HD\,181597 and HD\,15008 we can prove a good 
correspondence at 70\,$\mu$m, but have no means to do so at longer wavelengths,
since our measurements are not above the detection limit. 

We discuss now the origin of the excess emission for $\delta$\,Dra, 
$\theta$\,Umi, HD\,138265, HD\,39608, $\beta$\,Gem and $\eta$\,Dra.

\subsection{Stars with FIR excess}
\label{sect:starsfirexcess}

A FIR excess can be an intrinsic source property or be caused by confusing 
background sources, as already addressed in Sect.~\ref{sect:bgconfusion}. 
\citet{cohen05} and \citet{dehaes11} discuss possible chromospheric emission 
or thermal emission from an ionized wind which gives noticeable excess at 
sub-mm wavelengths, but may already set in at FIR wavelengths, as intrinsic 
emission mechanisms. \citet{groenewegen12} investigated the phenomenon of 
infrared excess around red giant branch stars assuming mass loss arising from 
chromospheric activity.

\begin{figure}[h!]
   \centering
   \includegraphics[width=0.48\textwidth]{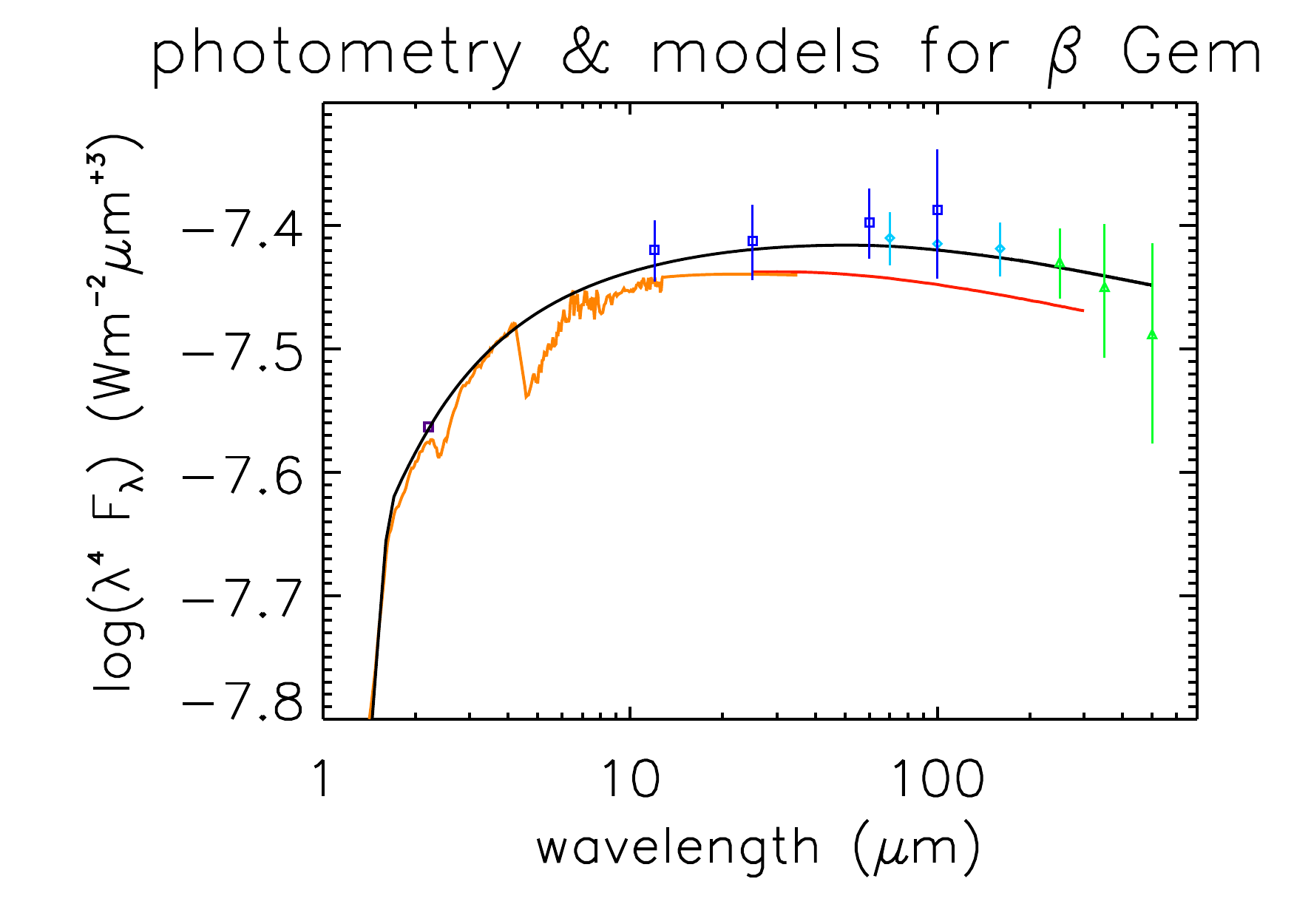}
      \caption{Investigation of the discrepancy of the 
               $\beta$\,Gem~\citet{cohen96} model and measured FIR photometry. 
               For a better zoom-in over a large wavelength range, 
               log$_{\rm 10}$($\lambda^4 \cdot f_{\rm \lambda}$) is displayed. The 
               orange and red lines are the~\citet{cohen95} absolutely 
               calibrated 1.2 -- 35\,$\mu$m spectral model and 
               the~\citet{cohen96} FIR extension, respectively. The black line 
               represents a scaled ($\times$0.169) fiducial star continuum 
               model of $\alpha$\,Boo~\citep{dehaes11}. The scaling factor has 
               been derived from the \citet{selby88} K$_{\rm n}$ narrow band
               photometry ratio 10$^{-0.4 \cdot (-1.14 + 3.07)}$ and the position 
               of the scaling wavelength (2.205\,$\mu$m) is indicated by the 
               violet square symbol. We note that the K$_{\rm n}$ zero point is 
               $\approx$3\% higher than that for the Cohen models (cf.\ file 
               header of $\alpha$\,Boo model (reference, see text) vs.\ 
               Table 1 in \citet{cohen92}). Dark-blue squares, light-blue 
               diamonds and green triangles represent {\em IRAS} FSC, PACS and 
               SPIRE photometry and their respective uncertainties. This 
               photometry has been colour-corrected for a 4500\,K blackbody 
               spectrum.}
         \label{fig:betageminvest}
   \end{figure}

One other aspect to consider in this context is possible source variability; 
we investigate this for the case of $\beta$\,Gem: The~\citet{cohen96} FIR 
model SED is an extension of an absolutely calibrated 1.2 -- 35\,$\mu$m 
model\footnotemark[5]\citet{cohen95}\footnotetext[5]{http://general-tools.cosmos.esa.int/iso/users/expl\_lib/ISO/wwwcal/\\isoprep/cohen/composites/}. 
In Fig.~\ref{fig:betageminvest} we show both model parts represented by the 
orange and red lines. No inconsistency between both parts can be recognised. 
Besides the PACS photometry we show the colour-corrected {\em IRAS} 
FSC~\citep{moshir89} and SPIRE PSC\footnotemark[6]~\citep{schulz17} 
\footnotetext[6]{http://irsa.ipac.caltech.edu/cgi-bin/Gator/nph-scan?submit=Select\&projshort=HERSCHEL; \\
for SPIRE colour correction, see SPIRE Handbook, Table 5.7 \\http://herschel.esac.esa.int/Docs/SPIRE/spire\_handbook.pdf}
photometry spanning the wavelength range from 12 to 500\,$\mu$m. All measured 
photometry was taken between 1983 and 2009 -- 2013 and is clearly above the 
model, meaning that variability of the source is an unlikely explanation, 
since a major part of the photometric input for the Cohen model was obtained 
in-between (but {\em IRAS} 12 and 25\,$\mu$m photometry was not considered).

Another explanation for an intrinsic FIR excess can be a residual dust disk 
from a stellar- or planetary-system-formation process. One of the first giant 
stars, for which an infrared excess was reported, is the K3 giant 
$\delta$\,And \citep{judge87}. The giant is the brightest star in a triple or 
quadruple system and is itself a spectroscopic binary with a companion that is 
most likely a main sequence K-type star \citep{bottom15}. \citet{judge87} 
argued that the infrared excess appears to be caused by a detached primordial 
dust shell around the giant. \citet{plets97} concluded for a larger sample of 
giants with infrared excess, that this phenomenon is most likely related to 
the Vega phenomenon around first-ascent giants.

\subsubsection{Intrinsic FIR excess}
\label{sect:intrinsicfirexcess}

$\beta$\,Gem is another good candidate for having a residual debris disk, 
since it is the host star of a confirmed~\citep{hatzes06} Jupiter-sized planet 
(HD\,62509\,b, M = 2.9$\pm$0.1\,M$_{\rm J}$, 
a = 1.69$\pm$0.03\,AU\footnotemark[7])\footnotetext[7]{The Extrasolar Planet 
Encyclopedia \\ http://exoplanet.eu/catalog/HD 62509\_b/}. A rough estimate
(assuming a Jupiter diameter and T = 300\,K) gives a contrast of $>$10$^{5}$ 
between star and planet, meaning that the planet cannot account for the FIR 
excess of $\approx$8\%.

The Cohen models of $\beta$\,Gem are based on T$_{\rm eff} \approx$ 4850\,K 
(see file headers of models with reference to~\citet{blackwell91}). The 
effective temperature T$_{\rm eff}$ of giant stars are determined in the 
ultraviolet to near-infrared wavelength range, either from photometric
indices~\citep[e.g.\ ][]{lyubimkov14}, colours and metallicities~\citep[e.g.\ ][]{alonso99},
or integrated fluxes and interferometric measurements of the stellar
diameters~\citep[e.g.\ ][]{dyck96}. Other references give T$_{\rm eff}$ = 
4850\,K~\citep{gray03} and 4946$\pm$18\,K~\citep{jofre15} and thus confirm the 
value used by Cohen. We have scaled a continuum model\footnotemark[8] of the
PACS fiducial standard star $\alpha$\,Boo, a K2III star\footnotetext[8]{This~model~can~be~found~under \\
http://archives.esac.esa.int/hsa/legacy/ADP/StellarModels/},
by calculating the \citet{selby88} K$_{\rm n}$ narrow band photometry ratio 
10$^{-0.4 \cdot (-1.14 - (-3.07))}$ = 0.169. The intention here is not to accurately
model the $\beta$\,Gem FIR photometry, but to demonstrate that the SED of the 
cooler source with T$_{\rm eff}$ = 4320\,K provides a better fit. This indicates
that there is an additional cooler FIR component beside the photospheric 
emission of the star. The shape of the SED $>$10\,$\mu$m given by the 
{\em IRAS}, PACS, and SPIRE photometry argues for a flat blackbody dust 
disk~\citep[see e.g.\ ][for a discussion of the dust disk SED shape depending 
on its geometry]{chiang97,beckwith99}.

%
%

$\alpha$\,Ari and 42\,Dra are also host stars of confirmed Jupiter-sized 
planets (alf\,Ari\,b, M = 1.8$\pm$0.2\,M$_{\rm J}$, 
a = 1.2\,AU~\citep{lee11}\footnotemark[9]\footnotetext[9]{The Extrasolar 
Planet Encyclopedia \\ http://exoplanet.eu/catalog/alf Ari\_b/}; 42\,Dra\,b, 
M = 3.88$\pm$0.85\,M$_{\rm J}$, 
a = 1.19\,AU~\citep{doellinger09}\footnotemark[10])
\footnotetext[10]{http://exoplanet.eu/catalog/42\_Dra\_b/}, but for these 
stars any possible debris disk emission is much fainter than for $\beta$\,Gem. 
The observed SED shape of $\alpha$\,Ari is a little bit shallower than the 
model prediction (cf.\ Table~\ref{table:photminiscan}), but the measurement 
and model uncertainties do not allow any detection. For 42\,Dra, no deviation 
from a pure photospheric emission SED can be found from our photometry. We 
therefore keep both stars in our standard star list. 

\begin{figure}[h!]
   \centering
   \includegraphics[width=0.48\textwidth]{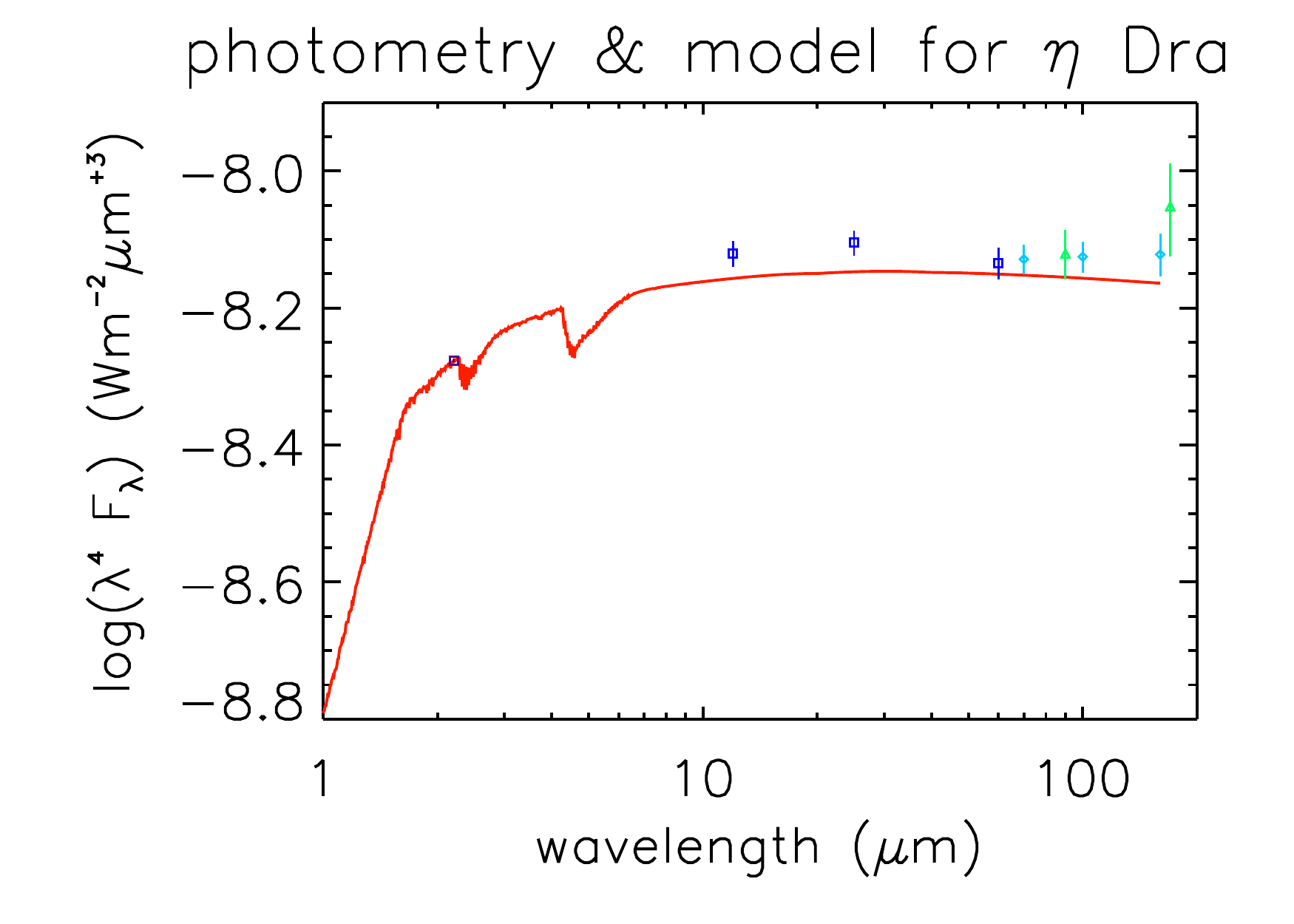}
      \caption{Investigation of the discrepancy of the 
               $\eta$\,Dra~\citet{hammersley98} model and measured FIR
               photometry. For a better zoom-in over a large wavelength range, 
               log$_{\rm 10}$($\lambda^4 \cdot f_{\rm \lambda}$) is displayed. The 
               red line is the model, absolutely calibrated at 2.208\,$\mu$m 
               (K$_{\rm n}$ magnitude = 0.62$\pm$0.005\,mag,
               c.f.~\citet{cohen99},~\citet{selby88}) as represented by the 
               violet square symbol. Dark-blue squares, light-blue diamonds, 
               and green triangles represent {\em IRAS} FSC, PACS, and ISOPHOT 
               HPDP photometry and their respective uncertainties. This 
               photometry has been colour-corrected for a 5000\,K blackbody 
               spectrum.}
         \label{fig:etadrainvest}
   \end{figure}

In Sect.~\ref{sect:bgconfusion} we have shown the 160\,$\mu$m map of 
$\eta$\,Dra (Fig.~\ref{fig:hd148387cirrusmaps}) as a representative example 
for cirrus confusion. Indeed, from Table~\ref{table:bgconfinvest},
$\frac{f_{\rm 160,corr}^{\rm 10\farcs7}}{f_{\rm 160,corr}^{\rm 5\farcs35}}$ =
1.17, which could fully explain $\frac{f_{160}}{f_{\rm model}}$ = 1.16 in 
Table~\ref{table:photminiscan} as an excess due to cirrus emission in the 
standard 10\farcs7-radius aperture. On the other hand, from 
Fig.~\ref{scanmap_comp_model}, we see that already the 70 and 100\,$\mu$m 
fluxes are off  by $+$6 and $+$7\%, respectively, with regard to the model.
We investigate this discrepancy in Fig.~\ref{fig:etadrainvest} by complementing
the PACS photometry with additional {\em IRAS} FSC~\citep{moshir89}
photometry and ISOPHOT~\citep{lemke96} HPDP (Highly Processed Data
Products photometry (c.f.\ Appendix.~\ref{sect:appd}, 
Table~\ref{table:isophothpdpphot}). It is obvious that all photometric 
measurements are consistently above the model, irrespective of whether the 
observations were obtained during the {\em IRAS}, {\em ISO}, or {\em Herschel} 
missions (1983 -- 2013). The excess is a clear hint of an additional emission 
component besides the photospheric emission of the star, whereby the rise in 
flux beyond 100\,$\mu$m is likely caused by cirrus emission. The G8 giant 
$\eta$\,Dra (also identified as CCDM\,J16239$+$6130\,A) has a K1V companion, 
CCDM\,J16239$+$6130\,B~\citep{dommanget02}, at a distance of 2\farcs5 and at a 
position angle of 71.6$^o$ NE. The origin of the excess emission could then be 
dust inside this binary system.

%
%

%
%
\begin{table*}[ht!]
\caption{Input data for the investigation of 160\,$\mu$m photometric flux 
contamination of faint stars by background confusion. The determination of 
B$_{\rm 160}^{ISM}$ is described in the text. The OBSID combinations of the 
deepest maps are used for this investigation. f$_{\rm 160,excess}$ is estimated 
as the difference of the measured f$_{160}$ minus model f$_{\rm model}$ flux from 
Table~\ref{table:photminiscan}. Listed fluxes f$_{\rm 160,corr}^{\rm aperture~radius}$
are not colour corrected. Uncertainties of the flux estimates include a 1\% 
uncertainty of the aperture correction to obtain the total flux, which is 
quadratically added to $\sigma_{\rm aper}$ (c.f.~Eq.~\ref{eq:sigaper}).
}             
\label{table:bgconfinvest}      
\centering                          
\begin{tabular}{r c c c l c c c c c}        
\hline\hline                 
            \noalign{\smallskip}
   HD   &   l   &    b    &  B$_{\rm 160}^{ISM}$ & OBSIDs & f$_{\rm 160,excess}$ & f$_{\rm 160,corr}^{\rm 5\farcs35}$ & f$_{\rm 160,corr}^{\rm 10\farcs7}$ & f$_{\rm 160,corr}^{\rm 14\farcs0}$ & $\frac{f_{\rm 160,corr}^{\rm 10\farcs7}}{f_{\rm 160,corr}^{\rm 5\farcs35}}$\\    
        & ($^o$)& ($^o$)  &  (MJy\,sr$^{-1}$)  &         &     (mJy)         &         (mJy)             &             (mJy)        &       (mJy)              \\
            \noalign{\smallskip} 
\hline                        
            \noalign{\smallskip}
148387  &  92.6  & $+$41.0 & 2.44$\pm$0.13 & 1342186146$+$..47$+$..55$+$..56 & 8.5 & 106.7$\pm$1.4 & 124.9$\pm$2.1 & 130.7$\pm$2.6 & 1.17 \\
180711  &  98.7  & $+$23.0 & 6.04$\pm$0.02 & 1342212497$+$.498$+$.499$+$.500 & 5.5 &  87.4$\pm$0.9 &  91.8$\pm$1.1 &  94.8$\pm$1.2 & 1.05 \\
        &        &         &               & 1342222147$+$..48$+$..49$+$..50 &     &  90.8$\pm$1.0 &  94.4$\pm$1.1 &  95.3$\pm$1.3 & 1.04 \\
        &        &         &               & 1342233571$+$..72$+$..73$+$..74 &     &  92.0$\pm$1.0 &  95.9$\pm$1.1 &  99.0$\pm$1.2 & 1.04 \\
        &        &         &               & 1342250093$+$..94$+$..95$+$..96 &     &  87.2$\pm$0.9 &  85.9$\pm$1.0 &  84.2$\pm$1.1 & 0.99 \\
        &        &         &               & 1342257987$+$..88$+$..89$+$..90 &     &  86.2$\pm$0.9 &  91.4$\pm$1.1 &  95.2$\pm$1.3 & 1.06 \\
139669  & 112.9  & $+$36.5 & 2.67$\pm$0.07 & 1342191982$+$..83               & 8.0 &  61.3$\pm$0.7 &  66.5$\pm$0.9 &  68.4$\pm$1.0 & 1.08 \\
138265  &  95.8  & $+$47.4 & 1.27$\pm$0.03 & 1342188841$+$42                 & 9.0 &  30.1$\pm$0.4 &  33.3$\pm$0.5 &  34.9$\pm$0.6 & 1.11 \\
 39608  & 269.6  & $-$30.9 & 5.22$\pm$0.17 & 1342198535$+$..36$+$..37$+$..38 & 6.0 &  10.7$\pm$0.2 &  13.5$\pm$0.3 &  15.9$\pm$0.4 & 1.26 \\
152222  &  98.5  & $+$36.7 & 2.57$\pm$0.15 & 1342227973$+$..74               &  -- &   8.2$\pm$0.2 &   7.0$\pm$0.3 &   5.9$\pm$0.4 & 0.85 \\
           \noalign{\smallskip}
\hline                                   
\end{tabular}
\end{table*}

\subsubsection{FIR background confusion}
\label{sect:firbackconf}

In Table~\ref{table:bgconfinvest} we have compiled crucial information for 
those sources whose 160\,$\mu$m fluxes may be contaminated by background 
confusion. All sources are at relatively high galactic latitudes in the range 
23$^o$ $\le$ |b| $\le$ 47$^o$. The 160\,$\mu$m brightness of the ISM, 
B$_{\rm 160}^{ISM}$, was derived from {\it AKARI}-FIS WIDE-L (140\,$\mu$m) 
all-sky survey maps\footnotemark[4]~\citep{doi15} in the following way:
\begin{equation}
B_{\rm 160}^{ISM} = \frac{B_{\rm WIDE-L}^{AKARI}- B_{\rm CFIRB}\,cc_{\nu^0}}{cc_{\nu^1.5\,BB(T=20\,K)}\,K_{\nu^1.5\,BB(T=20\,K)}^{FIS140-PACS160}} 
,\end{equation}
with B$_{\rm WIDE-L}^{AKARI}$ being the measured {\it AKARI} 140\,$\mu$m flux
(we highlight that we have transformed the original 6\,deg\,$\times$\,6\,deg 
maps in ecliptic coordinates to the equatorial coordinate system and 
re-centred to the central coordinates of the PACS maps), 
B$_{\rm CFIRB}$ = 1.08\,MJy/sr being the cosmic far-infrared background 
level~\citep[cf.\ ][]{juvela09}, cc$_{\nu^0}$ = 0.964 and 
cc$_{\nu^{1.5}\,BB(T=20\,K)}$ = 0.95 are {\it AKARI}-FIS WIDE-L colour-correction 
factors~\citep{shirahata09} for the indexed SEDs and 
K$_{\nu^{1.5}\,BB(T=20\,K)}^{FIS140--PACS160}$ = 0.954 is the reference 
wavelength-correction factor between {\it AKARI}-FIS WIDE-L and the PACS 
160\,$\mu$m filter~\citep{mueller11} for the modified blackbody SED 
$\nu^{1.5}$\,BB(T=20\,K), which is typical for IR cirrus emission according to
latest results~\citep{planckcollab14,bianchi17}. The listed surface brightness 
of B$_{\rm 160}^{ISM}$ is associated with the 15\arcsec\,$\times$\,15\arcsec
pixel covering the star position, the uncertainty was computed as the standard 
deviation of the eight neighboring pixel values with regard to the central one.
B$_{\rm 160}^{ISM}$ are between 1.3 and 6.0\,MJy\,sr$^{-1}$, with a gradient 
with |b|. 

\citet{kiss05} parameterized the sky confusion noise (1\,$\sigma$) for FIR
measurements with ISOPHOT depending on wavelength and background reference 
configuration geometry as
\begin{equation}
\frac{N_{\rm conf}^{\rm PHT}(\theta, k, \lambda)}{1\,mJy} = C_{0}(\theta, k, \lambda) + C_{1}(\theta, k, \lambda) \langle \frac{B(\lambda) - B_{\rm CFIRB}(\lambda)}{1\,MJy\,sr^{-1}} \rangle^{\eta(\theta, k, \lambda)}.
\end{equation}
The ISOPHOT C200 measurement configuration P/C/184\arcsec~in Table~4 
of~\citet{kiss05} is closest to our PACS mini-scan-map measurement 
configuration, except that aperture size and background ring radius have to be 
scaled down by a factor of $\approx$0.22 (92\arcsec~ISOPHOT C200 pixel size 
vs.\ 19\arcsec~PACS "pixel" size corresponding to a circular aperture with 
10\farcs7 radius and 184\arcsec~vs.\ 40\arcsec~background ring radius). This 
means that the PACS sky confusion noise, N$_{\rm conf}^{\rm PACS}$, has to be 
scaled down by a factor of 0.22$^{2.5}\,$~\citep{kiss05} due to the better 
spatial resolution of PACS. For computation of a point-source representative 
sky confusion noise we multiply with the aperture correction factor 
c$_{\rm aper}(160\,{\mu}m)$ = 1.56. 
\begin{equation}
\label{eq:pacsskyconf}
\frac{N_{\rm conf,PS}^{\rm PACS}}{1\,mJy} = 3.54\,10^{-2} \times [C_{0} + C_{1} \langle \frac{B(\lambda) - B_{\rm CFIRB}(\lambda)}{1\,MJy\,sr^{-1}} \rangle^{\eta}] 
.\end{equation}
Applied parameters are C$_{0}$ = 9.3$\pm$6.7\,mJy, C$_{1}$ = 3.37$\pm$1.01\,mJy,
and $\eta$ = 1.46$\pm$0.17. 

The C$_{0}$ term represents the confusion noise due to cosmic infrared 
background fluctuations and amounts to 0.33$\pm$0.24\,mJy. \citet{berta11} 
give a confusion noise $\sigma_{\rm c}$ = 0.92\,mJy from cosmological fields in 
the {\it Herschel}-PEP survey (obtained 
for $q = \frac{f_{\rm lim}}{\sigma_{\rm c}} = 5$), which corresponds to a C$_{0}$ 
value equal to 26.0\,mJy.

The $C_{1}\,\langle B_{\rm 160}^{ISM} \rangle^{\eta}$ term represents the cirrus 
confusion noise, which depends on the surface brightness of the emitting 
cirrus material. With the range 1.3 $\le B_{\rm 160}^{ISM} \le$ 
6.0\,MJy\,sr$^{-1}$, we predict a cirrus confusion noise 0.18$\pm$0.06$ \le 
N_{\rm cirrconf,PS}^{\rm PACS} \le$ 1.63$\pm$1.02\,mJy.

If we attribute the 160\,$\mu$m excess f$_{\rm 160,excess}$ in 
Table~\ref{table:bgconfinvest} fully to sky confusion and compare with our 
confusion noise prediction, we note the following: For our small sample of 
160\,$\mu$m excess stars we do not see any dependence on B$_{\rm 160}^{ISM}$, in 
particular for HD\,138265 with the lowest B$_{\rm 160}^{ISM}$ = 
1.27\,MJy\,sr$^{-1}$, the highest f$_{\rm 160,excess}$ = 9\,mJy is found; and
the 1\,$\sigma$ sky confusion noise derived via Eq.~\ref{eq:pacsskyconf}
underestimates the actually measured noise by factors 3 -- 25
(1.9 -- 21 accounting for the maximum positive uncertainty).

The confusion noise predictions are average numbers based on a statistical
analysis. Peaks and depressions in the sky noise can significantly deviate
from the average. The spatial resolution of the {\it AKARI} 140\,$\mu$m 
all-sky survey maps is $\approx$88\arcsec~\citep{takita15}. The PACS maps
reveal much finer structures. Their weight to the noise is much higher than to 
the average surface brightness. Therefore, calculating the cirrus noise from 
the surface brightness of a larger area will always underestimate the local 
cirrus noise. Another possibility is that the PSF of a discrete few-mJy source 
coincides - accidentally to a large percentage - with the PSF of the star. 
Differential number counts in cosmological fields, as in Fig.~7 
of~\citet{berta11}, suggest that there are 9.2\,$\times$\,10$^3$ background 
sources/deg$^2$ for f$_{\rm lim} \ge$ 3.5\,mJy, which gives 0.25 sources per 
photometric aperture of 10\farcs7, hence an already high likelihood that such 
a source can blend the photometry of our faint stars. We cannot exclude either 
that in some maps some amount of the 160\,$\mu$m excess is produced by the 
data reduction scheme itself by reducing the background level in some of the 
background reference areas (this can vary from map to map depending on the 
actual detector drift behaviour along the time-line and the level of 
reduction).  

To some extent a contribution by an underlying source can be disentangled via 
multi-aperture photometry which includes aperture sizes as small as the PSF 
FWHM. From multi-aperture photometry of the deepest maps (OBSIDs combinations 
are indicated in Table~\ref{table:bgconfinvest}) with aperture radii 5\farcs35, 
10\farcs7 (the default one for our photometric analysis), and 14\farcs0, we 
see that the flux increase is usually greater than the associated 
uncertainties, which is a hint of flux contribution by another source. As a 
reference, we also include the multi-aperture photometry of HD\,152222 which 
does not show any flux increase (rather a flux decrease due to increasing 
uncertainties in the background subtraction with larger aperture size). 

%
   \begin{figure*}[ht!]
   \centering
   \includegraphics[width=0.26\textwidth]{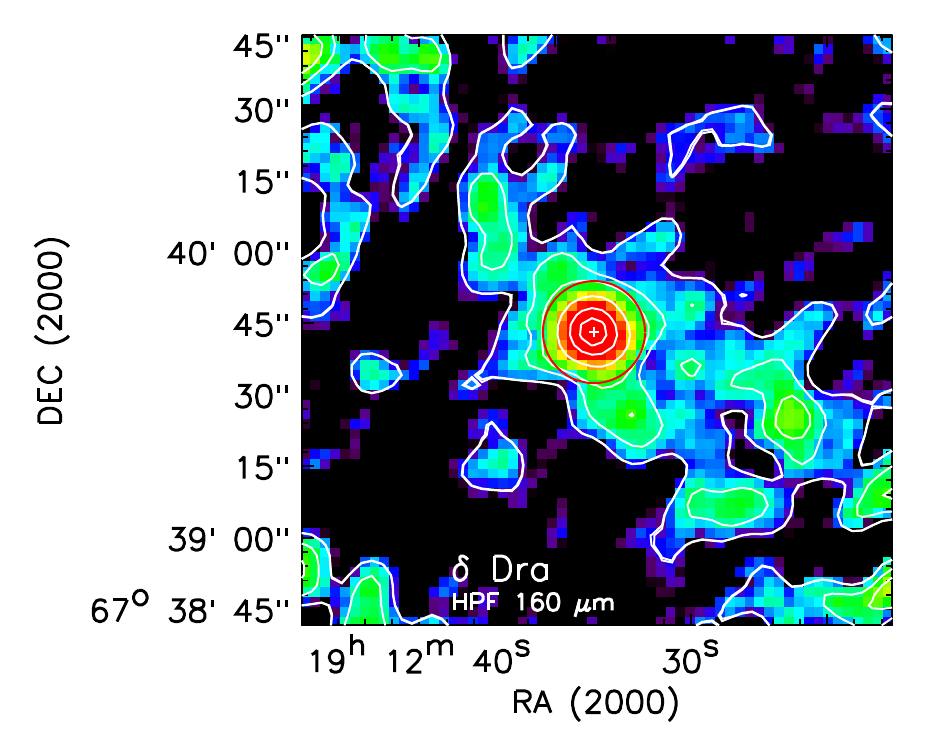}
   \includegraphics[width=0.26\textwidth]{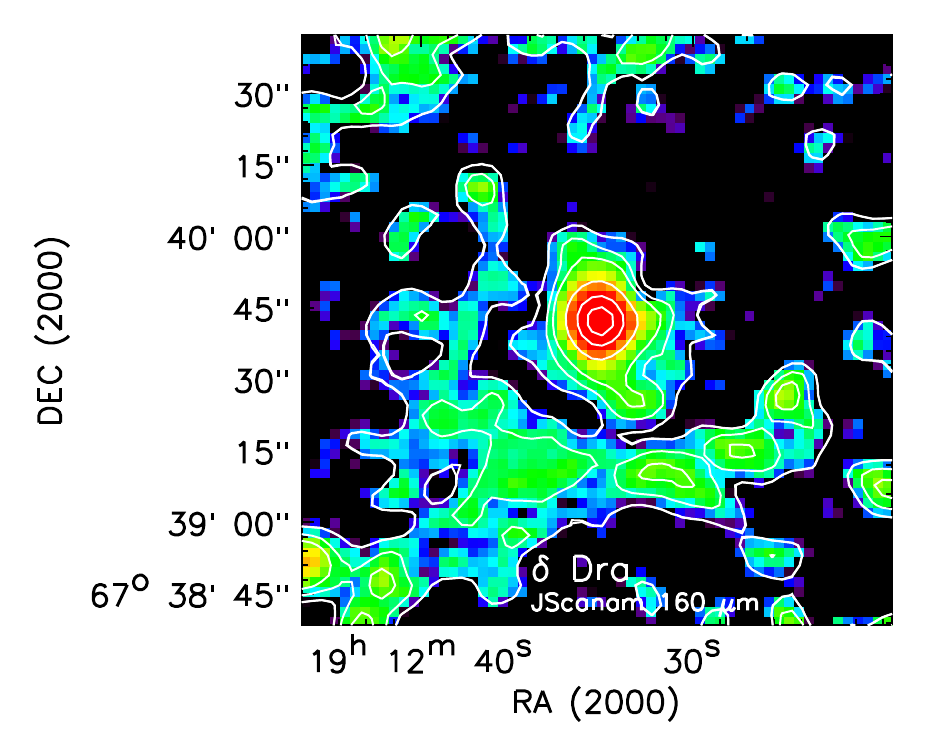}
   \includegraphics[width=0.24\textwidth]{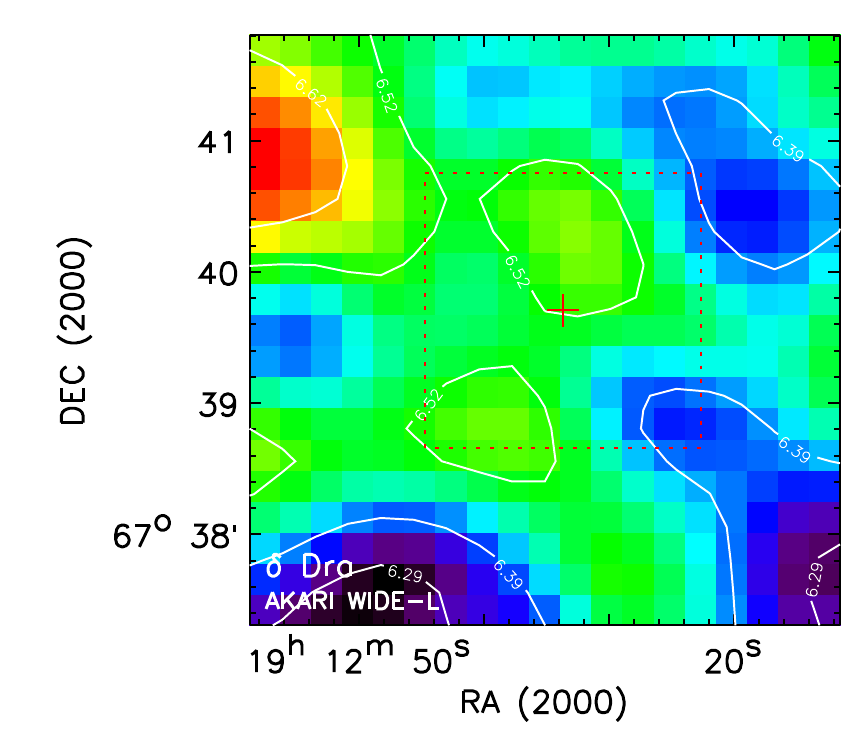} \\
   \includegraphics[width=0.26\textwidth]{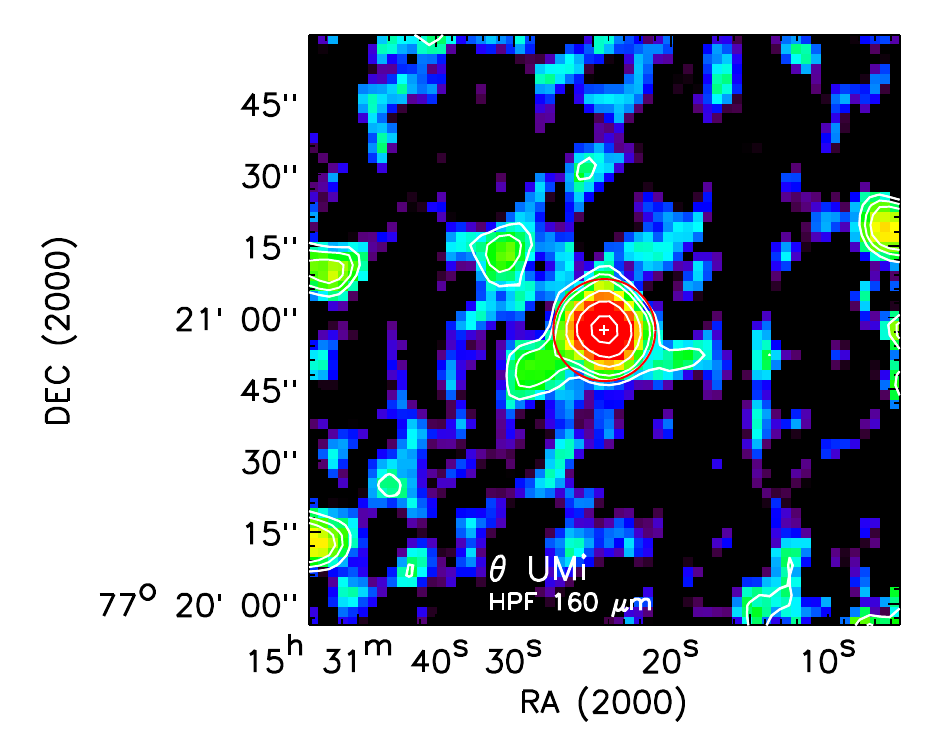}
   \includegraphics[width=0.26\textwidth]{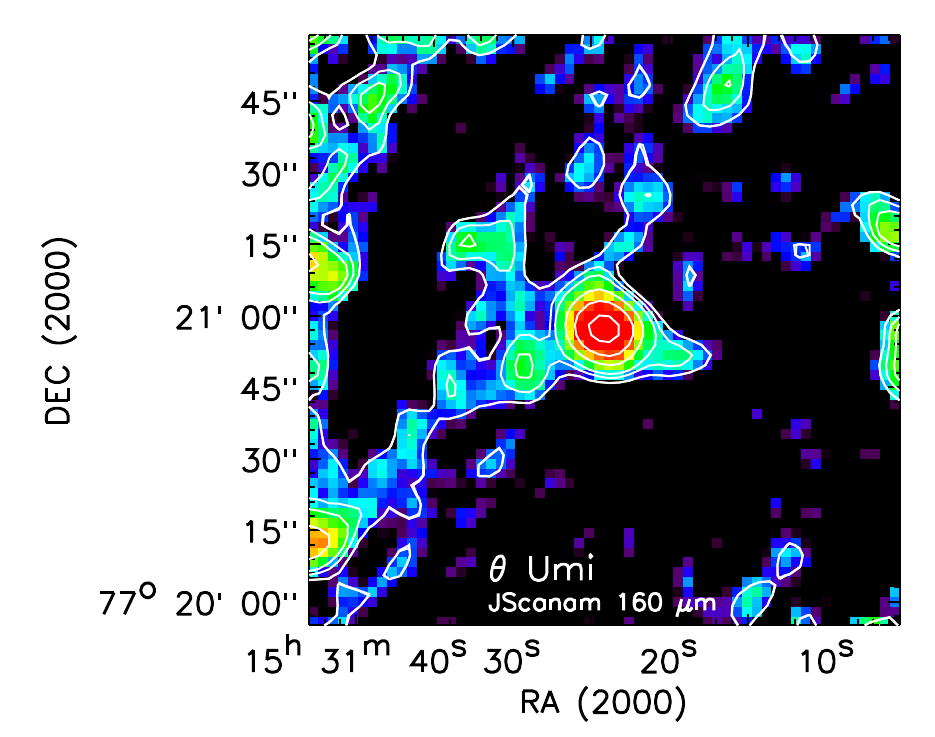}
   \includegraphics[width=0.24\textwidth]{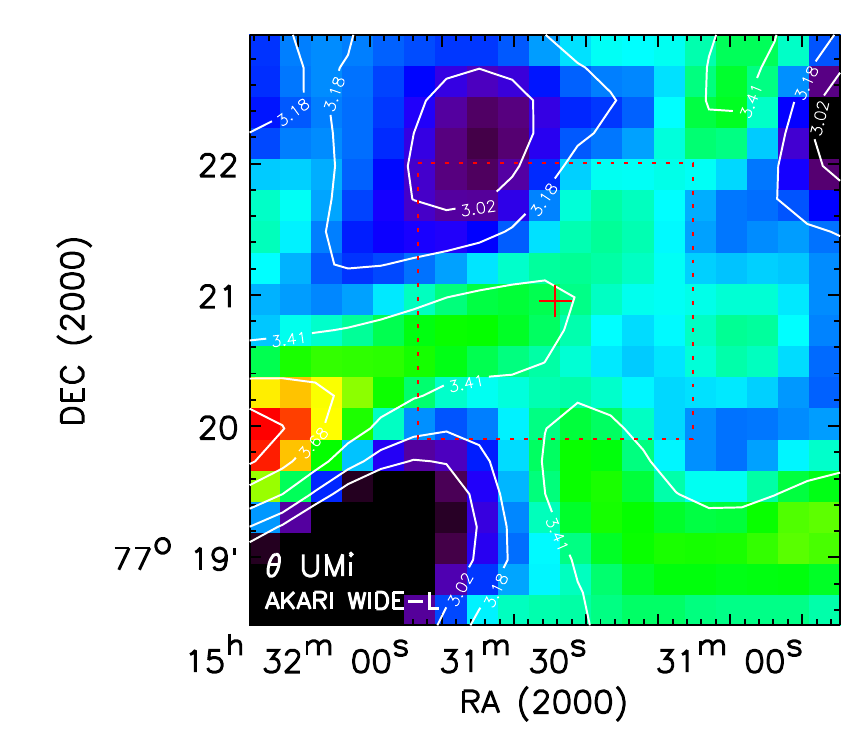} \\
   \includegraphics[width=0.26\textwidth]{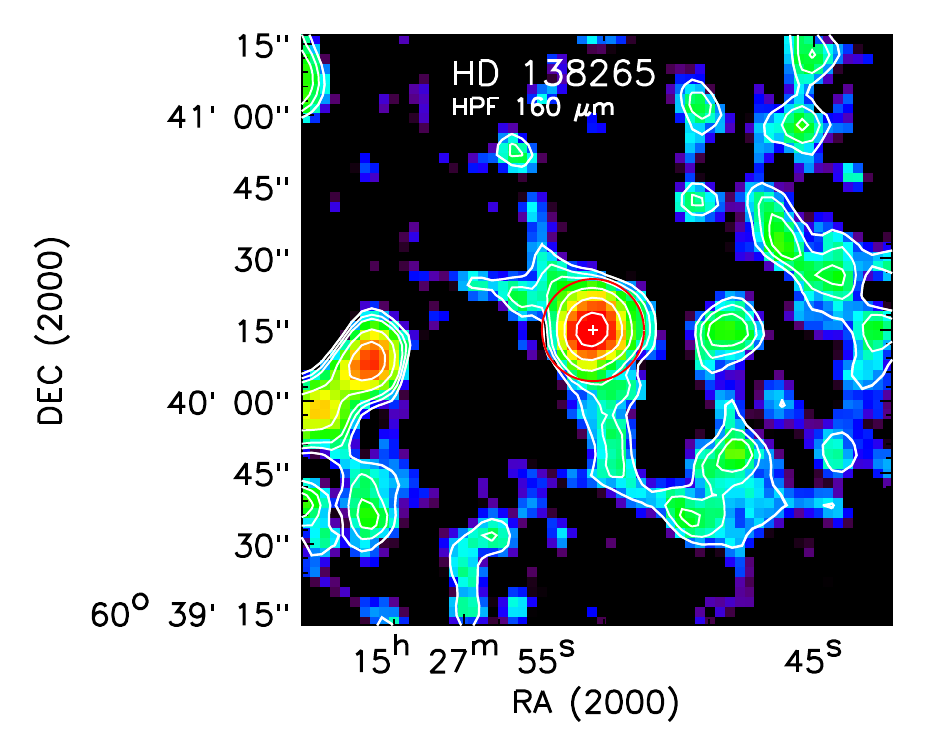}
   \includegraphics[width=0.26\textwidth]{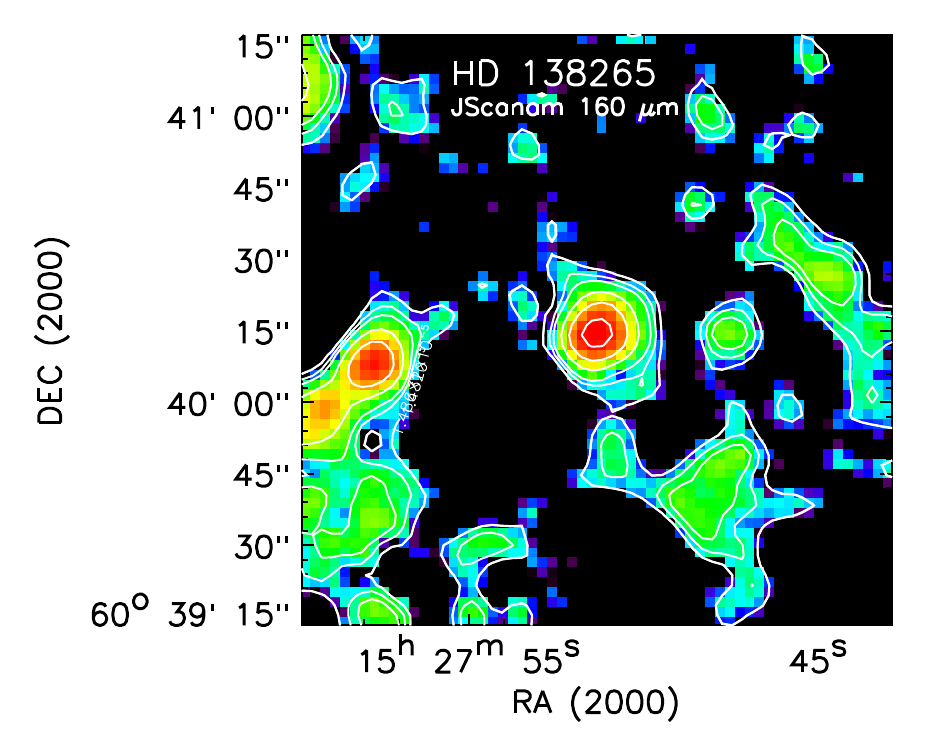}
   \includegraphics[width=0.24\textwidth]{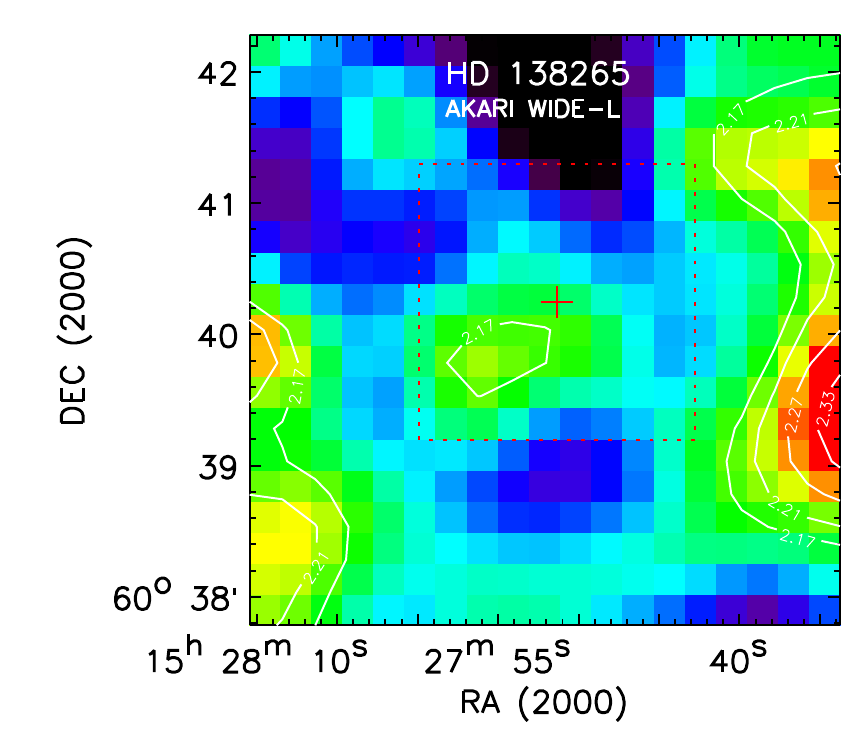} \\
   \includegraphics[width=0.26\textwidth]{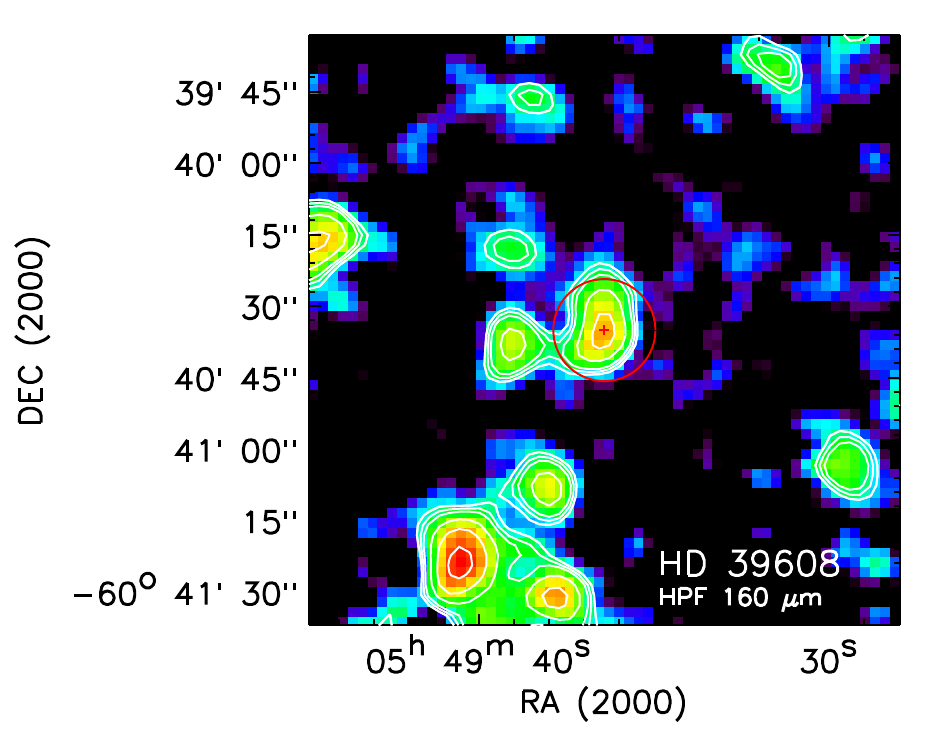}
   \includegraphics[width=0.26\textwidth]{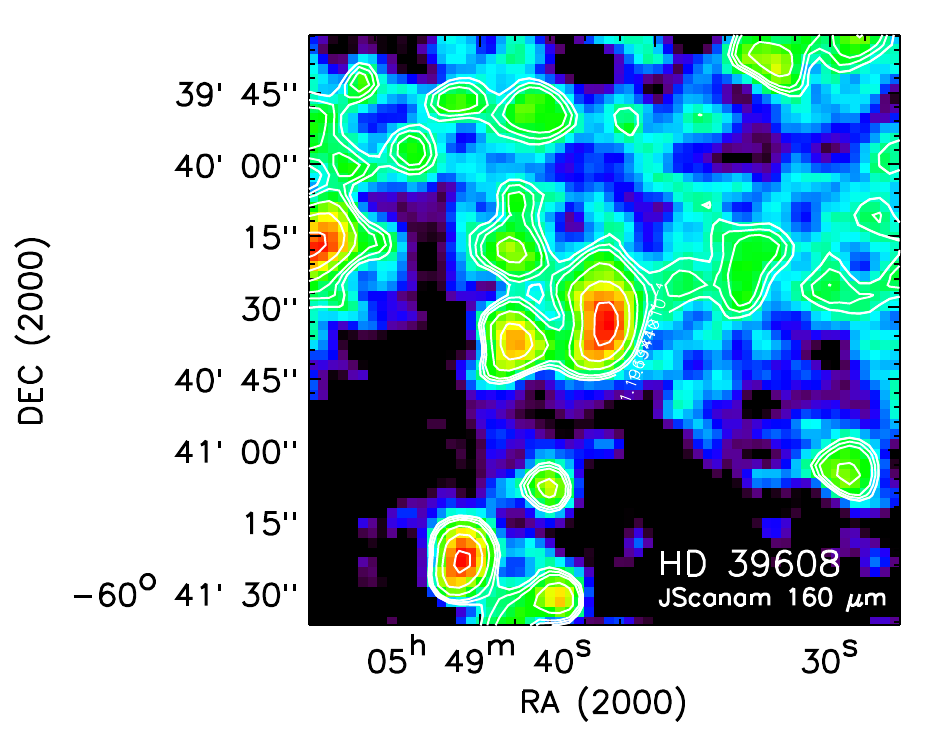}
   \includegraphics[width=0.24\textwidth]{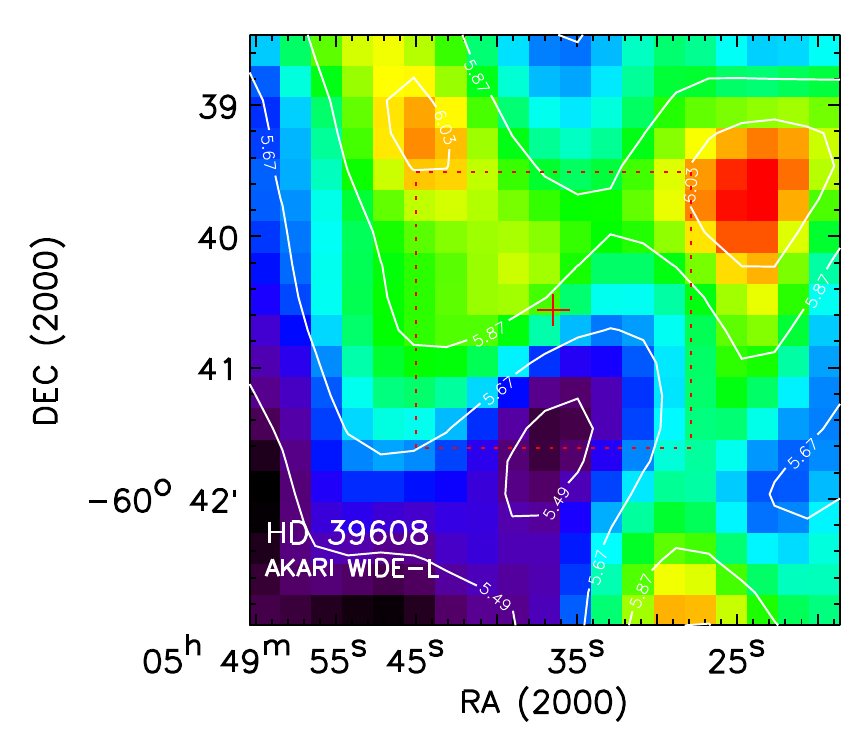} \\
   \includegraphics[width=0.26\textwidth]{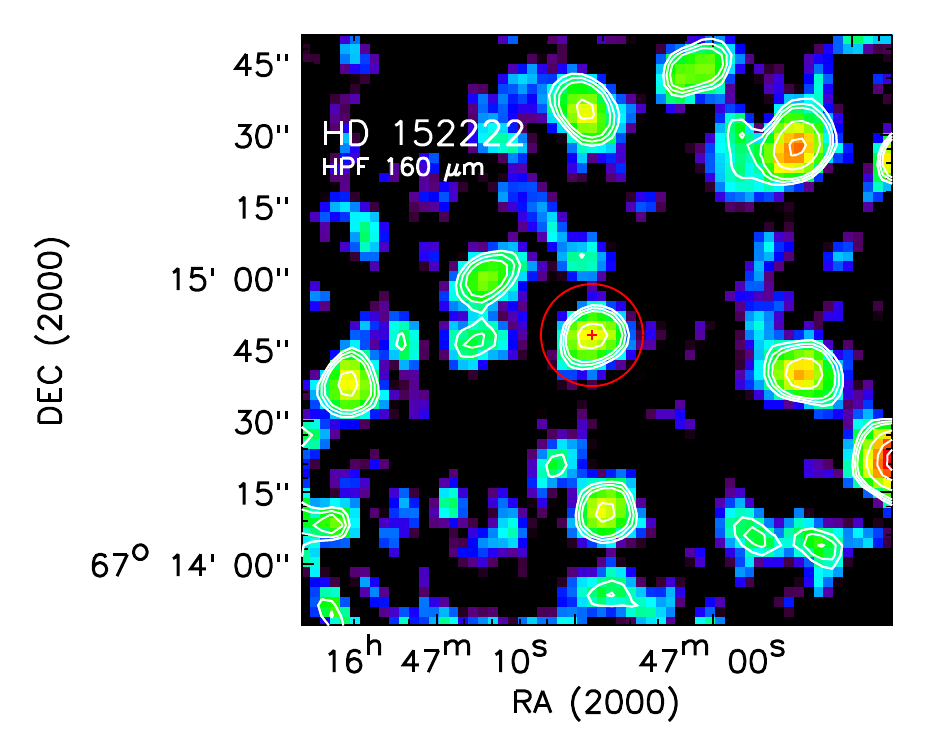}
   \includegraphics[width=0.26\textwidth]{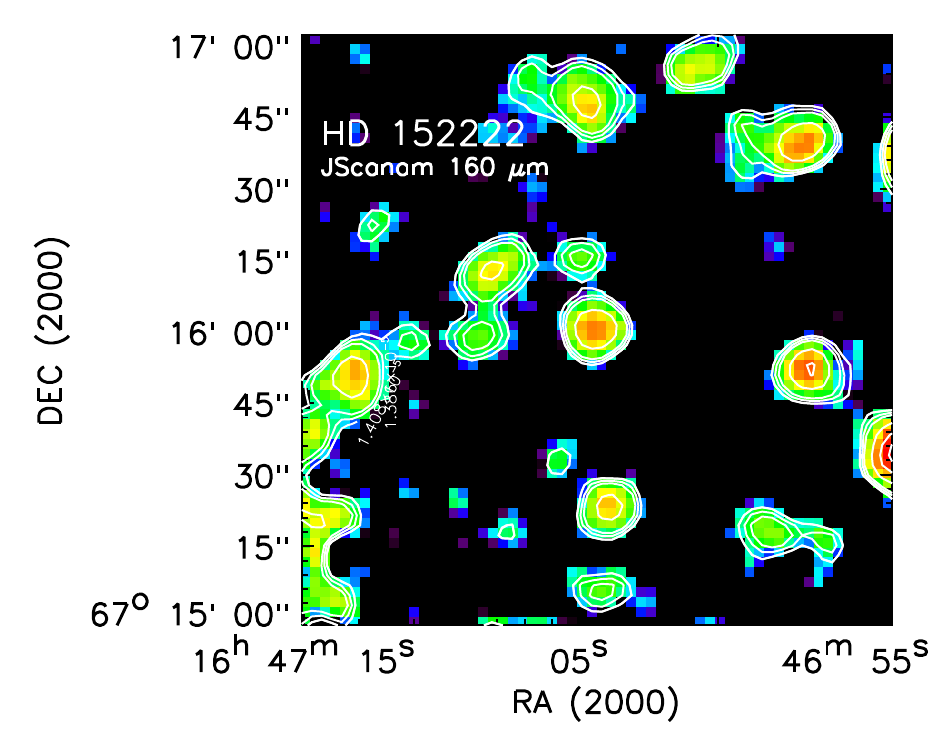}
   \includegraphics[width=0.24\textwidth]{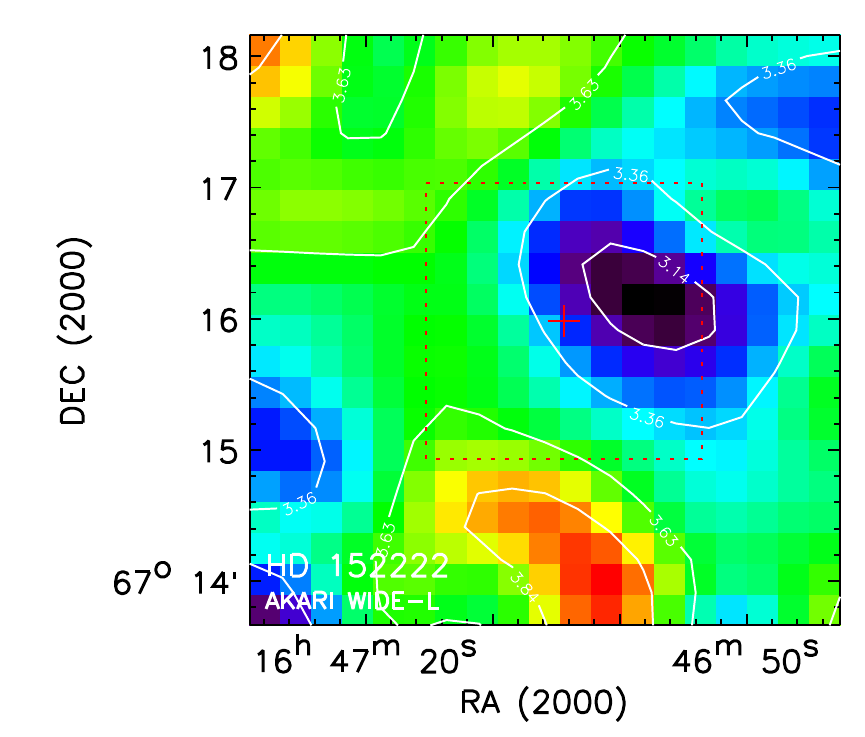}
      \caption{Investigation of the 160\,$\mu$m sky background structure around
               the sources $\delta$\,Dra (OD\,934), $\theta$\,UMi (OD\,160), 
               HD\,138265 (OD\,233), HD\,39608 (OD\,400) and HD\,152222 
               (OD\,843) from top to bottom. The deepest available maps were 
               used, see Table~\ref{table:bgconfinvest} for the OBSID 
               combination and Table~\ref{table:scanmapphotred10_1} for the 
               observation details. The left panel shows the high-pass filter 
               processed map used for the photometry, the photometric aperture 
               with 10\farcs7 radius is indicated by the red circle and a 
               small red or white cross in its centre. The middle panel shows 
               the JScanam processed map which should reproduce extended 
               emission more reliably. The JScanam map of $\delta$\,Dra shows 
               the superposition of all five sets of OBSIDs in 
               Table~\ref{table:bgconfinvest} (ODs 607, 751, 934, 1198 and 
               1328) as the deepest image of this field. The right panel shows 
               the {\em AKARI} WIDE-L (140\,$\mu$m) background emission around 
               the source (red cross), the {\em AKARI} map area is about four 
               times as large as the PACS map area, which is indicated by the
               red dashed square. 
              }
         \label{fig:bgconfusion}
   \end{figure*}

In Fig.~\ref{fig:bgconfusion} we investigate the nature of the 160\,$\mu$m
sky background structure, both on an absolute level and larger scale with the 
help of the {\it AKARI}-FIS WIDE-L (140\,$\mu$m) all-sky survey 
maps\footnotemark[4]~\citep{doi15} and on the PACS scale by parallel JScanam 
processing of the maps which tends to preserve more reliably small-scale 
structured extended emission, while the larger-scale background is subtracted.
In the following, we discuss the sources $\delta$\,Dra, $\theta$\,Umi, 
HD\,138265, HD\,39608 and HD\,152222 individually with regard to level and 
nature of their background confusion.

For $\delta$\,Dra the scan map photometry in Table~\ref{table:bgconfinvest} 
gives on average $\frac{f_{\rm 160,corr}^{\rm 10\farcs7}}{f_{\rm 160,corr}^{\rm
5\farcs35}}$ = 1.04, which is relatively consistent with 
$\frac{f_{160}}{f_{\rm model}}$ = 1.07 in Table~\ref{table:photminiscan}, given 
that also the 70 and 100\,$\mu$m fluxes are 2 -- 3\% above the model. The 
{\it AKARI}-map in Fig.~\ref{fig:bgconfusion} shows that the source is located 
at the wing of a cirrus knot. The JScanam map reveals filamentary structure 
around the source, which indicates that the small excess in the order of 4\% 
is likely by IR cirrus contamination. One out of the five cases investigated 
in Table~\ref{table:bgconfinvest} does not indicate any excess and the 
measured 10\farcs7 aperture flux is close to the model flux. This is an 
example of the ability of the scan map orientation to influence the structure 
of the source environment due to the high-pass filter running along the scan 
direction.
  
For $\theta$\,Umi the scan map photometry gives 
$\frac{f_{\rm 160,corr}^{\rm 10\farcs7}}{f_{\rm 160,corr}^{\rm 5\farcs35}}$ = 1.08, 
which is relatively consistent with $\frac{f_{160}}{f_{\rm model}}$ = 1.13 in 
Table~\ref{table:photminiscan}, meaning that the 160\,$\mu$m photometry in the
small aperture is quite close to the model flux. The {\it AKARI}-map in 
Fig.~\ref{fig:bgconfusion} shows that the source is located on a cirrus 
filament. The JScanam map reveals filamentary structure, too, coinciding in 
direction with the {\it AKARI}-map feature, which supports that the excess 
found for the default photometric aperture of 10\farcs7 is fully accounted for 
by IR cirrus contamination. 
  
HD\,138265 exhibits a noticeable FIR excess at 160$\mu$m only, which makes 
it a potential background-contaminated source, too. The {\it AKARI}-map in 
Fig.~\ref{fig:bgconfusion} shows that it is located on the wing of a small 
faint cirrus knot. The JScanam processed map indicates filamentary knotty 
structure mostly east, south, and west of the source, which fits to its 
location on the knot. The morphology of the filamentary structure resembles 
cirrus emission rather than compact sources. We derive 
$\frac{f_{\rm 160,corr}^{\rm 10\farcs7}}{f_{\rm 160,corr}^{\rm 5\farcs35}}$ = 1.11,
which only partially explains $\frac{f_{160}}{f_{\rm model}}$ = 1.36. In this 
case the background contribution may be more centered in the beam, meaning 
that it cannot be fully separated by the multi-aperture method.

%
%
\begin{table*}[ht!]
\caption{K-band and PACS 100\,$\mu$m photometry and a selection of stellar 
parameter information for the PACS fiducial primary standards (status: "f") and
PACS faint star primary standard candidates (status: "c") with nearly 
identical spectral type. Fiducial primary standards and related candidates are 
grouped together. V-band photometry is given for completeness and for the 
conversion of the K magnitude between the \citep{selby88} K$_{\rm n}$ narrow 
band photometric system and the Johnson K-band photometric system.}
\label{table:primeproperties}      
\begin{center}
    \begin{tabular}{l c c | c c c c c | c c}
   \hline\hline
            \noalign{\smallskip}
Name         & Status & SpType &  V  &      K       &K$_{\rm n}$\tablefootmark{p3}&K$_{\rm s,2MASS}$\tablefootmark{p4}&f$_{100}$\tablefootmark{p5}& T$_{\rm eff}$ & $\Theta_{\rm d}$ \\
             &        &        &(mag)&    (mag)     &   (mag)     &   (mag)        &   (mJy)   &    (K)      &   m\arcsec     \\
            \noalign{\smallskip}
\hline
            \noalign{\smallskip}
$\alpha$\,Boo&   f    & K2III  &-0.04&              &-3.07$\pm$0.01&               & 7509$\pm$375 & 4320$\pm$140\tablefootmark{s1}& 20.74$\pm$0.10\tablefootmark{s1} \\
$\alpha$\,Ari&   c    & K1IIIb & 2.01&-0.63$\pm$0.03\tablefootmark{p1}&              &               &  820$\pm$41  & 4636$\pm$13\tablefootmark{s2} &  6.90$\pm$0.07\tablefootmark{s5} \\
 42\,Dra     &   c    &K1.5III & 4.82& 1.95$\pm$0.05\tablefootmark{p2}&              &               &  73.1$\pm$3.8& 4446$\pm$12\tablefootmark{s2} &  2.03$\pm$0.03\tablefootmark{s5} \\
 HD\,159330\tablefootmark{1}  &   c    & K2III  & 6.21&              &              &2.787$\pm$0.206&  31.0$\pm$1.9&             &                \\
 HD\,152222  &   c    & K2III  & 7.03&              &              &3.654$\pm$0.238&  20.8$\pm$1.0&             &                \\
            \noalign{\smallskip}
\hline
            \noalign{\smallskip}
$\alpha$\,Tau&   f    & K5III  & 0.85&              &-2.94$\pm$0.01&               & 6909$\pm$345 & 3850$\pm$140\tablefootmark{s1}& 20.89$\pm$0.10\tablefootmark{s1} \\
$\gamma$\,Dra&   f    & K5III  & 2.23&              &-1.38$\pm$0.01&               & 1604$\pm$80  & 3960$\pm$140\tablefootmark{s1}&  9.94$\pm$0.05\tablefootmark{s1} \\
$\varepsilon$\,Lep& c & K4III  & 3.18&-0.20$\pm$0.03\tablefootmark{p1}&              &               &  568$\pm$28  & 4243$\pm$25\tablefootmark{s2} &  6.08$\pm$0.06\tablefootmark{s5} \\
 HD\,41047   &   c    & K5III  & 5.52&              &              &1.740$\pm$0.218&  96.7$\pm$4.9&             &  2.47$\pm$0.03\tablefootmark{s4} \\
$\theta$\,Umi\tablefootmark{1}&   c    & K5III  & 4.98& 1.33$\pm$0.05\tablefootmark{p2}&              &        &  144$\pm$8.8 &             &  2.97$\pm$0.04\tablefootmark{s5} \\
HD\,138265\tablefootmark{1}   &   c    & K5III  & 5.90& 2.38$\pm$0.04\tablefootmark{p2}&              &        & 56.5$\pm$2.9 & 3758$\pm$166\tablefootmark{s3} &  2.06$\pm$0.04\tablefootmark{s3} \\
            \noalign{\smallskip}
\hline
            \noalign{\smallskip}
$\beta$\,And &   f    & M0III  & 2.06&              &-1.93$\pm$0.01&               & 2737$\pm$137 & 3880$\pm$140\tablefootmark{s1}& 13.03$\pm$0.06\tablefootmark{s1} \\
$\omega$\,Cap&   c    & M0III  & 4.12& 0.21$\pm$0.03\tablefootmark{p1}&              &               &  413$\pm$21  & 3760$\pm$150\tablefootmark{s4} &  5.16$\pm$0.06\tablefootmark{s5} \\  
            \noalign{\smallskip}
\hline
    \end{tabular} \\
\end{center}
$\tablefoottext{1}$~Star is a proven reliable standard only up to 100\,$\mu$m due to background confusion \\
$\tablefoottext{p1}$~\citet[][catalogue of stellar photometry in Johnson's 11-colour system]{ducati02} \\
$\tablefoottext{p2}$~\citet[][two-micron sky survey]{neugebauer69} \\
$\tablefoottext{p3}$~\citet{selby88} \\
$\tablefoottext{p4}$~\citet[][2MASS all-sky catalogue of point sources]{cutri03}~~~~~Note: For K$_{\rm s,2MASS} <$ 4\,mag photometric uncertainties are high, because fluxes were estimated from a fit to the wings of the saturated stellar profile~\citep{skrutskie06}.  \\
$\tablefoottext{p5}$~For the fiducial standards we use the continuum model\footnotemark[8] flux with an uncertainty of 5\%; for the candidate stars we use the flux from scan map photometry. \\
$\tablefoottext{s1}$~\citet{dehaes11} \\
$\tablefoottext{s2}$~\citet{jofre15} \\
$\tablefoottext{s3}$~\citet{baines10} \\
$\tablefoottext{s4}$~\citet{tsuji81} \\
$\tablefoottext{s5}$~\citet{cohen99}
\end{table*}

HD\,39608 shows the by far strongest 160\,$\mu$m excess, with 
$\frac{f_{160}}{f_{\rm model}}$ = 2.20. Our multi-aperture photometry gives only 
$\frac{f_{\rm 160,corr}^{\rm 10\farcs7}}{f_{\rm 160,corr}^{\rm 5\farcs35}}$ = 1.26.
The {\it AKARI}-map in Fig.~\ref{fig:bgconfusion} shows that it is located on 
the wings of two brighter cirrus knots with a depression south of it. The 
JScanam map reveals that there is extended filamentary structure mainly north 
of the source, in agreement with the larger-scale feature of the 
{\it AKARI}-map. This also affects the area where the background is determined. 
Given that the source is one of the faintest in our sample, with an expected 
photospheric flux of only about 6\,mJy, any inaccuracy in the background 
determination has a severe impact on the resulting source flux. Furthermore,
the source looks elongated in the north-south direction, which indicates
contaminating emission inside the measurement aperture. From 
Fig.~\ref{scanmap_comp_model}, we see that there is already a noticeable 
excess of 19\% at 100\,$\mu$m. Unfortunately we cannot investigate this 
properly on a JScanam processed map, since there exists only one map in one 
scan direction (the cross-scan map was erroneously executed in the 70\,$\mu$m 
filter). HD\,39608 has the second strongest ISM sky background 
B$_{\rm 160}^{ISM}$ (c.f.\ Table~\ref{table:bgconfinvest}), meaning that it is 
very likely that already at 100\,$\mu$m there can be significant sky 
background contamination. The deep combined 70\,$\mu$m map shows an elongated 
emission feature underneath the source, too.

For comparison we also show the environment of HD\,152222 in 
Fig.~\ref{fig:bgconfusion}, which is only slightly brighter than HD\,39608. 
The {\it AKARI}-map shows that it is located outside a cirrus ridge close to a 
depression in the cirrus emission. The JScanam map reveals that the area 
around it is also crowded, but the sources are discrete compact sources, which 
argues for an extragalactic nature, and besides the star itself appears 
isolated inside the measurement aperture. The derived flux is quite consistent 
with the model flux which argues against a systematic background underestimate 
in this source flux range. 

HD\,39608 is hence no longer qualified as a potential calibration standard. 
$\theta$\,Umi and HD\,138265 can be considered as suitable standards up to 
100\,$\mu$m.

\section{Establishment of new faint FIR primary standards}
\label{sect:primestandardestablish}

Primary flux standards are used for absolute flux calibration. Their SED is 
assumed to be known and stable or predictable. Absolute calibration of these 
sources is achieved either by a direct method, like comparison against a 
blackbody source, or by a indirect method, for example\ stellar or planetary 
atmosphere models. \citet{deustua13} give a detailed description of absolute 
calibration of astronomical flux standards. Primary flux standards in the 
far-infrared wavelength range are, with decreasing brightness, 
planets~\citep{mueller16}, asteroids~\citep{mueller14}, and 
stars~\citep{dehaes11}, which are all calibrated via the indirect method and
verified by independently calibrated multi-wavelength flux measurements. The 
best achievable uncertainties are currently 5 -- 7\%.

The \citet{cohen96} models of $\alpha$\,Ari, $\varepsilon$\,Lep, $\omega$\,Cap, 
$\delta$\,Dra and HD\,41047 are well confirmed by our PACS photometry and are 
thus adequate representations of the stellar FIR photospheric emission. These 
stars together with 42\,Dra and HD\,152222 are good candidates to establish 
fainter FIR primary standards. This list is complemented by $\theta$\,Umi, 
HD\,159330, and HD\,138265 for which we can confirm a reliable FIR spectrum 
only up to 100\,$\mu$m due to neighbouring source- or cirrus confusion at 
longer wavelengths.

As already discussed earlier, the~\citet{cohen96} models are FIR extensions of 
absolutely calibrated 1.2 -- 35\,$\mu$m template 
spectra~\citet{cohen95,cohen99}. Another set of models ranging from 
0.7\,$\mu$m to 7\,cm was developed by~\citet{dehaes11} for the 
{\em Herschel}-PACS fiducial primary standards. Several of our faint primary 
standard candidates have the same or similar spectral type as one of the PACS 
primary standard stars. As a first model approximation we can scale these 
fiducial standard star models to the flux levels of our primary standard 
candidates. For an accurate model one would have to run a flux model code 
taking into account the stellar parameter information of each star, a project 
which is beyond the scope of this paper. 

$\delta$\,Dra with spectral type G9III has no suitable counterpart among the 
{\em Herschel}-PACS fiducial primary standards, since the earliest spectral 
type is K2III. But we note that it was modelled earlier 
by~\citep{decin03} as {\em ISO}-SWS calibrator. We do not include $\delta$\,Dra 
in Table~\ref{table:primeproperties}, but refer to Table~3 in~\citet{decin03} 
which gives its stellar properties.

In Table~\ref{table:primeproperties} we have compiled photometry and stellar 
properties of the fiducial primary standards and our primary standard 
candidates which match in spectral type. \citet{jofre15} provide essential 
stellar parameters for $\alpha$\,Ari, $\varepsilon$\,Lep, and 42\,Dra, 
therefore a dedicated flux model code could be run.

In Table~\ref{table:primecomp} we compile the K-magnitude ratio and the 
100\,$\mu$m flux ratio for matching pairs of fiducial primary standards and 
primary standard candidates. From both ratios we compute the scale factor for 
the fiducial star model as a weighted mean. We apply the following 
transformations between the different K-band photometric systems:  
(1) V-K = -0.020 + 0.989 $\times$ (V-K$_{\rm n}$) \citep{selby88}
and (2) K = K$_{\rm s,2MASS}$+0.044 \citep{bessel05}.
Scale factors are between 0.356 and 0.0026, that is,\ the faintest primary
standard candidate, HD\,152222, is about 360 times fainter than the related 
fiducial primary standard $\alpha$\,Boo and still about 80 times fainter than 
the faintest fiducial primary standard, $\gamma$\,Dra. 
Table~\ref{table:primecomp} also gives the percentage of the uncertainty of 
the scaling. We estimate the uncertainty due to variation in stellar 
parameters, such as the effective temperature, by scaling the fiducial model
of $\alpha$\,Tau to the level of $\gamma$\,Dra, which are both K5III stars. 
The difference over the wavelength range 2 -- 250\,$\mu$m is less than 0.8\%. 
We therefore adopt an uncertainty of 1\% due to variations in stellar 
parameters. Given, that the fiducial primary standards have an absolute 
accuracy of 5\%, then the absolute uncertainty of the scaled model 
approximation, as listed in the last column of Table~\ref{table:primecomp}, 
can be determined as the sum of the three uncertainty terms described above. 

\begin{figure*}[ht!]
   \centering
   \includegraphics[width=0.99\textwidth]{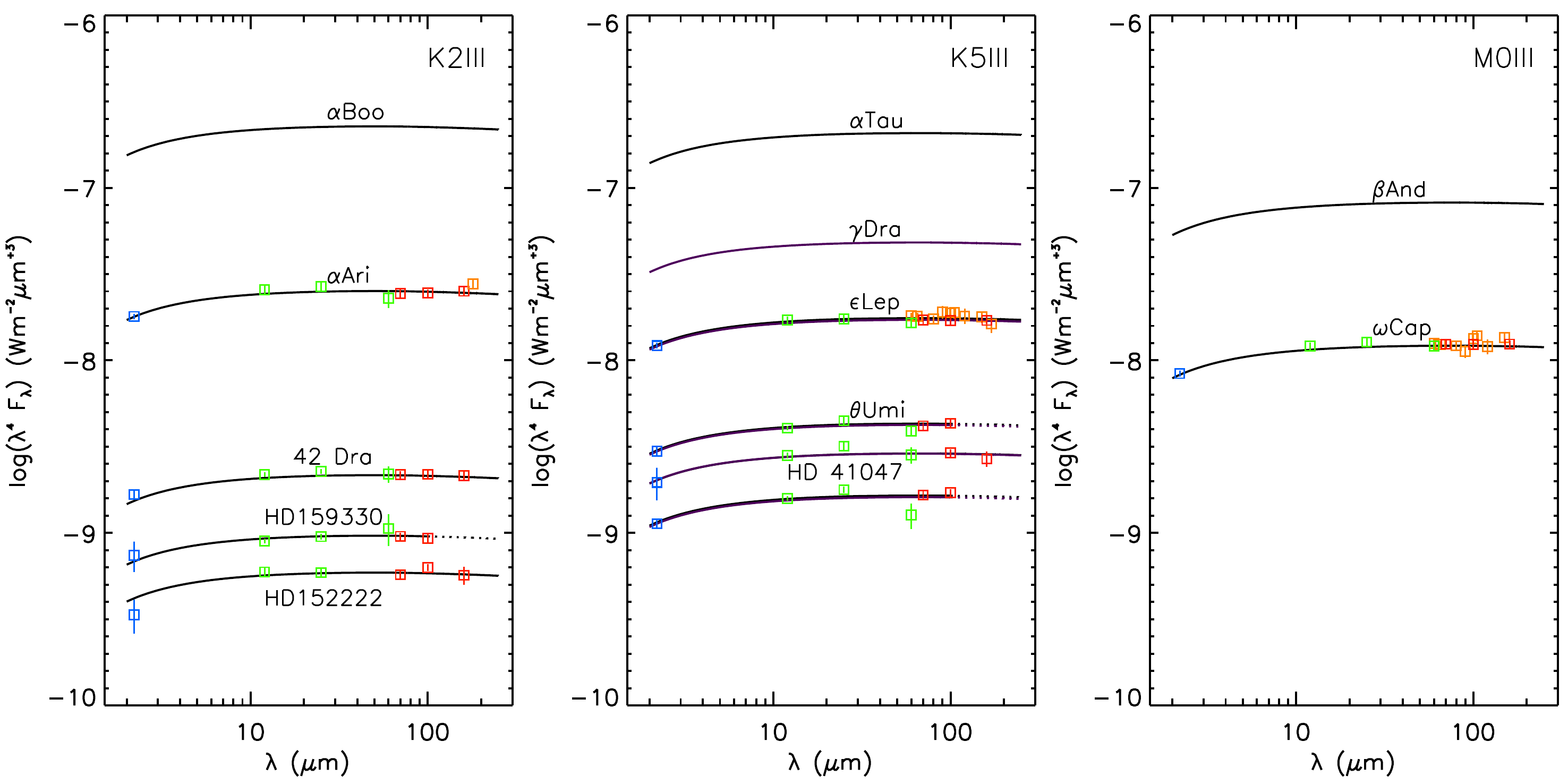}
      \caption{Scaling of PACS fiducial star continuum models (black and
               purple lines) to the flux level of the primary standard
               candidates applying the scale factors of 
               Table~\ref{table:primecomp}. For a better zoom-in over a large
               wavelength range, log$_{\rm 10}$($\lambda^4 \cdot f_{\rm \lambda}$)
               is displayed. Blue squares are the K-band photometry, green 
               squares are colour-corrected {\em IRAS} FSC 
               photometry~\citep{moshir89}, orange squares are ISOPHOT HPDP 
               photometry~\citep[][see Appendix~\ref{sect:appd}]{lemke96}
               and red squares are PACS photometry. In the middle panel the 
               scaled models of both $\alpha$\,Tau (black) and $\gamma$\,Dra 
               (purple) are plotted. Dashed parts of the SEDs of HD\,159330, 
               $\theta$\,Umi, and HD\,138265 indicate that these stars are 
               proven reliable standards only up to 100\,$\mu$m due to 
               background confusion.
              }
         \label{fiducial_star_scale}
   \end{figure*}

Figure~\ref{fiducial_star_scale} shows the scaling of the fiducial standard 
star models to the flux levels of our primary standard candidates and a 
verification with available photometry. K-band and PACS photometry are 
supplemented by {\em IRAS} FSC photometry~\citep{moshir89} and in some cases 
by ISOPHOT HDPD photometry~\citep[][see Appendix~\ref{sect:appd}]{lemke96}. 

%
%
\begin{table*}[ht!]
\caption{Determination of the scale factor to adjust the related fiducial 
 primary standard star model to the flux level of the candidate primary 
 standard. The last two columns list the uncertainty percentage of the scale 
 factor and the absolute uncertainty of the scaled model approximation, the 
 latter being the Gaussian error propagation of the scale uncertainty, 
 1\% uncertainty in the stellar parameters and 5\% uncertainty of the models. 
}             
\label{table:primecomp}      
\begin{center}
    \begin{tabular}{l l c c c c c}
   \hline\hline
            \noalign{\smallskip}
Primary candidate & Primary standard&    Kmag ratio   &  f$_{100}$ ratio & Scale factor & \% scale uncert. & \% abs. uncert.\\
            \noalign{\smallskip}
\hline
            \noalign{\smallskip}
$\alpha$\,Ari   & $\alpha$\,Boo  & 1.11$\pm$0.04\,10$^{-1}$ & 1.09$\pm$0.11\,10$^{-1}$ & 1.11$\pm$0.01\,10$^{-1}$ &  0.9 &  5.2 \\
 42\,Dra        & $\alpha$\,Boo  & 1.03$\pm$0.06\,10$^{-2}$ & 9.73$\pm$1.00\,10$^{-3}$ & 1.01$\pm$0.03\,10$^{-2}$ &  3.0 &  5.9 \\
HD\,152222      & $\alpha$\,Boo  & 2.07$\pm$0.46\,10$^{-3}$ & 2.77$\pm$0.27\,10$^{-3}$ & 2.59$\pm$0.31\,10$^{-3}$ & 12.0 & 13.0 \\
HD\,159330\tablefootmark{1}      & $\alpha$\,Boo  & 4.58$\pm$0.92\,10$^{-3}$ & 4.13$\pm$0.46\,10$^{-3}$ & 4.22$\pm$0.18\,10$^{-3}$ &  4.3 &  6.7 \\
            \noalign{\smallskip}
\hline
            \noalign{\smallskip}
$\varepsilon$\,Lep& $\alpha$\,Tau& 8.49$\pm$0.31\,10$^{-2}$ & 8.22$\pm$0.82\,10$^{-2}$ & 8.46$\pm$0.09\,10$^{-2}$ &  1.1 &  5.2 \\
$\varepsilon$\,Lep& $\gamma$\,Dra& 3.56$\pm$0.13\,10$^{-1}$ & 3.54$\pm$0.35\,10$^{-1}$ & 3.56$\pm$0.007\,10$^{-1}$&  0.2 &  5.1 \\
HD\,41047       &  $\alpha$\,Tau & 1.37$\pm$0.29\,10$^{-2}$ & 1.40$\pm$0.14\,10$^{-2}$ & 1.39$\pm$0.01\,10$^{-2}$ &  0.7 &  5.1 \\
HD\,41047       &  $\gamma$\,Dra & 5.73$\pm$1.21\,10$^{-2}$ & 6.03$\pm$0.61\,10$^{-2}$ & 5.97$\pm$0.12\,10$^{-2}$ &  2.0 &  5.5 \\
$\theta$\,Umi\tablefootmark{1}   &  $\alpha$\,Tau & 2.07$\pm$0.12\,10$^{-2}$ & 2.08$\pm$0.23\,10$^{-2}$ & 2.07$\pm$0.004\,10$^{-2}$&  0.2 &  5.1 \\
$\theta$\,Umi\tablefootmark{1}   &  $\gamma$\,Dra & 8.71$\pm$0.48\,10$^{-2}$ & 8.98$\pm$1.00\,10$^{-2}$ & 8.76$\pm$0.11\,10$^{-2}$ &  1.3 &  5.3 \\
HD\,138265\tablefootmark{1}      &  $\alpha$\,Tau & 7.88$\pm$0.36\,10$^{-3}$ & 8.18$\pm$0.83\,10$^{-3}$ & 7.93$\pm$0.11\,10$^{-3}$ &  1.4 &  5.3 \\
HD\,138265\tablefootmark{1}      &  $\gamma$\,Dra & 3.31$\pm$0.15\,10$^{-2}$ & 3.52$\pm$0.36\,10$^{-2}$ & 3.34$\pm$0.07\,10$^{-2}$ &  2.2 &  5.6 \\
            \noalign{\smallskip}
\hline
            \noalign{\smallskip}
$\omega$\,Cap   & $\beta$\,And   & 1.48$\pm$0.06\,10$^{-1}$ & 1.51$\pm$0.15\,10$^{-1}$ & 1.48$\pm$0.01\,10$^{-1}$ &  0.7 &  5.1 \\
            \noalign{\smallskip}
\hline
    \end{tabular} \\
\end{center}
$\tablefoottext{1}$~Star is a proven reliable standard only up to 100\,$\mu$m due to background confusion.
\end{table*}

For $\alpha$\,Ari, $\varepsilon$\,Lep, $\omega$\,Cap, HD\,41047 and 42\,Dra
the derived absolute photometric uncertainty is in the range 5--6\%, hence 
they are well suited new FIR primary standards, which are about 2 -- 20 times 
fainter than our faintest fiducial primary standard $\gamma$\,Dra. Only for 
the faintest star, HD\,152222, does the higher uncertainty of the scaling 
factor result in a derived absolute uncertainty of 13\%. A major driver for 
this high scaling uncertainty is the high uncertainty of the publicly 
available K-band photometry (cf.\ Table~\ref{table:primeproperties}, footnote 
p4), meaning that the K-magnitude ratio and the 100\,$\mu$m flux ratio do not 
match well. From Fig.~\ref{fiducial_star_scale} it is obvious that a more 
accurate K-band photometry would certainly help to bring this star into a 
similar absolute photometric uncertainty range to the brighter ones. 

For $\theta$\,Umi, HD\,138265 and HD\,159330, which are substantially affected
by neighbouring source- or cirrus confusion at 160\,$\mu$m 
(Sects.~\ref{sect:bgconfusion} and~\ref{sect:starsfirexcess}), clean photometry 
can be obtained up to 100\,$\mu$m with a telescope of angular resolution 
similar to {\it Herschel}, which leads to an equally good absolute photometric 
uncertainty in the range 5 -- 7\%. Also, for HD\,159330, improved K-band 
photometry can further reduce its resulting absolute photometric uncertainty.
We therefore keep these three sources as reliable standards up to 100\,$\mu$m, 
but with the strong caveat not to use them beyond this wavelength. Only with a 
considerably higher angular resolution than {\it Herschel} could the confusion 
issues of these sources be overcome at 160\,$\mu$m.

HD\,138265, HD\,159330, and HD\,152222 will be observable with the
{\it James Webb Space Telescope} MIRI Imager at 20\,$\mu$m in bright source 
mode with the 64 $\times$ 64 sub-array~\citep{bouchet15}.

%
%

\section{Conclusions}

The PACS faint star sample with 14 giant and 3 dwarf stars has allowed a 
comprehensive sensitivity assessment of the PACS photometer and provided 
accurate photometry for detailed SED investigation and establishment of a set 
of faint FIR primary standard candidates for use by future space missions.

For PACS scan maps, the recommended scientific observation mode for the PACS 
photometer, we have described a consistent method for how to derive S/Ns, 
based on a robust noise measurement with the help of a flux histogram 
restricted to the applicable map coverage value range. The comparison with the 
S/N predictions of the exposure time calculation tool in the {\em Herschel} 
Observatory Planning Tool HSpot has resulted in very good consistency, proving 
the tools for PACS photometry observation planning as very reliable. We have 
demonstrated that the underlying assumptions of the tool, that the S/N scales 
linearly with flux and with the square root of the observing time, are valid 
over large ranges of flux and time. A restriction appears for the 160\,$\mu$m 
filter, where source confusion often limits the gain in S/N with increasing 
observing time. We could also show that scan maps obtained with the 
recommended scan speed of 20\arcsec/s yield a higher S/N than scan maps with 
10\arcsec/s, the scan speed favoured pre-flight.

We have shown that in the case of faint sources, small aperture sizes (with a 
radius of the size of the PSF FWHM) reduces background noise inside the 
aperture and optimizes the accuracy of the flux determination.

We have obtained reliable photometry for 11 stars in all three PACS filters
(at 70, 100, 160\,$\mu$m). For one further star we have obtained reliable 100 
and 160\,$\mu$m photometry. For one more star we have obtained reliable 70 and 
100\,$\mu$m photometry only, 160\,$\mu$m photometry being limited here by 
confusion of neighbouring sources. For two other stars we have obtained 
reliable photometry only at 70\,$\mu$m, a detection at longer wavelengths 
being limited by sensitivity limitations and confusion noise. The two faintest 
sources observed in chop/nod mode have not been detected at all despite high 
repetition factors of the basic chop/nod pattern. The non-detection is likely 
due to a not-yet-perfect knowledge of the optimum observing strategy early in 
the mission. Faintest fluxes, for which the photometry has still good quality, 
are about 10 -- 20\,mJy for the scan map observations and 30\,mJy for the 
available chop/nod observations.

For the faintest star at 160\,$\mu$m with reliable photometry in all three 
filters, HD\,152222, we have conducted an inter-comparison of the high-pass 
filter (HPF) photometry from the deepest map with the results of three 
additional {\it Herschel} mapper softwares, namely JScanam, Scanamorphos and 
Unimap. All four mappers allow us to obtain sound photometry in all three 
filters. We have identified the level of qualitative consistency as well as 
some systematic differences with regard to photometry, noise, and beam profiles 
among the four mappers. A more systematic and quantitative photometric 
performance comparison of the four mappers will be the subject of a dedicated 
publication. 

For the 12 stars with reliable photometry out to 160\,$\mu$m we can prove that 
7 stars are consistent with models or flux predictions for pure photospheric 
emission. $\delta$\,Dra has a slight 160\,$\mu$m excess due to cirrus 
contamination of the order of 4\%, but this is still within the overall 
uncertainty margin. Two stars show excess emission over the whole 
($>$10\,$\mu$m) FIR range. For $\beta$\,Gem (Pollux), which is the host star 
of a confirmed Jupiter-sized exoplanet, we conclude from our photometry 
results that it has in addition a flat blackbody dust disk. The G8 giant 
$\eta$\,Dra has a K1 dwarf companion, therefore the origin of the excess 
emission likely arises from dust inside this binary system. For three stars 
with 160\,$\mu$m fluxes below 60\,mJy we find 160\,$\mu$m excesses in the 
order 6 to 9\,mJy. Investigation of the 160\,$\mu$m absolute sky brightness 
with the help of {\it AKARI}-maps, the filamentary emission structure in the 
environment of the source on the PACS maps, and multi-aperture PACS photometry 
strongly support an explanation of this excess as being due to sky background 
confusion. This is a combination of cirrus confusion affecting the background 
subtraction and faint underlying objects inside the photometric aperture 
around the star affecting the source profile. The faintest star at 70\,$\mu$m 
with reliable photometry in all three filters, HD\,39608, is heavily affected 
by sky confusion noise from 100\,$\mu$m onwards and has therefore to be 
excluded as a primary standard candidate.

The seven stars with pure photospheric emission over the full PACS wavelength 
range, $\alpha$\,Ari, $\varepsilon$\,Lep, $\omega$\,Cap, $\delta$\,Dra, 
HD\,41047, 42\,Dra and HD\,152222, are promising primary standard candidates. 
The stars $\theta$\,Umi, HD 138265 and HD 159330 prove to be good primary 
standard candidates, too, but only up to 100\,$\mu$m due to significant source 
confusion at 160\,$\mu$m at the spatial resolution of PACS. For three of the 
new primary standard candidates essential stellar parameters are known, 
meaning that a dedicated flux model code could be run. As a good model 
approximation for nine of our primary standard candidates we can scale the 
continuum flux models of four PACS fiducial standards with the same or quite 
similar spectral type. Only for $\delta$\,Dra is there no suitable counterpart 
among the fiducial standard stars. This allows us to establish a set of five 
FIR primary standard candidates up to 160\,$\mu$m, which are 2 -- 20 times 
fainter than the faintest PACS fiducial standard ($\gamma$\,Dra) with absolute 
accuracy of $<$6\%. The accuracy for the faintest primary standard candidate, 
HD152222 (80 times fainter than $\gamma$\,Dra), is currently limited to 13\% 
by the accuracy of the existing K-band photometry. A set of three primary 
standard candidates up to 100\,$\mu$m with an absolute accuracy of $<$7\% 
complements the list of proven flux standards.

\begin{acknowledgements}
      PACS has been developed by a consortium of institutes led by 
      MPE (Germany) and including UVIE (Austria); KUL, CSL, IMEC (Belgium); 
      CEA, OAMP (France); MPIA (Germany); IFSI, OAP/AOT, OAA/CAISMI, LENS, 
      SISSA (Italy); IAC (Spain). This development has been supported by the 
      funding agencies BMVIT (Austria), ESA-PRODEX (Belgium), 
      CEA/CNES (France), DLR (Germany), ASI (Italy), and CICYT/MCYT (Spain).
      ZB acknowledges funding by DLR for this work. TM receives funding
      from the European Union’s Horizon 2020 Research and Innovation Programme,
      under Grant Agreement no.\ 687378. This research has made use of the
      SIMBAD data base and the VizieR catalogue access tool, operated at CDS, 
      Strasbourg, France. This research has made use of SAOImage DS9,
      developed by Smithsonian Astrophysical Observatory.
      This research has made use of the NASA/IPAC Infrared Science Archive, 
      which is operated by the Jet Propulsion Laboratory, California Institute 
      of Technology, under contract with the National Aeronautics and Space 
      Administration. We thank the referee for constructive comments.
\end{acknowledgements}

\bibliographystyle{aa}

\clearpage

\begin{appendix}

\section{Scan map photometry}
\label{sect:appa}

\subsection{Comparison of different aperture sizes for optimization
  of photometric aperture}
\label{sect:compapersize}

%

%
%
\begin{table}[ht!]
\caption{Comparison of mini scan map photometry for different aperture sizes. 
phot\_l is the photometry with the large aperture sizes 12"/12"/22", phot\_s
is the photometry with the small aperture sizes 5.6"/6.8"/10.7". Stellar 
fluxes f$_{\rm star}$ are determined as the colour-corrected weighted average of 
aperture corrected fluxes f$_{\rm tot}$ from \# of individual scan maps. Model 
fluxes are from Table~\ref{table:sources}. Quoted uncertainties are the 
weighted standard deviations for \#$\ge$2 and $\sigma_{\rm aper,corr}$ 
(Eq.~\ref{eq:sigapercorr}) for \#=1.
}             
\label{table:photminiscanapercom}      
\centering                          
\begin{tabular}{r r r c c c c}        
\hline\hline                 
            \noalign{\smallskip}
  Star             & Filter & \# &      phot\_ l     &      phot\_s  &  Model \\
                   &($\mu$m)&    &       (mJy)      &       (mJy)   &  (mJy) \\
            \noalign{\smallskip}
\hline                        
            \noalign{\smallskip}
$\beta$\,Gem       &   70   &  8 & 2649.4$\pm$4.1 & 2648.7$\pm$1.2 & 2457. \\
                   &  100   &  8 & 1287.1$\pm$2.4 & 1283.8$\pm$1.4 & 1190. \\
                   &  160   & 16 &  497.4$\pm$5.8 &  496.9$\pm$2.1 & 455.9 \\
$\alpha$\,Ari      &   70   &  8 & 1668.3$\pm$3.7 & 1663.7$\pm$3.1 & 1707. \\
                   &  100   &  8 &  831.6$\pm$4.3 &  820.2$\pm$1.8 & 831.4 \\
                   &  160   & 16 &  336.0$\pm$3.0 &  328.0$\pm$1.9 & 321.0 \\
$\varepsilon$\,Lep &   70   &  8 & 1157.1$\pm$6.5 & 1165.6$\pm$2.3 & 1182. \\
                   &  100   & 12 &  568.8$\pm$2.7 &  568.4$\pm$1.6 & 576.2 \\
                   &  160   & 18 &  217.5$\pm$6.5 &  223.5$\pm$2.6 & 222.7 \\
$\omega$\,Cap      &   70   &  8 &  839.0$\pm$2.9 &  845.2$\pm$1.7 & 857.7 \\
                   &  100   &  8 &  414.3$\pm$1.8 &  412.8$\pm$1.4 & 418.0 \\   
                   &  160   & 16 &  168.6$\pm$4.0 &  160.5$\pm$1.9 & 161.5 \\
$\eta$\,Dra        &   70   &  8 &  517.8$\pm$13.0&  506.0$\pm$2.6 & 479.5 \\
                   &  100   &  8 &  237.9$\pm$9.1 &  249.5$\pm$4.5 & 232.6 \\
                   &  160   & 16 &  116.4$\pm$7.3 &   98.4$\pm$5.1 &  89.4 \\
$\delta$\,Dra      &   70   & 12 &  433.6$\pm$2.1 &  436.0$\pm$1.1 & 428.9 \\
                   &  100   & 10 &  214.0$\pm$1.6 &  214.2$\pm$1.4 & 207.7 \\
                   &  160   & 22 &   89.0$\pm$5.1 &   85.4$\pm$2.2 &  79.6 \\
$\theta$\,Umi      &   70   &  4 &  278.7$\pm$5.7 &  284.2$\pm$1.5 & 286.2 \\
                   &  100   &  2 &  128.0$\pm$2.4 &  144.2$\pm$5.0 & 139.5 \\
                   &  160   &  2 &   72.8$\pm$10.4&   62.3$\pm$2.5 &  53.9 \\
HD\,41047          &  100   &  2 &   99.4$\pm$2.3 &   96.7$\pm$0.7 &  95.4 \\
                   &  160   &  2 &   52.8$\pm$4.1 &   35.7$\pm$5.6 &  36.9 \\
42\,Dra            &   70   &  4 &  146.5$\pm$1.8 &  147.8$\pm$0.9 & 153.7 \\
                   &  100   &  4 &   75.1$\pm$1.7 &   73.1$\pm$0.9 &  75.3 \\
                   &  160   &  8 &   31.2$\pm$1.4 &   28.2$\pm$1.0 &  29.4 \\
HD\,138265         &   70   &  4 &  109.2$\pm$2.6 &  112.8$\pm$1.0 & 115.9 \\
                   &  100   &  6 &   57.2$\pm$1.6 &   56.5$\pm$0.5 &  56.8 \\
                   &  160   &  5 &   31.2$\pm$6.1 &   30.8$\pm$1.0 &  22.2 \\
HD\,159330         &   70   &  4 &   60.7$\pm$4.7 &   64.8$\pm$1.1 &  64.2 \\
                   &  100   &  6 &   32.9$\pm$2.3 &   31.0$\pm$1.1 &  31.5 \\
HD\,152222         &   70   &  4 &   35.1$\pm$1.0 &   39.0$\pm$0.9 &  39.4 \\
                   &  100   &  2 &   22.4$\pm$0.7 &   20.8$\pm$0.1 &  19.3 \\
                   &  160   &  4 &    6.2$\pm$2.2 &    7.4$\pm$0.8 &   7.5 \\
HD\,39608          &   70   &  3 &   29.8$\pm$3.4 &   30.5$\pm$1.1 &  30.9 \\
                   &  100   &  1 &   18.1$\pm$2.2 &   17.9$\pm$0.9 &  15.1 \\
                   &  160   &  2 &   18.2$\pm$2.6 &   12.1$\pm$0.7 &   5.9 \\
HD\,181597         &   70   &  2 &   25.9$\pm$3.3 &   29.1$\pm$2.0 &  28.0 \\      
$\delta$\,Hyi      &   70   &  1 &    7.7$\pm$2.4 &   22.2$\pm$1.5 &  22.9 \\
          \noalign{\smallskip}
\hline                                   
\end{tabular}
\end{table}

\subsection{Dependence on applied mapper software}
\label{sect:mappersoftware}

For the faintest star at 160\,$\mu$m with reliable photometry in all three 
filters, HD\,152222, we conduct an inter-comparison of the high-pass filter 
(HPF) photometry from the deepest map with the results of three additional 
{\em Herschel} mapper softwares, namely JScanam~\citep{gracia15}, 
Scanamorphos\footnotemark[11]\footnotetext[11]{{\em Herschel} 
user contributed software  \\
https://www.cosmos.esa.int/web/herschel/user-contributed-software}~\citep{roussel13} 
and Unimap\footnotemark[11]~\citep{piazzo15}. The data analysis was done by 
applying the standard HIPE ipipe (interactive pipeline) scripts of these 
mappers\footnotemark[12]\footnotetext[12]{{\it Herschel} data processing 
overview  \\
https://www.cosmos.esa.int/web/herschel/data-processing-overview \\ in 
particular PACS Data Reduction Guide Photometry} and selecting the same output
pixel sizes as defined in Table~\ref{table:scanmapparamsphotnoise}. 

%
   \begin{figure*}[ht!]
   \centering
   \includegraphics[width=0.24\textwidth]{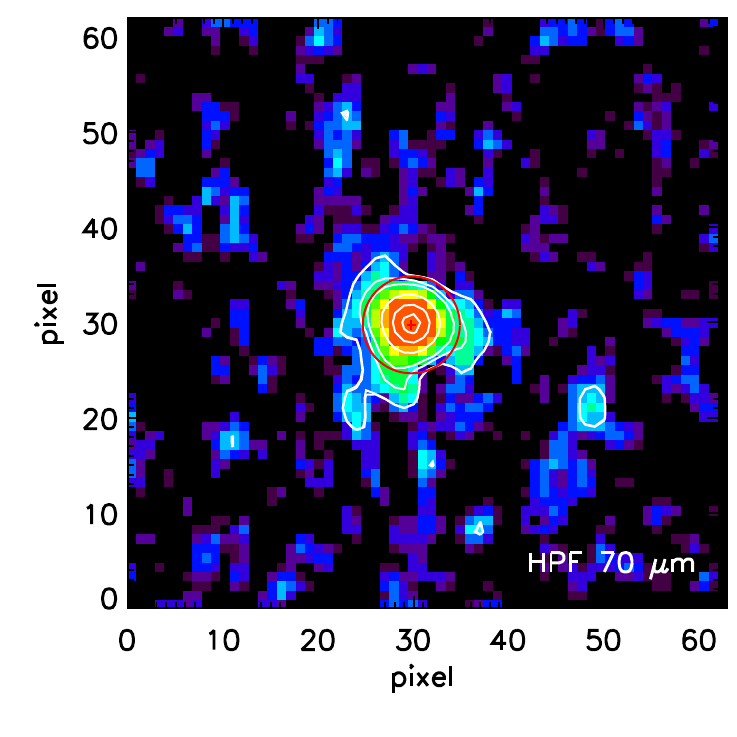}
   \includegraphics[width=0.24\textwidth]{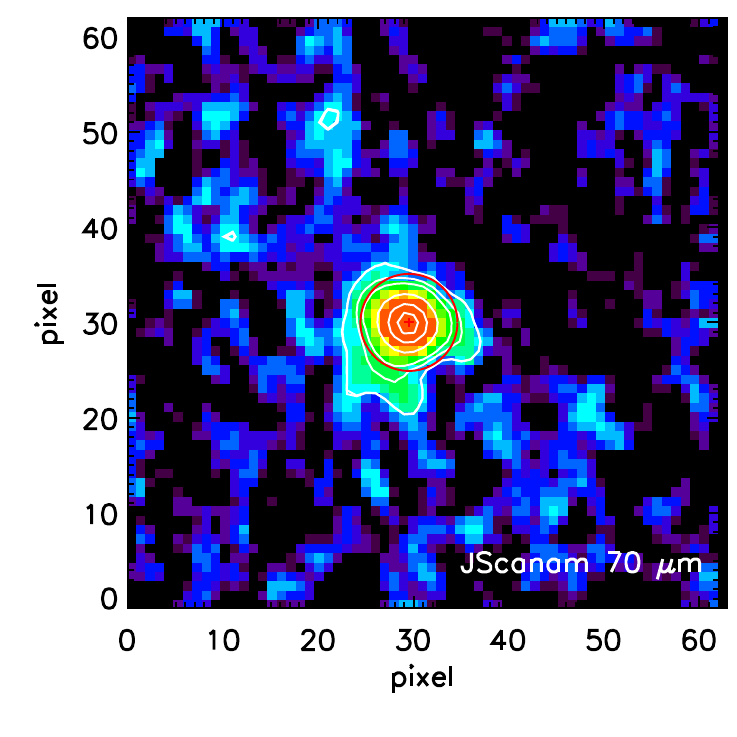}
   \includegraphics[width=0.24\textwidth]{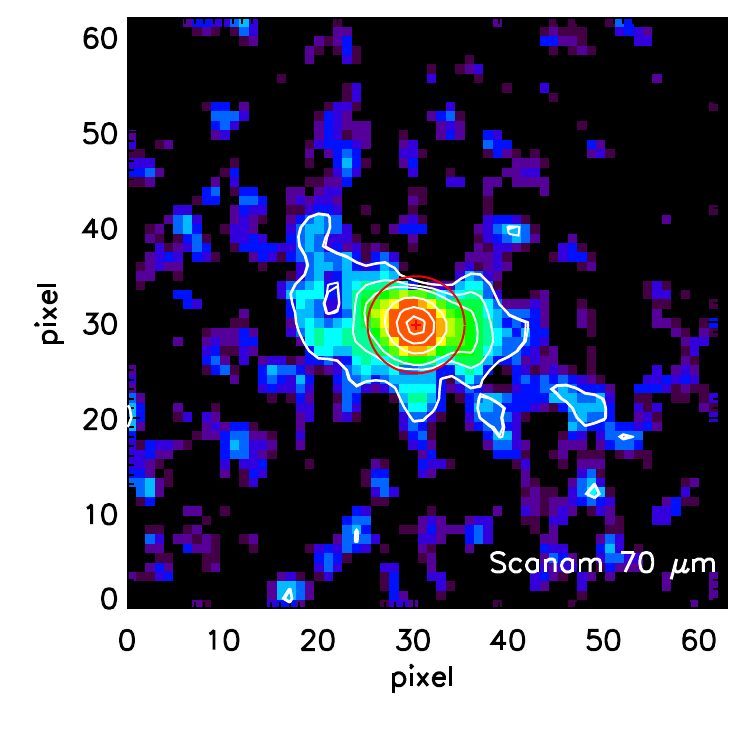}
   \includegraphics[width=0.24\textwidth]{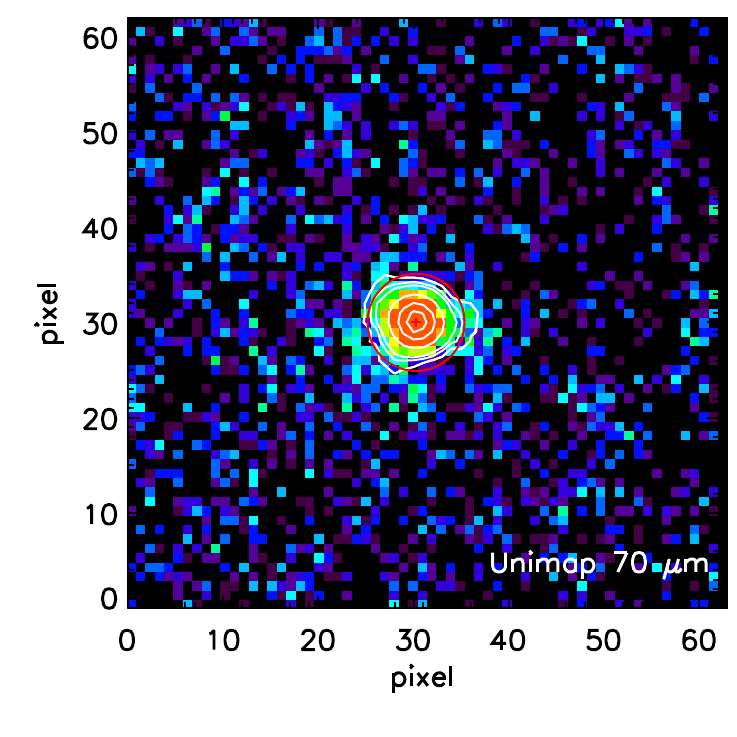}
   \includegraphics[width=0.24\textwidth]{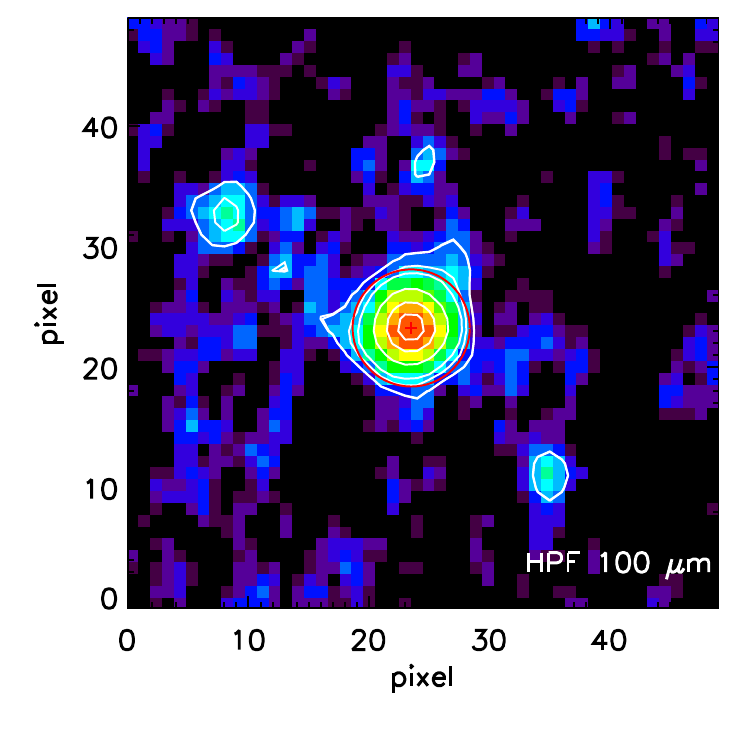}
   \includegraphics[width=0.24\textwidth]{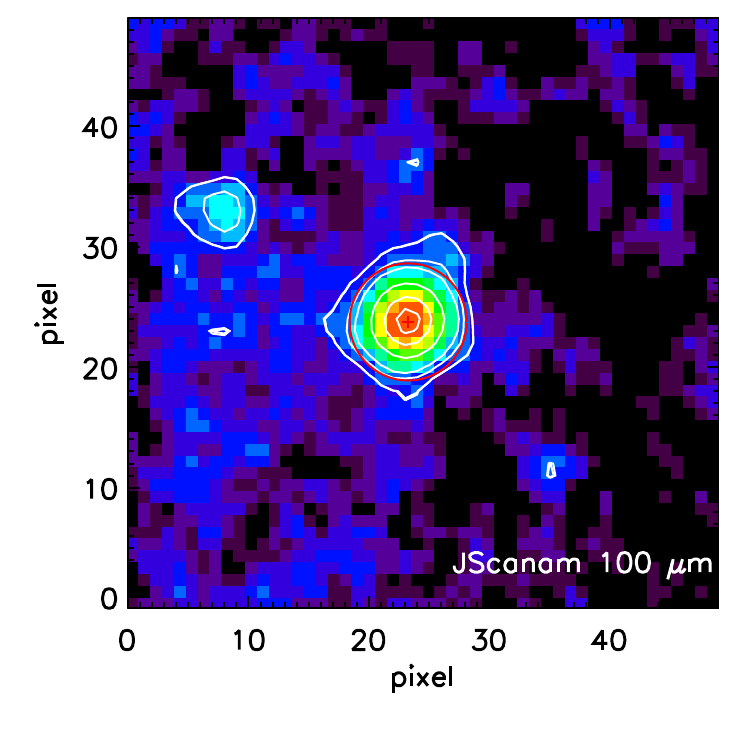}
   \includegraphics[width=0.24\textwidth]{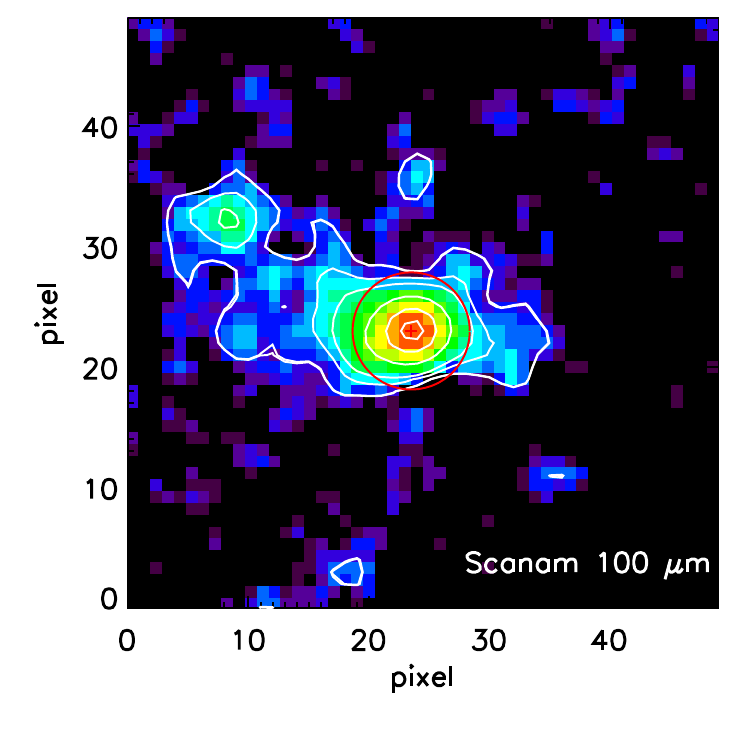}
   \includegraphics[width=0.24\textwidth]{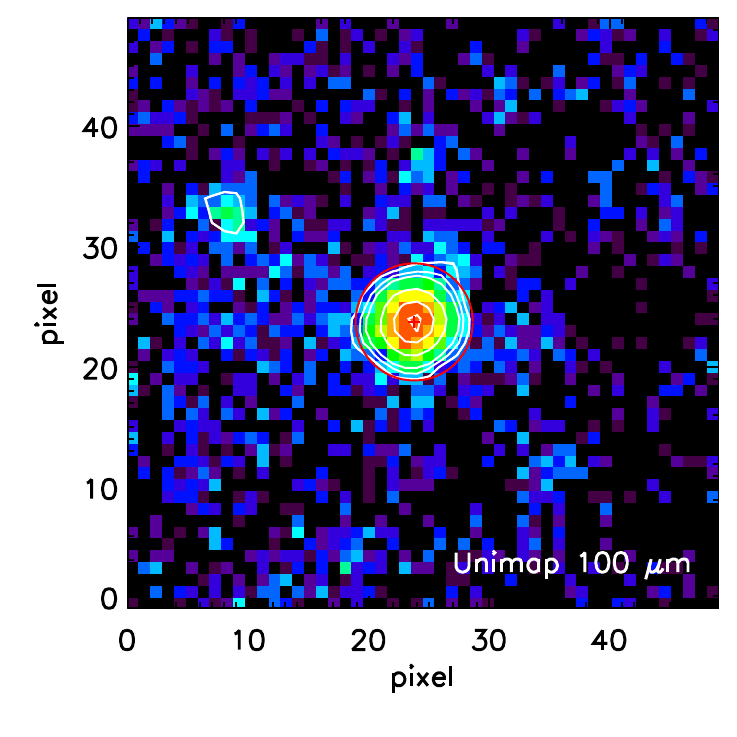}
   \includegraphics[width=0.24\textwidth]{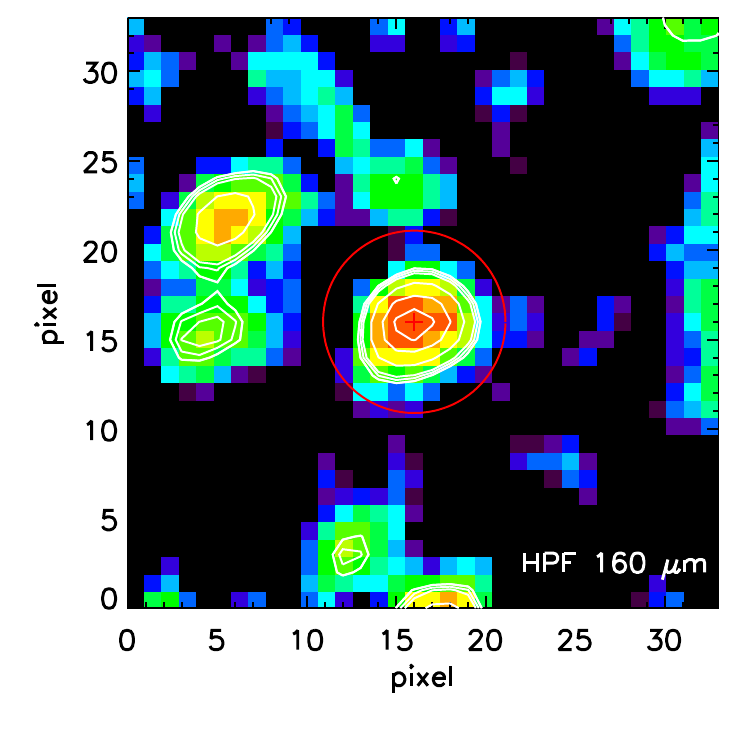}
   \includegraphics[width=0.24\textwidth]{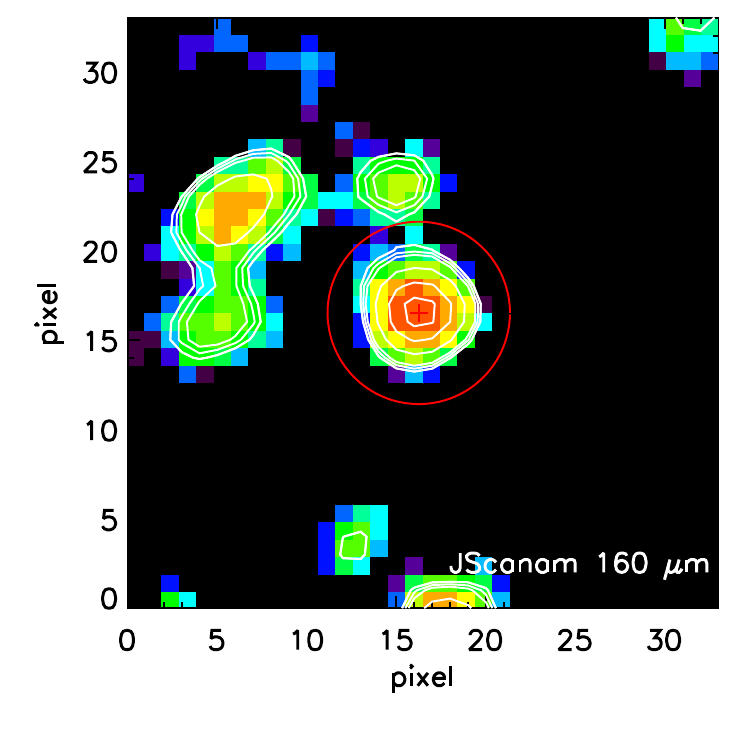}
   \includegraphics[width=0.24\textwidth]{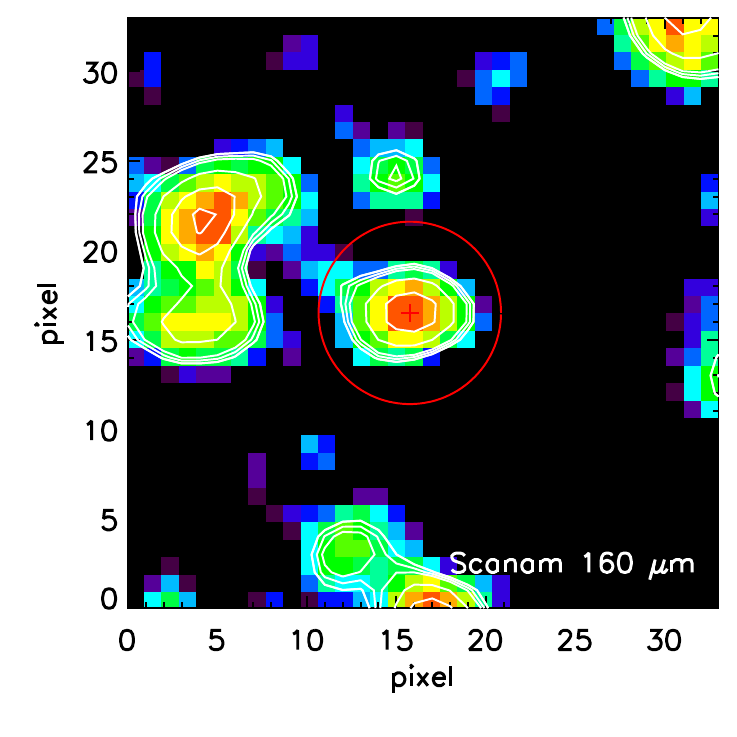}
   \includegraphics[width=0.24\textwidth]{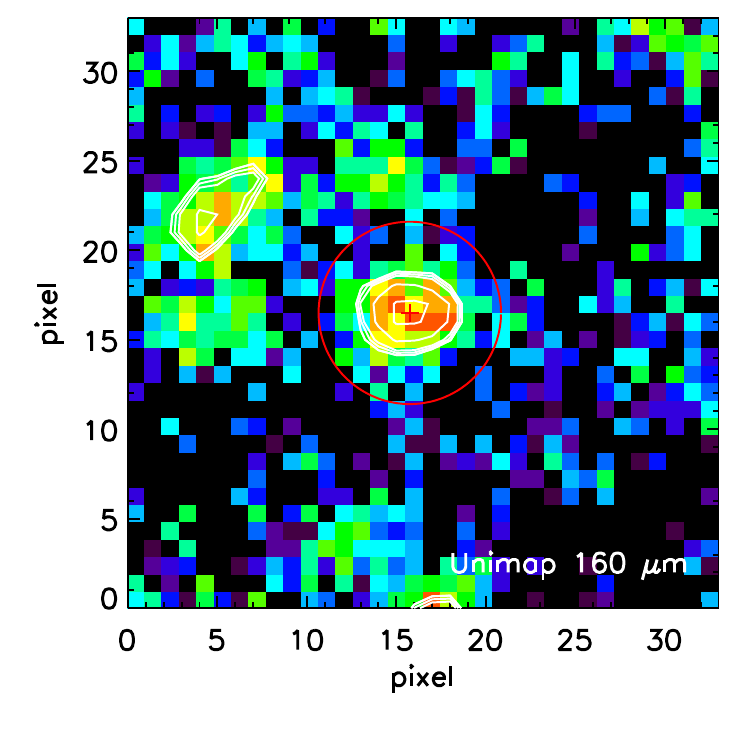}
      \caption{Inter-comparison of HD\,152222 photometric maps for different
               mapper softwares for 70, 100, and 160\,$\mu$m (top to bottom). 
               First column: HPF (High Pass Filter, default reduction scheme 
               of this work), second column: JScanam, third column: 
               Scanamorphos and 4th column: Unimap. Used OBSIDs are the 
               combinations of 1342240702$+$03 at 70\,$\mu$m and 
               1342227973$+$74 at 100 and 160\,$\mu$m. The red circle 
               indicates the photometric aperture.
              }
         \label{fig:hd152222mappercomp}
   \end{figure*}

For the Scanamorphos processing release version 25 of the software was applied, 
the "mini-map" option was selected, and the software was set to correct for 
the PACS distortion flat-field. For the JScanam processing version 14.2.0 
(analogue to HIPE version) was applied and the "galactic" option was switched 
on. For the Unimap processing, version 6.5.3 was applied with the parameter 
{\sf pixelNoise} (gain to apply to the estimated pixel noise in the GLS pixel 
noise compensation\footnotemark[12]) set to zero. For the comparison with the 
other mappers we used the weighted GLS (Generalized Least-Squares) L2.5 map 
product, corresponding to the FITS XTENSION "Image".

%
%
\begin{table}[ht!]
\caption{Comparison of the photometric results of HD\,152222 from different 
mapper softwares: HPF (High Pass Filter, default reduction scheme of this work),
JScanam~\citep{gracia15}, Scanamorphos~\citep{roussel13} and 
Unimap~\citep{piazzo15}. Used OBSIDs are the combinations of 1342240702$+$03 
at 70\,$\mu$m and 1342227973$+$74 at 100 and 160\,$\mu$m. Listed fluxes are 
the colour-corrected total stellar fluxes f$_{\rm star}$. 
}             
\label{table:compphot_mappers}      
\centering                          
\begin{tabular}{r r r c c c}        
\hline\hline                 
            \noalign{\smallskip}
Mapper & Filter  & r$_{\rm aper}$ & f$_{\rm star}$ & $\sigma_{\rm aper}$ &$\frac{f_{\rm star}^{mapper}}{f_{\rm star}^{HPF}}$ \\ 
         & ($\mu$m)&(\arcsec)& (mJy)&  (mJy)&  \\ 
            \noalign{\smallskip} 
\hline                        
            \noalign{\smallskip}
HPF          &  70 &  5.6 & 37.7 &$\pm$0.21 &  --  \\ 
             & 100 &  6.8 & 20.7 &$\pm$0.15 &  --  \\ 
             & 160 & 10.7 &  6.5 &$\pm$0.27 &  --  \\ 
JScanam      &  70 &  5.6 & 39.3 &$\pm$0.24 & 1.04$\pm$0.02 \\ 
             & 100 &  6.8 & 20.5 &$\pm$0.16 & 0.99$\pm$0.02 \\ 
             & 160 & 10.7 &  6.5 &$\pm$0.31 & 1.00$\pm$0.14 \\ 
Scanamorphos &  70 &  5.6 & 35.5 &$\pm$0.19 & 0.94$\pm$0.02 \\ 
             & 100 &  6.8 & 17.6 &$\pm$0.13 & 0.85$\pm$0.02 \\ 
             & 160 & 10.7 &  4.2 &$\pm$0.28 & 0.65$\pm$0.11 \\ 
Unimap       &  70 &  5.6 & 39.9 &$\pm$0.45 & 1.06$\pm$0.03 \\ 
             & 100 &  6.8 & 20.5 &$\pm$0.23 & 0.99$\pm$0.03 \\ 
             & 160 & 10.7 &  7.5 &$\pm$0.49 & 1.15$\pm$0.19 \\ 
           \noalign{\smallskip}
\hline                                   
\end{tabular}
\end{table}

Since our analysis is restricted to one map in each filter, we do not intend
to give a full quantitative performance assessment of the four mapper 
softwares, but rather point out some qualitative findings for these faint star
map products.

Fig.~\ref{fig:hd152222mappercomp} shows the inter-comparison of the maps 
produced by the default High Pass Filter (HPF) mapper software and the three 
other mapper softwares for the three PACS filters. The star can be clearly 
identified as the central source in all three filters for all four mappers.

Table~\ref{table:compphot_mappers} lists the colour-corrected total stellar 
fluxes f$_{\rm star}$ derived from aperture photometry with 5\farcs6, 6\farcs8 
and 10\farcs7 aperture radius at 70, 100 and 160\,$\mu$m, respectively. 

We also list the noise inside the measurement aperture, $\sigma_{\rm aper}$
(cf.\ Eq.~\ref{eq:sigaper}), which was determined from all maps with the 
histogram method described in Sect.~\ref{sect:noise_sn_determination}. We note 
that here only this noise term can be used for inter-comparison, not the one 
corrected for correlated noise, $\sigma_{\rm aper,corr}$, since correlated noise 
correction factors f$_{\rm corr}$ were only derived for the high-pass filtered 
data reduction (one may argue that the final corrected noise 
$\sigma_{\rm aper,corr}$ should be about the same for all mappers, since it is 
mainly determined by the detector noise). A $\sigma_{\rm tot}$ for the total 
flux can be calculated as $\sigma_{\rm tot} = c_{\rm aper} \times \sigma_{\rm aper}$.

The noise determined from the JScanam maps is slightly larger than that of the 
HPF maps. This finding indicates that the noise correlation is slightly less 
for the JScanam mapper. The noise determined from the Scanamorphos maps is 
slightly smaller than that of the HPF maps, indicating a slightly higher noise 
correlation. The noise determined from the Unimap maps is a factor of 
1.5 -- 2.1 higher than that of the HPF maps. This is explained by Unimap using 
the Generalized Least-Squares (GLS) algorithm to remove the correlated 
$\frac{1}{f}$-noise~\citep{piazzo15}. The Unimap noise is hence closer to the 
real noise level, and the above scaling factors do not exceed the correlated 
noise correction factors f$_{\rm corr}$ to be applied to the HPF noise (cf.\
Eq.~\ref{eq:sigapercorr} and Table~\ref{table:scanmapparamsphotnoise})
for calculation of the correlation-free noise level.   

At 70\,$\mu$m, with an expected source flux in the order of 40\,mJy, the 
fluxes of all four mappers correspond to each other within 4 -- 6\%. At 
100\,$\mu$m, with an expected source flux in the order of 20\,mJy, the 
correspondence is still better than 15\%. At 160\,$\mu$m, with an expected 
source flux of only $\approx$8\,mJy and S/N $\lesssim$ 10, the scatter is 
naturally larger. Jscanam photometry shows the best correspondence with the 
HPF photometry, being within 4\% for all filters. This can be expected, since 
both mappers use the same projection algorithm {\sf photProject()}. Unimap 
photometry shows the second best correspondence with HPF photometry, with the 
tendency that the Unimap fluxes are larger (at 70 and 160\,$\mu$m).
Scanamorphos photometry gives systematically smaller fluxes than HPF 
photometry, with the deviation increasing with wavelength and a 160\,$\mu$m 
flux which is noticeably off.

The PACS photometric calibration scheme~\citep{balog14} was established with 
HPF analysis, in particular also the derivation of the  aperture photometry 
correction factors c$_{\rm aper}$ from the PACS Point Spread 
Function~\citep{lutz15}, by determining the Encircled Energy Fractions with 
radius. Therefore, one aspect affecting the aperture photometry depending on 
the selected mapper software was not considered in the evaluation scheme 
described above, namely the shape of the point spread function. From the 
inspection of the stellar intensity profiles and their close surrounding in 
Fig.~\ref{fig:hd152222mappercomp}, in particular from the 70 and 
100\,$\mu$m images, it is obvious that there are systematic differences in the 
resulting profile shapes of the star depending on the applied mapper. The HPF 
processing shows the typical tri-lobe pattern of the PACS point spread 
function~\citep{lutz15}. The JScanam processing shows the closest appearance. 
The Unimap profiles look slightly sharper and show less pronounced lobes. The 
Scanamorphos profiles on the other hand appear somewhat less concentrated than 
the HPF profiles and with more pronounced lobes. 

This means that for the other mappers adapted aperture correction factors 
should be applied, which alters the photometric results somewhat. The ratios 
$\frac{f_{\rm tot,cc}^{mapper}}{f_{\rm tot,cc}^{HPF}}$ in 
Table~\ref{table:compphot_mappers} should not be used as general scaling 
factors between the various mappers, since they are based on the evaluation of 
a single map of a very faint star implying quite some uncertainty. Accurate 
scaling factors for photometry with the various mappers do not exist yet and 
will be determined on the basis of the high signal-to-noise fiducial standard 
observational database in a forthcoming paper (Balog et al., 2018, in 
preparation). 

\subsection{Photometry results of individual measurements}
\label{sect:individualscanphot}

Individual photometric results for the 70, 100, and 160\,$\mu$m filters are 
compiled in Tables~\ref{table:scanmapphotblue6_1} to~\ref{table:scanmapphotred10_1}.
The applied radius for the photometric aperture was 5.6, 6.8 and 10.7\arcsec 
for the 70, 100 and 160\,$\mu$m filter, respectively. The number of output 
pixels (1\farcs1, 1\farcs4, and 2\farcs1 size, respectively) inside this photometric 
aperture is N$_{\rm aper}$= 81.42, 74.12, and 81.56, respectively. The corresponding 
correction factors for correlated noise are f$_{\rm corr}$ = 3.13, 2.76, and 
4.12, respectively. Aperture correction factors are c$_{\rm aper}$ = 1.61, 1.56 
and 1.56 for the 70, 100 and 160\,$\mu$m filter, respectively. Proper motion 
correction was applied throughout.

The tables contain the following information: Col.~1: Unique observational 
identifier (OBSID) of the PACS observation; Col.~2: Herschel Observational Day 
(OD); Col.~3: Target name; Col.~4: Applied gain (G) of the PACS bolometer 
electronics: h(igh)/l(ow); Col.~5: Scan speed: low = 10\,\arcsec/s, medium = 
20\,\arcsec/s, high = 60\,\arcsec/s; Col.~6: Number of repetitions (rep.) of 
the basic scan map with the parameters given in next column; Col.~7: 
Parameters of the scan map: scan leg length(\arcsec) / scan leg separation 
(\arcsec) / number of scan legs; Col.~8: Scan angle of the map, in case of 
co-added maps all angles of the individual maps are given; Col.~9: Measured 
flux inside the photometric aperture of this filter, f$_{\rm aper}$; Col.~10: 
Noise per pixel, $\sigma_{\rm pix}$; Col.~11: Noise corrected for correlated 
noise inside the measurement aperture, $\sigma_{\rm aper,corr}$, according to 
Eq.~\ref{eq:sigapercorr}. Col.~12: Achieved signal-to-noise ratio according to 
Eq.~\ref{eq:sn_meas}; Col.~13: Stellar flux f$_{\rm star}$ according to 
Eq.~\ref{eq:apercolcorr}; Cols.~14 - 16: Maximum and minimum Full Width (W) 
Half Maximum (in \arcsec) of the source PSF of the source PSF and its 
uncertainty determined by an elliptical fit to the intensity profile.

\begin{sidewaystable*}[h!]
\caption{Scan map photometry measurements in the blue (70\,$\mu$m filter).
Processing proceeded from SPG~v13.1.0 level 1 products with HIPE version 15 
build 165. 
}             
\label{table:scanmapphotblue6_1}      
\centering                            
\begin{tabular}{r r r r r r r c r r r r r r r r r}        
\hline\hline                 
            \noalign{\smallskip}
 OBSID     &   OD   &   Target     & G & Speed  & Rep. & Map params  & ScanAngles &f$_{\rm aper}$&$\sigma_{\rm pix}$&$\sigma_{\rm aper,corr}$& S/N &f$_{\rm star}$&  W$_{max}$ & W$_{min}$ & $\Delta$\,W\\
           &        &              &    &(\arcsec/s)&  &(\arcsec/\arcsec/no.)&(deg)& (mJy)     &      (mJy)     &       (mJy)        &     &   (mJy)    &  (\arcsec) & (\arcsec) &   (\arcsec)    \\
            \noalign{\smallskip}
\hline
            \noalign{\smallskip}
1342217348 &  684   &$\beta$\,Gem  & h & 20 &  3 & 180/4/10 &   70   & 1674.5 & 0.052 & 1.45 & 1152 & 2649.1 & 5.61 & 5.42 & 0.03 \\
1342217349 &  684   &$\beta$\,Gem  & h & 20 &  3 & 180/4/10 &  110   & 1672.7 & 0.052 & 1.48 & 1134 & 2646.3 & 5.74 & 5.19 & 0.03 \\
1342217348$+$49& 684&$\beta$\,Gem  & h & 20 &  6 & 180/4/10 &70$+$110& 1667.0 & 0.034 & 0.97 & 1728 & 2637.2 & 5.55 & 5.34 & 0.05 \\
1342230120 &  872   &$\beta$\,Gem  & h & 20 &  3 & 180/4/10 &   70   & 1676.4 & 0.052 & 1.47 & 1143 & 2652.1 & 5.53 & 5.44 & 0.03 \\
1342230121 &  872   &$\beta$\,Gem  & h & 20 &  3 & 180/4/10 &  110   & 1675.0 & 0.050 & 1.42 & 1183 & 2649.9 & 5.73 & 5.24 & 0.03 \\
1342230120$+$21& 872&$\beta$\,Gem  & h & 20 &  6 & 180/4/10 &70$+$110& 1669.4 & 0.034 & 0.97 & 1720 & 2641.1 & 5.58 & 5.31 & 0.04 \\
1342242772 & 1051   &$\beta$\,Gem  & h & 20 &  3 & 180/4/10 &   70   & 1678.4 & 0.057 & 1.60 & 1051 & 2655.2 & 5.52 & 5.41 & 0.03 \\
1342242773 & 1051   &$\beta$\,Gem  & h & 20 &  3 & 180/4/10 &  110   & 1671.3 & 0.055 & 1.55 & 1075 & 2644.1 & 5.71 & 5.17 & 0.03 \\
1342242772$+$73&1051&$\beta$\,Gem  & h & 20 &  6 & 180/4/10 &70$+$110& 1671.2 & 0.039 & 1.10 & 1519 & 2644.0 & 5.55 & 5.25 & 0.04 \\
1342252881 & 1244   &$\beta$\,Gem  & h & 20 &  3 & 180/4/10 &   70   & 1672.6 & 0.051 & 1.45 & 1154 & 2646.1 & 5.52 & 5.48 & 0.03 \\
1342252882 & 1244   &$\beta$\,Gem  & h & 20 &  3 & 180/4/10 &  110   & 1673.6 & 0.049 & 1.38 & 1211 & 2647.7 & 5.72 & 5.32 & 0.03 \\
1342252881$+$82&1244&$\beta$\,Gem  & h & 20 &  6 & 180/4/10 &70$+$110& 1670.3 & 0.034 & 0.96 & 1734 & 2642.4 & 5.48 & 5.34 & 0.05 \\
1342212864 &  614   &$\alpha$\,Ari & h & 20 &  3 & 180/4/10 &   70   & 1056.5 & 0.050 & 1.42 &  743 & 1671.5 & 5.59 & 5.25 & 0.03 \\
1342212865 &  614   &$\alpha$\,Ari & h & 20 &  3 & 180/4/10 &  110   & 1055.3 & 0.051 & 1.44 &  733 & 1669.5 & 5.51 & 5.40 & 0.04 \\
1342212864$+$65& 614&$\alpha$\,Ari & h & 20 &  6 & 180/4/10 &70$+$110& 1053.6 & 0.035 & 0.99 & 1067 & 1666.9 & 5.41 & 5.28 & 0.05 \\
1342237398 &  974   &$\alpha$\,Ari & h & 20 &  3 & 180/4/10 &   70   & 1048.2 & 0.050 & 1.41 &  742 & 1658.3 & 5.61 & 5.37 & 0.03 \\
1342237399 &  974   &$\alpha$\,Ari & h & 20 &  3 & 180/4/10 &  110   & 1049.1 & 0.050 & 1.41 &  744 & 1659.7 & 5.51 & 5.44 & 0.04 \\
1342237398$+$99& 974&$\alpha$\,Ari & h & 20 &  6 & 180/4/10 &70$+$110& 1045.9 & 0.034 & 0.96 & 1087 & 1654.7 & 5.47 & 5.35 & 0.05 \\
1342248029 & 1157   &$\alpha$\,Ari & h & 20 &  3 & 180/4/10 &   70   & 1058.3 & 0.050 & 1.41 &  749 & 1674.2 & 5.61 & 5.27 & 0.03 \\
1342248030 & 1157   &$\alpha$\,Ari & h & 20 &  3 & 180/4/10 &  110   & 1052.9 & 0.052 & 1.48 &  714 & 1665.7 & 5.54 & 5.52 & 0.04 \\
1342248029$+$30&1157&$\alpha$\,Ari & h & 20 &  6 & 180/4/10 &70$+$110& 1052.7 & 0.034 & 0.97 & 1082 & 1665.4 & 5.48 & 5.30 & 0.05 \\
1342259801 & 1344   &$\alpha$\,Ari & h & 20 &  3 & 180/4/10 &   70   & 1051.3 & 0.052 & 1.46 &  719 & 1663.1 & 5.61 & 5.30 & 0.03 \\
1342259802 & 1344   &$\alpha$\,Ari & h & 20 &  3 & 180/4/10 &  110   & 1041.5 & 0.050 & 1.42 &  735 & 1647.7 & 5.51 & 5.50 & 0.04 \\
1342259801$+$02&1344&$\alpha$\,Ari & h & 20 &  6 & 180/4/10 &70$+$110& 1043.3 & 0.036 & 1.01 & 1031 & 1650.4 & 5.49 & 5.38 & 0.05 \\
1342205202 &  502   &$\varepsilon$\,Lep&h&20&  1 & 180/4/10 &   70   &  738.9 & 0.091 & 2.58 &  287 & 1169.0 & 5.55 & 5.42 & 0.03 \\
1342205203 &  502   &$\varepsilon$\,Lep&h&20&  1 & 180/4/10 &  110   &  743.1 & 0.086 & 2.43 &  306 & 1175.5 & 5.73 & 5.24 & 0.03 \\
1342205202$+$03& 502&$\varepsilon$\,Lep&h&20&  2 & 180/4/10 &70$+$110&  738.9 & 0.058 & 1.64 &  451 & 1168.9 & 5.53 & 5.34 & 0.05 \\
1342227297 &  833   &$\varepsilon$\,Lep&h&20&  1 & 180/4/10 &   70   &  734.9 & 0.085 & 2.39 &  307 & 1162.6 & 5.64 & 5.30 & 0.03 \\
1342227298 &  833   &$\varepsilon$\,Lep&h&20&  1 & 180/4/10 &  110   &  740.9 & 0.087 & 2.44 &  303 & 1172.1 & 5.52 & 5.38 & 0.03 \\
1342227297$+$98& 833&$\varepsilon$\,Lep&h&20&  2 & 180/4/10 &70$+$110&  735.5 & 0.057 & 1.60 &  459 & 1163.6 & 5.51 & 5.34 & 0.05 \\
1342241333 & 1034   &$\varepsilon$\,Lep&h&20&  1 & 180/4/10 &   70   &  736.6 & 0.086 & 2.44 &  302 & 1165.4 & 5.57 & 5.42 & 0.03 \\
1342241334 & 1034   &$\varepsilon$\,Lep&h&20&  1 & 180/4/10 &  110   &  734.2 & 0.088 & 2.48 &  296 & 1161.4 & 5.68 & 5.28 & 0.03 \\
1342241333$+$34&1034&$\varepsilon$\,Lep&h&20&  2 & 180/4/10 &70$+$110&  732.8 & 0.059 & 1.66 &  442 & 1159.3 & 5.54 & 5.38 & 0.05 \\
1342263904 & 1377   &$\varepsilon$\,Lep&h&20&  1 & 180/4/10 &   70   &  730.4 & 0.084 & 2.37 &  308 & 1155.5 & 5.60 & 5.28 & 0.03 \\
1342263905 & 1377   &$\varepsilon$\,Lep&h&20&  1 & 180/4/10 &  110   &  735.6 & 0.082 & 2.32 &  317 & 1163.7 & 5.54 & 5.30 & 0.03 \\
1342263904$+$05&1377&$\varepsilon$\,Lep&h&20&  2 & 180/4/10 &70$+$110&  731.2 & 0.056 & 1.57 &  465 & 1156.7 & 5.43 & 5.28 & 0.05 \\
            \noalign{\smallskip}
\hline                                   
\end{tabular}
\end{sidewaystable*}

\addtocounter{table}{-1}

\begin{sidewaystable*}[h!]
\caption{Scan map photometry measurements in the blue (70\,$\mu$m filter) 
continued.
}             
\label{table:scanmapphotblue6_2}      
\centering                          
\begin{tabular}{r r r r r r r c r r r r r r r r r}        
\hline\hline                 
            \noalign{\smallskip}
 OBSID     &   OD   &   Target     & G & Speed  & Rep. & Map params  & ScanAngles &f$_{\rm aper}$&$\sigma_{\rm pix}$&$\sigma_{\rm aper,corr}$& S/N &f$_{\rm star}$&  W$_{max}$ & W$_{min}$ & $\Delta$\,W\\
           &        &              &    &(\arcsec/s)&  &(\arcsec/\arcsec/no.)&(deg)& (mJy)     &      (mJy)     &       (mJy)        &     &   (mJy)    &  (\arcsec) & (\arcsec) &   (\arcsec)    \\
            \noalign{\smallskip}
\hline
            \noalign{\smallskip}
1342218544 &  697   &$\omega$\,Cap & h & 20 & 3 & 180/4/10 &   70   & 535.8 & 0.050 & 1.42 & 376 & 847.6 & 5.70 & 5.36 & 0.03 \\
1342218545 &  697   &$\omega$\,Cap & h & 20 & 3 & 180/4/10 &  110   & 532.8 & 0.048 & 1.36 & 392 & 842.9 & 5.48 & 5.34 & 0.03 \\
1342218544$+$45& 697&$\omega$\,Cap & h & 20 & 6 & 180/4/10 &70$+$110& 532.4 & 0.033 & 0.94 & 569 & 842.2 & 5.55 & 5.42 & 0.04 \\
1342231259 &  889   &$\omega$\,Cap & h & 20 & 3 & 180/4/10 &   70   & 535.7 & 0.050 & 1.41 & 380 & 847.5 & 5.61 & 5.25 & 0.03 \\
1342231260 &  889   &$\omega$\,Cap & h & 20 & 3 & 180/4/10 &  110   & 532.8 & 0.049 & 1.38 & 387 & 842.9 & 5.51 & 5.35 & 0.04 \\
1342231259$+$60& 889&$\omega$\,Cap & h & 20 & 6 & 180/4/10 &70$+$110& 532.3 & 0.033 & 0.94 & 564 & 842.0 & 5.50 & 5.29 & 0.04 \\
1342243780 & 1058   &$\omega$\,Cap & h & 20 & 3 & 180/4/10 &   70   & 536.1 & 0.049 & 1.39 & 386 & 848.1 & 5.68 & 5.33 & 0.03 \\
1342243781 & 1058   &$\omega$\,Cap & h & 20 & 3 & 180/4/10 &  110   & 538.8 & 0.050 & 1.40 & 384 & 852.4 & 5.52 & 5.50 & 0.04 \\
1342243780$+$81&1058&$\omega$\,Cap & h & 20 & 6 & 180/4/10 &70$+$110& 535.7 & 0.034 & 0.96 & 558 & 847.5 & 5.54 & 5.40 & 0.05 \\
1342254209 & 1266   &$\omega$\,Cap & h & 20 & 3 & 180/4/10 &   70   & 529.0 & 0.051 & 1.44 & 367 & 836.9 & 5.65 & 5.27 & 0.03 \\
1342254210 & 1266   &$\omega$\,Cap & h & 20 & 3 & 180/4/10 &  110   & 532.5 & 0.051 & 1.43 & 374 & 842.5 & 5.48 & 5.37 & 0.03 \\
1342254209$+$10&1266&$\omega$\,Cap & h & 20 & 6 & 180/4/10 &70$+$110& 528.6 & 0.032 & 0.90 & 586 & 836.3 & 5.51 & 5.30 & 0.04 \\
1342186142 &  160   &$\eta$\,Dra   & h & 10 & 1 & 120/5/9  &   80   & 327.5 & 0.101 & 2.86 & 115 & 518.1 & 5.43 & 5.18 & 0.04 \\
1342186143 &  160   &$\eta$\,Dra   & h & 10 & 1 & 120/5/9  &  100   & 318.8 & 0.097 & 2.74 & 117 & 504.3 & 5.35 & 5.30 & 0.04 \\
1342186142$+$43& 160&$\eta$\,Dra   & h & 10 & 2 & 120/5/9  &80$+$100& 323.1 & 0.071 & 2.01 & 161 & 511.2 & 5.28 & 5.14 & 0.05 \\
1342186144 &  160   &$\eta$\,Dra   & h & 10 & 1 &  90/5/9  &   60   & 320.5 & 0.113 & 3.19 & 101 & 507.0 & 5.37 & 5.10 & 0.05 \\
1342186145 &  160   &$\eta$\,Dra   & h & 10 & 1 &  90/5/9  &  120   & 316.5 & 0.110 & 3.11 & 102 & 500.7 & 5.33 & 5.17 & 0.04 \\
1342186144$+$45& 160&$\eta$\,Dra   & h & 10 & 2 &  90/5/9  &60$+$120& 317.6 & 0.075 & 2.11 & 151 & 502.5 & 5.20 & 5.11 & 0.05 \\
1342186146 &  160   &$\eta$\,Dra   & h & 20 & 1 &  90/5/9  &   70   & 311.7 & 0.132 & 3.71 & 83.9& 493.1 & 5.48 & 5.24 & 0.05 \\
1342186147 &  160   &$\eta$\,Dra   & h & 20 & 1 &  90/5/9  &  110   & 314.4 & 0.127 & 3.58 & 87.7& 497.4 & 5.54 & 5.40 & 0.04 \\
1342186146$+$47& 160&$\eta$\,Dra   & h & 20 & 2 &  90/5/9  &70$+$110& 311.6 & 0.084 & 2.38 & 131 & 492.9 & 5.38 & 5.33 & 0.05 \\
1342186148 &  160   &$\eta$\,Dra   & h & 10 & 1 & 150/4/9  &   85   & 321.1 & 0.091 & 2.58 & 125 & 508.0 & 5.23 & 5.05 & 0.05 \\
1342186149 &  160   &$\eta$\,Dra   & h & 10 & 1 & 150/4/9  &   95   & 322.1 & 0.090 & 2.53 & 127 & 509.5 & 5.26 & 5.26 & 0.04 \\
1342186148$+$49& 160&$\eta$\,Dra   & h & 10 & 2 & 150/4/9  &85$+$95 & 320.3 & 0.063 & 1.77 & 181 & 506.8 & 5.21 & 5.09 & 0.05 \\
1342189191 &  244   &$\delta$\,Dra & h & 20 & 1 & 240/4/8  &  117   & 275.7 & 0.095 & 2.69 & 103 & 436.2 & 5.59 & 5.28 & 0.04 \\
1342189192 &  244   &$\delta$\,Dra & h & 20 & 1 & 240/4/8  &   63   & 270.4 & 0.096 & 2.71 & 99.9& 427.8 & 5.49 & 5.22 & 0.03 \\
1342189191$+$92& 244&$\delta$\,Dra & h & 20 & 2 & 240/4/8  &63$+$117& 272.9 & 0.070 & 1.98 & 138 & 431.7 & 5.44 & 5.22 & 0.04 \\
1342212499 &  607   &$\delta$\,Dra & h & 20 & 3 & 180/4/10 &   70   & 278.4 & 0.052 & 1.45 & 191 & 440.4 & 5.58 & 5.31 & 0.04 \\
1342212500 &  607   &$\delta$\,Dra & h & 20 & 3 & 180/4/10 &  110   & 277.9 & 0.053 & 1.50 & 185 & 439.6 & 5.74 & 5.29 & 0.03 \\
1342212499$+$500& 244&$\delta$\,Dra& h & 20 & 6 & 180/4/10 &70$+$110& 277.3 & 0.035 & 1.00 & 278 & 438.8 & 5.59 & 5.31 & 0.05 \\
1342222149 &  751   &$\delta$\,Dra & h & 20 & 3 & 180/4/10 &   70   & 278.2 & 0.053 & 1.50 & 185 & 440.1 & 5.71 & 5.27 & 0.03 \\
1342222150 &  751   &$\delta$\,Dra & h & 20 & 3 & 180/4/10 &  110   & 274.4 & 0.049 & 1.38 & 199 & 434.1 & 5.38 & 5.30 & 0.04 \\
1342222149$+$50& 751&$\delta$\,Dra & h & 20 & 6 & 180/4/10 &70$+$110& 275.3 & 0.035 & 0.97 & 283 & 435.5 & 5.46 & 5.35 & 0.05 \\
1342233571 &  934   &$\delta$\,Dra & h & 20 & 3 & 180/4/10 &   70   & 277.2 & 0.050 & 1.42 & 195 & 438.6 & 6.06 & 5.23 & 0.03 \\
1342233572 &  934   &$\delta$\,Dra & h & 20 & 3 & 180/4/10 &  110   & 275.4 & 0.051 & 1.43 & 192 & 435.7 & 5.46 & 5.34 & 0.03 \\
1342233571$+$72& 934&$\delta$\,Dra & h & 20 & 6 & 180/4/10 &70$+$110& 275.7 & 0.036 & 1.00 & 275 & 436.2 & 5.64 & 5.35 & 0.05 \\
1342250095 & 1198   &$\delta$\,Dra & h & 20 & 3 & 180/4/10 &   70   & 275.2 & 0.052 & 1.47 & 187 & 435.3 & 5.55 & 5.45 & 0.04 \\
1342250096 & 1198   &$\delta$\,Dra & h & 20 & 3 & 180/4/10 &  110   & 271.3 & 0.050 & 1.42 & 191 & 429.1 & 5.66 & 5.06 & 0.03 \\
1342250095$+$96&1198&$\delta$\,Dra & h & 20 & 6 & 180/4/10 &70$+$110& 272.8 & 0.035 & 0.99 & 275 & 431.5 & 5.53 & 5.29 & 0.04 \\
1342257989 & 1328   &$\delta$\,Dra & h & 20 & 3 & 180/4/10 &   70   & 276.6 & 0.049 & 1.39 & 198 & 437.6 & 5.60 & 5.39 & 0.03 \\
1342257990 & 1328   &$\delta$\,Dra & h & 20 & 3 & 180/4/10 &  110   & 273.8 & 0.050 & 1.40 & 195 & 433.1 & 5.63 & 5.34 & 0.03 \\
1342257989$+$90&1328&$\delta$\,Dra & h & 20 & 6 & 180/4/10 &70$+$110& 274.6 & 0.035 & 0.97 & 282 & 434.4 & 5.49 & 5.34 & 0.05 \\
\hline                                   
\end{tabular}
\end{sidewaystable*}

\addtocounter{table}{-1}

\begin{sidewaystable*}[h!]
\caption{Scan map photometry measurements in the blue (70\,$\mu$m filter) 
continued.
}             
\label{table:scanmapphotblue6_3}      
\centering                          
\begin{tabular}{r r r r r r r c r r r r r r r r r}        
\hline\hline                 
            \noalign{\smallskip}
 OBSID     &   OD   &   Target     & G & Speed  & Rep. & Map params  & ScanAngles &f$_{\rm aper}$&$\sigma_{\rm pix}$&$\sigma_{\rm aper,corr}$& S/N &f$_{\rm star}$&  W$_{max}$ & W$_{min}$ & $\Delta$\,W\\
           &        &              &    &(\arcsec/s)&  &(\arcsec/\arcsec/no.)&(deg)& (mJy)     &      (mJy)     &       (mJy)        &     &   (mJy)    &  (\arcsec) & (\arcsec) &   (\arcsec)    \\
            \noalign{\smallskip}
\hline
            \noalign{\smallskip}
1342184574 &  139   &$\theta$\,Umi & h &  10   &   1  & 210/15/4 &   90   & 173.4 & 0.126 & 3.57 & 48.6 & 274.4 & 5.09 & 4.97 & 0.07 \\
1342184575 &  139   &$\theta$\,UMi & h &  20   &   1  & 210/15/4 &   90   & 174.7 & 0.152 & 4.30 & 40.6 & 276.5 & 5.33 & 5.21 & 0.06 \\
1342191982 &  300   &$\theta$\,Umi & h &  20   &   6  & 150/4/10 &   70   & 179.6 & 0.038 & 1.06 &  170 & 284.2 & 5.67 & 5.29 & 0.03 \\
1342191983 &  300   &$\theta$\,Umi & h &  20   &   6  & 150/4/10 &  110   & 180.5 & 0.038 & 1.07 &  169 & 285.5 & 5.47 & 5.45 & 0.04 \\
1342191982$+$83& 300&$\theta$\,Umi & h &  20   &  12  & 150/4/10 &70$+$110& 179.5 & 0.025 & 0.70 &  255 & 284.0 & 5.48 & 5.42 & 0.05 \\
1342203111 &  460   &42\,Dra       & h &  20   &  20  & 180/4/10 &   70   &  94.6 & 0.020 & 0.57 &  166 & 149.6 & 5.50 & 5.50 & 0.04 \\
1342203112 &  460   &42\,Dra       & h &  20   &  20  & 180/4/10 &  110   &  93.4 & 0.020 & 0.55 &  170 & 147.7 & 5.76 & 5.12 & 0.03 \\
1342203111$+$12& 460&42\,Dra       & h &  20   &  40  & 180/4/10 &70$+$110&  94.0 & 0.013 & 0.38 &  249 & 148.8 & 5.58 & 5.27 & 0.05 \\
1342243730 & 1057   &42\,Dra       & h &  20   &  20  & 180/4/10 &   70   &  92.0 & 0.019 & 0.55 &  167 & 145.6 & 5.60 & 5.31 & 0.03 \\
1342243731 & 1057   &42\,Dra       & h &  20   &  20  & 180/4/10 &  110   &  93.8 & 0.020 & 0.56 &  168 & 148.3 & 5.57 & 5.40 & 0.03 \\
1342243730$+$31&1057&42\,Dra       & h &  20   &  40  & 180/4/10 &70$+$110&  92.6 & 0.013 & 0.37 &  248 & 146.5 & 5.46 & 5.37 & 0.04 \\
1342185446 &  146   &HD\,138265    & h &  20   &   1  & 210/4/20 &   90   &  74.2 & 0.071 & 1.99 & 37.3 & 117.4 & 5.54 & 5.37 & 0.06 \\
1342185447 &  146   &HD\,138265    & h &  10   &   1  & 210/4/20 &   90   &  70.9 & 0.057 & 1.60 & 44.2 & 112.2 & 5.28 & 5.12 & 0.07 \\
1342188841 &  233   &HD\,138265    & h &  20   &  21  & 240/4/8  &   63   &  72.3 & 0.023 & 0.64 &  113 & 114.3 & 5.50 & 5.46 & 0.04 \\
1342188842 &  233   &HD\,138265    & h &  20   &  21  & 240/4/8  &  117   &  70.3 & 0.021 & 0.59 &  119 & 111.2 & 5.76 & 5.05 & 0.03 \\
1342188841$+$42& 233&HD\,138265    & h &  20   &  42  & 240/4/8  &63$+$117&  71.0 & 0.016 & 0.44 &  160 & 112.3 & 5.59 & 5.27 & 0.04 \\
1342186163 &  160   &HD\,159330    & h &  10   &   1  & 120/3/21 &   80   &  40.1 & 0.066 & 1.77 & 22.7 &  63.4 & 5.62 & 5.07 & 0.14 \\
1342186164 &  160   &HD\,159330    & h &  10   &   1  & 120/3/21 &  100   &  43.5 & 0.066 & 1.76 & 24.7 &  68.9 & 5.80 & 5.50 & 0.11 \\
1342186163$+$64& 160&HD\,159330    & h &  10   &   2  & 120/3/21 &80$+$100&  41.2 & 0.044 & 1.18 & 34.8 &  65.2 & 5.70 & 5.32 & 0.08 \\
1342213585 &  628   &HD\,159330    & h &  20   &   6  & 180/4/10 &   70   &  40.0 & 0.036 & 0.96 & 41.7 &  63.2 & 5.41 & 5.37 & 0.06 \\
1342213586 &  628   &HD\,159330    & h &  20   &   6  & 180/4/10 &  110   &  41.4 & 0.037 & 0.98 & 42.2 &  65.4 & 5.86 & 5.21 & 0.05 \\
1342213585$+$86& 628&HD\,159330    & h &  20   &  12  & 180/4/10 &70$+$110&  40.5 & 0.024 & 0.63 & 64.0 &  64.1 & 5.58 & 5.27 & 0.06 \\
1342191964 &  300   &HD\,152222    & h &  20   &   6  & 150/4/10 &   70   &  26.0 & 0.036 & 1.02 & 25.4 &  41.1 & 5.56 & 5.24 & 0.08 \\
1342191965 &  300   &HD\,152222    & h &  20   &   6  & 150/4/10 &  110   &  25.2 & 0.038 & 1.07 & 23.6 &  39.9 & 5.55 & 5.50 & 0.09 \\
1342191964$+$65& 300&HD\,152222    & h &  20   &  12  & 150/4/10 &70$+$110&  25.5 & 0.025 & 0.71 & 36.0 &  40.3 & 5.45 & 5.30 & 0.07 \\
1342240702 & 1028   &HD\,152222    & h &  20   &   6  & 180/4/10 &   70   &  23.6 & 0.037 & 1.03 & 22.9 &  37.3 & 5.17 & 5.13 & 0.08 \\
1342240703 & 1028   &HD\,152222    & h &  20   &   6  & 180/4/10 &  110   &  24.0 & 0.035 & 0.99 & 24.3 &  38.0 & 5.84 & 5.02 & 0.08 \\
1342240702$+$03&1028&HD\,152222    & h &  20   &  12  & 180/4/10 &70$+$110&  23.8 & 0.023 & 0.66 & 36.3 &  37.7 & 5.34 & 5.07 & 0.07 \\
1342198535 &  400   &HD\,39608     & h &  20   &  10  & 180/4/10 &   70   &  21.3 & 0.028 & 0.79 & 26.9 &  33.6 & 6.01 & 5.26 & 0.08 \\
1342198536 &  400   &HD\,39608     & h &  20   &  10  & 180/4/10 &  110   &  19.3 & 0.028 & 0.80 & 24.0 &  30.6 & 5.84 & 5.53 & 0.08 \\
1342198538 &  400   &HD\,39608     & h &  20   &  35  & 180/4/10 &  110   &  18.6 & 0.015 & 0.43 & 43.1 &  29.5 & 5.40 & 5.18 & 0.06 \\
1342198535$+$36$+$38&400&HD\,39608 & h &  20   &  55  & 180/4/10 &70$+$110&  19.2 & 0.011 & 0.32 & 59.5 &  30.4 & 5.54 & 5.27 & 0.06 \\
1342185450 &  146   &HD\,181597    & h &  10   &   9  & 210/4/20 &   90   &  17.4 & 0.019 & 0.53 & 32.6 &  27.5 & 5.25 & 4.74 & 0.08 \\
1342185451 &  146   &HD\,181597    & h &  20   &  10  & 210/4/20 &   90   &  19.9 & 0.023 & 0.64 & 30.9 &  31.5 & 5.51 & 5.34 & 0.09 \\
1342189130 &  241   &$\delta$\,Hyi & h &  20   &   9  & 240/4/8  &   63   &  14.0 & 0.033 & 0.92 & 15.3 &  22.2 & 6.01 & 5.39 & 0.15 \\
\hline                                   
\end{tabular}
\end{sidewaystable*}

\clearpage

\begin{sidewaystable*}[h!]
\caption{Scan map photometry measurements in the green (100\,$\mu$m filter).
Processing proceeded from SPG~v13.1.0 level 1 products with HIPE version 15 
build 165. 
}
\label{table:scanmapphotgreen6_1}      
\centering                          
\begin{tabular}{r r r r r r r c r r r r r r r r r}        
\hline\hline                 
            \noalign{\smallskip}
 OBSID     &   OD   &   Target     & G & Speed  & Rep. & Map params  & ScanAngles &f$_{\rm aper}$&$\sigma_{\rm pix}$&$\sigma_{\rm aper,corr}$& S/N &f$_{\rm star}$&  W$_{max}$ & W$_{min}$ & $\Delta$\,W\\
           &        &              &    &(\arcsec/s)&  &(\arcsec/\arcsec/no.)&(deg)& (mJy)     &      (mJy)     &       (mJy)        &     &   (mJy)    &  (\arcsec) & (\arcsec) &   (\arcsec)    \\
            \noalign{\smallskip}
\hline
            \noalign{\smallskip}
1342217351 &  684   &$\beta$\,Gem  & h &  20 &  3 & 180/4/10 &   70   & 848.6 & 0.085 & 2.01 & 422 & 1282.9 & 6.76 & 6.52 & 0.04 \\
1342217352 &  684   &$\beta$\,Gem  & h &  20 &  3 & 180/4/10 &  110   & 847.4 & 0.082 & 1.95 & 435 & 1281.0 & 6.91 & 6.44 & 0.03 \\
1342217351$+$52& 684&$\beta$\,Gem  & h &  20 &  6 & 180/4/10 &70$+$110& 845.7 & 0.059 & 1.39 & 609 & 1278.4 & 6.79 & 6.44 & 0.05 \\
1342230122 &  872   &$\beta$\,Gem  & h &  20 &  3 & 180/4/10 &   70   & 851.3 & 0.078 & 1.85 & 459 & 1287.0 & 6.72 & 6.52 & 0.03 \\
1342230123 &  872   &$\beta$\,Gem  & h &  20 &  3 & 180/4/10 &  110   & 847.2 & 0.080 & 1.90 & 446 & 1280.8 & 6.95 & 6.42 & 0.03 \\
1342230122$+$23& 872&$\beta$\,Gem  & h &  20 &  6 & 180/4/10 &70$+$110& 846.9 & 0.052 & 1.25 & 680 & 1280.3 & 6.79 & 6.41 & 0.05 \\
1342242770 & 1051   &$\beta$\,Gem  & h &  20 &  3 & 180/4/10 &   70   & 844.8 & 0.084 & 1.99 & 424 & 1277.1 & 6.74 & 6.54 & 0.04 \\
1342242771 & 1051   &$\beta$\,Gem  & h &  20 &  3 & 180/4/10 &  110   & 850.7 & 0.079 & 1.87 & 455 & 1286.0 & 6.88 & 6.49 & 0.04 \\
1342242770$+$71&1051&$\beta$\,Gem  & h &  20 &  6 & 180/4/10 &70$+$110& 845.0 & 0.055 & 1.31 & 647 & 1277.5 & 6.74 & 6.49 & 0.05 \\
1342252879 & 1244   &$\beta$\,Gem  & h &  20 &  3 & 180/4/10 &   70   & 851.7 & 0.083 & 1.96 & 434 & 1287.5 & 6.78 & 6.52 & 0.03 \\
1342252880 & 1244   &$\beta$\,Gem  & h &  20 &  3 & 180/4/10 &  110   & 851.4 & 0.075 & 1.77 & 481 & 1287.1 & 6.92 & 6.43 & 0.03 \\
1342252879$+$80&1244&$\beta$\,Gem  & h &  20 &  6 & 180/4/10 &70$+$110& 850.2 & 0.055 & 1.30 & 655 & 1285.3 & 6.75 & 6.41 & 0.05 \\
1342212862 &  614   &$\alpha$\,Ari & h &  20 &  3 & 180/4/10 &   70   & 538.9 & 0.079 & 1.87 & 288 &  814.7 & 6.84 & 6.52 & 0.04 \\
1342212863 &  614   &$\alpha$\,Ari & h &  20 &  3 & 180/4/10 &  110   & 545.9 & 0.081 & 1.93 & 282 &  825.3 & 6.78 & 6.63 & 0.04 \\
1342212862$+$63& 614&$\alpha$\,Ari & h &  20 &  6 & 180/4/10 &70$+$110& 540.5 & 0.055 & 1.31 & 413 &  817.1 & 6.73 & 6.56 & 0.05 \\
1342237396 &  974   &$\alpha$\,Ari & h &  20 &  3 & 180/4/10 &   70   & 547.5 & 0.080 & 1.89 & 289 &  827.7 & 6.87 & 6.50 & 0.04 \\
1342237397 &  974   &$\alpha$\,Ari & h &  20 &  3 & 180/4/10 &  110   & 540.2 & 0.078 & 1.84 & 293 &  816.6 & 6.80 & 6.57 & 0.03 \\
1342237396$+$97& 974&$\alpha$\,Ari & h &  20 &  6 & 180/4/10 &70$+$110& 543.2 & 0.055 & 1.30 & 418 &  821.1 & 6.72 & 6.46 & 0.05 \\
1342248027 & 1157   &$\alpha$\,Ari & h &  20 &  3 & 180/4/10 &   70   & 540.1 & 0.084 & 1.98 & 272 &  816.6 & 6.82 & 6.50 & 0.03 \\
1342248028 & 1157   &$\alpha$\,Ari & h &  20 &  3 & 180/4/10 &  110   & 539.5 & 0.085 & 2.02 & 267 &  815.6 & 6.73 & 6.59 & 0.04 \\
1342248027$+$28&1157&$\alpha$\,Ari & h &  20 &  6 & 180/4/10 &70$+$110& 538.4 & 0.056 & 1.32 & 407 &  813.9 & 6.77 & 6.54 & 0.05 \\
1342259799 & 1344   &$\alpha$\,Ari & h &  20 &  3 & 180/4/10 &   70   & 544.6 & 0.081 & 1.92 & 283 &  823.4 & 6.83 & 6.52 & 0.04 \\
1342259800 & 1344   &$\alpha$\,Ari & h &  20 &  3 & 180/4/10 &  110   & 543.4 & 0.082 & 1.95 & 278 &  821.5 & 6.71 & 6.68 & 0.04 \\
1342259799$+$800&1344&$\alpha$\,Ari& h &  20 &  6 & 180/4/10 &70$+$110& 539.0 & 0.055 & 1.31 & 413 &  814.8 & 6.71 & 6.54 & 0.05 \\
1342190969 &  284   &$\varepsilon$\,Lep&h&20 &  1 & 150/4/10 &   70   & 377.5 & 0.139 & 3.29 & 115 &  570.6 & 6.72 & 6.48 & 0.05 \\
1342190970 &  284   &$\varepsilon$\,Lep&h&20 &  1 & 150/4/10 &  110   & 372.9 & 0.144 & 3.42 & 109 &  563.7 & 6.70 & 6.63 & 0.04 \\
1342190969$+$70& 284&$\varepsilon$\,Lep&h&20 &  2 & 150/4/10 &70$+$110& 374.3 & 0.099 & 2.34 & 160 &  565.8 & 6.63 & 6.49 & 0.05 \\
1342205204 &  502   &$\varepsilon$\,Lep&h&20 &  3 & 180/4/10 &   70   & 379.2 & 0.078 & 1.85 & 205 &  573.2 & 6.76 & 6.55 & 0.04 \\
1342205205 &  502   &$\varepsilon$\,Lep&h&20 &  3 & 180/4/10 &  110   & 373.6 & 0.081 & 1.93 & 193 &  564.8 & 6.90 & 6.44 & 0.04 \\
1342205204$+$05& 502&$\varepsilon$\,Lep&h&20 &  6 & 180/4/10 &70$+$110& 375.2 & 0.053 & 1.25 & 299 &  567.3 & 6.76 & 6.44 & 0.05 \\
1342214208 &  640   &$\varepsilon$\,Lep&h&20 &  3 & 180/4/10 &   70   & 371.6 & 0.077 & 1.82 & 204 &  561.8 & 6.79 & 6.47 & 0.04 \\
1342214209 &  640   &$\varepsilon$\,Lep&h&20 &  3 & 180/4/10 &  110   & 373.4 & 0.081 & 1.91 & 195 &  564.6 & 6.76 & 6.58 & 0.04 \\
1342214208$+$09& 640&$\varepsilon$\,Lep&h&20 &  6 & 180/4/10 &70$+$110& 370.9 & 0.054 & 1.28 & 290 &  560.7 & 6.73 & 6.54 & 0.05 \\
1342227299 &  833   &$\varepsilon$\,Lep&h&20 &  3 & 180/4/10 &   70   & 377.8 & 0.077 & 1.84 & 206 &  571.1 & 6.90 & 6.50 & 0.04 \\
1342227300 &  833   &$\varepsilon$\,Lep&h&20 &  3 & 180/4/10 &  110   & 374.5 & 0.077 & 1.83 & 205 &  566.1 & 6.78 & 6.54 & 0.04 \\
1342227299$+$300&833&$\varepsilon$\,Lep&h&20 &  6 & 180/4/10 &70$+$110& 375.0 & 0.051 & 1.21 & 310 &  566.8 & 6.80 & 6.50 & 0.05 \\
1342241335 & 1034   &$\varepsilon$\,Lep&h&20 &  3 & 180/4/10 &   70   & 383.0 & 0.078 & 1.86 & 206 &  579.0 & 6.85 & 6.68 & 0.04 \\
1342241336 & 1034   &$\varepsilon$\,Lep&h&20 &  3 & 180/4/10 &  110   & 379.9 & 0.082 & 1.94 & 196 &  574.3 & 6.82 & 6.44 & 0.04 \\
1342241335$+$36&1034&$\varepsilon$\,Lep&h&20 &  6 & 180/4/10 &70$+$110& 379.7 & 0.054 & 1.28 & 296 &  574.0 & 6.81 & 6.62 & 0.05 \\
1342263902 & 1377   &$\varepsilon$\,Lep&h&20 &  3 & 180/4/10 &   70   & 375.2 & 0.078 & 1.86 & 202 &  567.3 & 6.84 & 6.53 & 0.04 \\
1342263903 & 1377   &$\varepsilon$\,Lep&h&20 &  3 & 180/4/10 &  110   & 372.6 & 0.077 & 1.83 & 203 &  563.2 & 6.67 & 6.57 & 0.04 \\
1342263902$+$03&1377&$\varepsilon$\,Lep&h&20 &  6 & 180/4/10 &70$+$110& 373.1 & 0.053 & 1.25 & 298 &  564.0 & 6.68 & 6.47 & 0.05 \\
            \noalign{\smallskip}
\hline                                   
\end{tabular}
\end{sidewaystable*}

\addtocounter{table}{-1}

\begin{sidewaystable*}[h!]
\caption{Scan map photometry measurements in the green (100\,$\mu$m filter) 
continued.
}             
\label{table:scanmapphotgreen6_2}      
\centering                          
\begin{tabular}{r r r r r r r c r r r r r r r r r}        
\hline\hline                 
            \noalign{\smallskip}
 OBSID     &   OD   &   Target     & G & Speed  & Rep. & Map params  & ScanAngles &f$_{\rm aper}$&$\sigma_{\rm pix}$&$\sigma_{\rm aper,corr}$& S/N &f$_{\rm star}$&  W$_{max}$ & W$_{min}$ & $\Delta$\,W\\
           &        &              &    &(\arcsec/s)&  &(\arcsec/\arcsec/no.)&(deg)& (mJy)     &      (mJy)     &       (mJy)        &     &   (mJy)    &  (\arcsec) & (\arcsec) &   (\arcsec)    \\
            \noalign{\smallskip}
\hline
            \noalign{\smallskip}
1342218546 &  697   &$\omega$\,Cap & h & 20 & 3 & 180/4/10 &   70   & 271.3 & 0.084 & 1.98 & 137 & 410.1 & 6.75 & 6.44 & 0.04 \\
1342218547 &  697   &$\omega$\,Cap & h & 20 & 3 & 180/4/10 &  110   & 270.5 & 0.078 & 1.85 & 146 & 409.0 & 6.69 & 6.65 & 0.04 \\
1342218546$+$47& 697&$\omega$\,Cap & h & 20 & 6 & 180/4/10 &70$+$110& 270.4 & 0.055 & 1.31 & 207 & 408.7 & 6.60 & 6.54 & 0.05 \\
1342231261 &  889   &$\omega$\,Cap & h & 20 & 3 & 180/4/10 &   70   & 276.7 & 0.084 & 1.98 & 139 & 418.3 & 6.82 & 6.53 & 0.04 \\
1342231262 &  889   &$\omega$\,Cap & h & 20 & 3 & 180/4/10 &  110   & 272.0 & 0.081 & 1.93 & 141 & 411.2 & 6.77 & 6.50 & 0.04 \\
1342231261$+$62& 889&$\omega$\,Cap & h & 20 & 6 & 180/4/10 &70$+$110& 273.5 & 0.057 & 1.35 & 203 & 413.4 & 6.75 & 6.46 & 0.05 \\
1342243778 & 1058   &$\omega$\,Cap & h & 20 & 3 & 180/4/10 &   70   & 272.9 & 0.077 & 1.82 & 150 & 412.5 & 6.78 & 6.54 & 0.04 \\
1342243779 & 1058   &$\omega$\,Cap & h & 20 & 3 & 180/4/10 &  110   & 275.8 & 0.081 & 1.93 & 143 & 416.9 & 6.86 & 6.59 & 0.04 \\
1342243778$+$79&1058&$\omega$\,Cap & h & 20 & 6 & 180/4/10 &70$+$110& 273.4 & 0.053 & 1.25 & 219 & 413.3 & 6.74 & 6.54 & 0.05 \\
1342254207 & 1266   &$\omega$\,Cap & h & 20 & 3 & 180/4/10 &   70   & 275.4 & 0.077 & 1.82 & 151 & 416.4 & 6.79 & 6.44 & 0.04 \\
1342254208 & 1266   &$\omega$\,Cap & h & 20 & 3 & 180/4/10 &  110   & 270.0 & 0.076 & 1.81 & 149 & 408.1 & 6.65 & 6.49 & 0.04 \\
1342254207$+$08&1266&$\omega$\,Cap & h & 20 & 6 & 180/4/10 &70$+$110& 271.5 & 0.050 & 1.20 & 227 & 410.4 & 6.68 & 6.44 & 0.05 \\
1342186151 &  160   &$\eta$\,Dra   & h & 10 & 1 & 120/5/9  &   80   & 160.4 & 0.160 & 3.79 &42.3 & 242.6 & 6.72 & 6.30 & 0.08 \\
1342186152 &  160   &$\eta$\,Dra   & h & 10 & 1 & 120/5/9  &  100   & 171.3 & 0.159 & 3.78 &45.3 & 258.9 & 6.79 & 6.54 & 0.09 \\
1342186151$+$52& 160&$\eta$\,Dra   & h & 10 & 2 & 120/5/9  &80$+$100& 165.5 & 0.108 & 2.57 &64.5 & 250.2 & 6.79 & 6.43 & 0.08 \\
1342186153 &  160   &$\eta$\,Dra   & h & 10 & 1 &  90/5/9  &   60   & 180.3 & 0.188 & 4.46 &40.4 & 272.6 & 7.28 & 6.22 & 0.08 \\
1342186154 &  160   &$\eta$\,Dra   & h & 10 & 1 &  90/5/9  &  120   & 158.9 & 0.177 & 4.21 &37.8 & 240.2 & 6.60 & 6.57 & 0.09 \\
1342186153$+$54& 160&$\eta$\,Dra   & h & 10 & 2 &  90/5/9  &60$+$120& 169.0 & 0.124 & 2.94 &57.4 & 255.6 & 6.85 & 6.35 & 0.08 \\
1342186155 &  160   &$\eta$\,Dra   & h & 20 & 1 &  90/5/9  &   70   & 164.0 & 0.214 & 5.07 &32.3 & 247.9 & 6.91 & 6.51 & 0.08 \\
1342186156 &  160   &$\eta$\,Dra   & h & 20 & 1 &  90/5/9  &  110   & 169.2 & 0.197 & 4.68 &36.1 & 255.9 & 6.99 & 6.83 & 0.09 \\
1342186155$+$56& 160&$\eta$\,Dra   & h & 20 & 2 &  90/5/9  &70$+$110& 167.0 & 0.145 & 3.43 &48.6 & 252.5 & 6.77 & 6.57 & 0.08 \\
1342186157 &  160   &$\eta$\,Dra   & h & 10 & 1 & 150/4/9  &   85   & 167.5 & 0.147 & 3.49 &48.0 & 253.3 & 6.70 & 6.59 & 0.11 \\
1342186158 &  160   &$\eta$\,Dra   & h & 10 & 1 & 150/4/9  &   95   & 154.1 & 0.146 & 3.46 &44.5 & 232.9 & 6.31 & 6.24 & 0.07 \\
1342186157$+$58& 160&$\eta$\,Dra   & h & 10 & 2 & 150/4/9  &85$+$95 & 160.2 & 0.099 & 2.34 &68.4 & 242.2 & 6.49 & 6.46 & 0.09 \\
1342212497 &  607   &$\delta$\,Dra & h & 20 & 3 & 180/4/10 &   70   & 141.7 & 0.081 & 1.94 &73.4 & 214.2 & 6.74 & 6.46 & 0.05 \\
1342212498 &  607   &$\delta$\,Dra & h & 20 & 3 & 180/4/10 &  110   & 143.2 & 0.081 & 1.92 &74.4 & 216.5 & 6.75 & 6.53 & 0.05 \\
1342212497$+$98& 607&$\delta$\,Dra & h & 20 & 6 & 180/4/10 &70$+$110& 141.7 & 0.053 & 1.25 & 113 & 214.3 & 6.72 & 6.48 & 0.05 \\
1342222147 &  751   &$\delta$\,Dra & h & 20 & 3 & 180/4/10 &   70   & 140.6 & 0.081 & 1.92 &73.1 & 212.5 & 6.76 & 6.43 & 0.05 \\
1342222148 &  751   &$\delta$\,Dra & h & 20 & 3 & 180/4/10 &  110   & 138.1 & 0.081 & 1.91 &72.2 & 208.8 & 6.56 & 6.51 & 0.05 \\
1342222147$+$48& 751&$\delta$\,Dra & h & 20 & 6 & 180/4/10 &70$+$110& 138.8 & 0.057 & 1.36 & 102 & 209.8 & 6.67 & 6.44 & 0.05 \\
1342233573 &  934   &$\delta$\,Dra & h & 20 & 3 & 180/4/10 &   70   & 139.2 & 0.080 & 1.89 &73.6 & 210.4 & 6.92 & 6.38 & 0.05 \\
1342233574 &  934   &$\delta$\,Dra & h & 20 & 3 & 180/4/10 &  110   & 138.3 & 0.077 & 1.84 &75.2 & 209.0 & 6.57 & 6.54 & 0.05 \\
1342233573$+$74& 934&$\delta$\,Dra & h & 20 & 6 & 180/4/10 &70$+$110& 138.8 & 0.054 & 1.29 & 108 & 209.8 & 6.73 & 6.41 & 0.05 \\
1342250093 & 1198   &$\delta$\,Dra & h & 20 & 3 & 180/4/10 &   70   & 143.1 & 0.079 & 1.88 &76.3 & 216.4 & 6.72 & 6.64 & 0.05 \\
1342250094 & 1198   &$\delta$\,Dra & h & 20 & 3 & 180/4/10 &  110   & 141.4 & 0.078 & 1.86 &76.0 & 213.7 & 6.76 & 6.47 & 0.05 \\
1342250093$+$94&1198&$\delta$\,Dra & h & 20 & 6 & 180/4/10 &70$+$110& 142.0 & 0.055 & 1.31 & 109 & 214.6 & 6.63 & 6.52 & 0.06 \\
1342257987 & 1328   &$\delta$\,Dra & h & 20 & 3 & 180/4/10 &   70   & 144.7 & 0.079 & 1.86 &77.6 & 218.7 & 6.84 & 6.47 & 0.05 \\
1342257988 & 1328   &$\delta$\,Dra & h & 20 & 3 & 180/4/10 &  110   & 146.6 & 0.080 & 1.90 &77.0 & 221.6 & 6.90 & 6.79 & 0.05 \\
1342257987$+$88&1328&$\delta$\,Dra & h & 20 & 6 & 180/4/10 &70$+$110& 145.1 & 0.051 & 1.21 & 119 & 219.3 & 6.82 & 6.60 & 0.05 \\
\hline                                   
\end{tabular}
\end{sidewaystable*}

\clearpage

\addtocounter{table}{-1}

\begin{sidewaystable*}[h!]
\caption{Scan map photometry measurements in the green (100\,$\mu$m filter) 
continued.
}             
\label{table:scanmapphotgreen6_3}      
\centering                          
\begin{tabular}{r r r r r r r c r r r r r r r r r}        
\hline\hline                 
            \noalign{\smallskip}
 OBSID     &   OD   &   Target     & G & Speed  & Rep. & Map params  & ScanAngles &f$_{\rm aper}$&$\sigma_{\rm pix}$&$\sigma_{\rm aper,corr}$& S/N &f$_{\rm star}$&  W$_{max}$ & W$_{min}$ & $\Delta$\,W\\
           &        &              &    &(\arcsec/s)&  &(\arcsec/\arcsec/no.)&(deg)& (mJy)     &      (mJy)     &       (mJy)        &     &   (mJy)    &  (\arcsec) & (\arcsec) &   (\arcsec)    \\
            \noalign{\smallskip}
\hline
            \noalign{\smallskip}
1342184585 &  139   &$\theta$\,Umi & h &  10   &   1  & 210/15/4 &   90   & 98.1 & 0.204 & 4.84 & 20.3 & 148.3 & 6.74 & 6.38 & 0.19 \\
1342184586 &  139   &$\theta$\,Umi & h &  20   &   1  & 210/15/4 &   90   & 91.6 & 0.246 & 5.84 & 15.7 & 138.4 & 7.48 & 6.15 & 0.16 \\
1342190972 &  284   &HD\,41047     & h &  20   &   3  & 150/4/10 &   70   & 64.4 & 0.082 & 1.94 & 33.3 &  97.4 & 6.77 & 6.37 & 0.09 \\
1342190973 &  284   &HD\,41047     & h &  20   &   3  & 150/4/10 &  110   & 63.5 & 0.084 & 1.99 & 31.8 &  95.9 & 6.81 & 6.59 & 0.10 \\
1342190972$+$73 & 284 &HD\,41047   & h &  20   &   6  & 150/4/10 &70$+$110& 63.7 & 0.059 & 1.40 & 45.5 &  96.2 & 6.70 & 6.42 & 0.07 \\
1342203113 &  460   &42\,Dra       & h &  20   &  20  & 180/4/10 &   70   & 49.0 & 0.034 & 0.80 & 61.1 &  74.1 & 6.77 & 6.59 & 0.07 \\
1342203114 &  460   &42\,Dra       & h &  20   &  20  & 180/4/10 &  110   & 46.7 & 0.033 & 0.78 & 59.9 &  70.6 & 6.84 & 6.33 & 0.06 \\
1342203113$+$14& 460&42\,Dra       & h &  20   &  40  & 180/4/10 &70$+$110& 47.9 & 0.023 & 0.54 & 89.4 &  72.4 & 6.68 & 6.53 & 0.07 \\
1342243732 & 1057   &42\,Dra       & h &  20   &  20  & 180/4/10 &   70   & 48.7 & 0.031 & 0.74 & 65.6 &  73.6 & 6.78 & 6.57 & 0.06 \\
1342243733 & 1057   &42\,Dra       & h &  20   &  20  & 180/4/10 &  110   & 49.1 & 0.031 & 0.74 & 65.9 &  74.2 & 6.75 & 6.54 & 0.05 \\
1342243732$+$33&1057&42\,Dra       & h &  20   &  40  & 180/4/10 &70$+$110& 48.8 & 0.021 & 0.49 & 99.5 &  73.7 & 6.72 & 6.49 & 0.06 \\
1342184302 &  132   &HD\,138265    & h &  10   &   3  & 210/4/25 &   90   & 37.6 & 0.048 & 1.15 & 32.8 &  56.8 & 6.91 & 6.51 & 0.10 \\
1342184303 &  132   &HD\,138265    & h &  20   &   3  & 210/4/25 &   90   & 37.3 & 0.058 & 1.38 & 27.1 &  56.4 & 6.86 & 6.46 & 0.10 \\
1342185448 &  146   &HD\,138265    & h &  20   &   1  & 210/4/20 &   90   & 39.8 & 0.106 & 2.51 & 15.8 &  60.2 & 7.00 & 6.94 & 0.19 \\
1342185449 &  146   &HD\,138265    & h &  10   &   1  & 210/4/20 &   90   & 38.1 & 0.095 & 2.25 & 17.0 &  57.6 & 6.70 & 6.27 & 0.14 \\
1342191986 &  300   &HD\,138265    & h &  20   &   6  & 150/4/10 &   70   & 36.7 & 0.064 & 1.52 & 24.2 &  55.4 & 6.83 & 6.24 & 0.09 \\
1342191987 &  300   &HD\,138265    & h &  20   &   6  & 150/4/10 &  110   & 36.9 & 0.056 & 1.34 & 27.6 &  55.8 & 6.74 & 6.73 & 0.11 \\
1342191986$+$87& 300&HD\,138265    & h &  20   &  12  & 150/4/10 &70$+$110& 36.4 & 0.044 & 1.04 & 35.0 &  55.0 & 6.68 & 6.40 & 0.08 \\
1342186160 &  160   &HD\,159330    & h &  10   &   1  & 120/3/21 &   80   & 18.0 & 0.104 & 2.46 &  7.3 &  27.3 & 6.83 & 5.93 & 0.31 \\
1342186161 &  160   &HD\,159330    & h &  10   &   1  & 120/3/21 &  100   & 17.0 & 0.109 & 2.60 &  6.5 &  25.7 & 5.19 & 5.01 & 0.39 \\
1342186160$+$61& 160&HD\,159330    & h &  10   &   2  & 120/3/21 &80$+$100& 17.3 & 0.072 & 1.72 & 10.1 &  26.2 & 6.22 & 6.03 & 0.26 \\
1342188839 &  233   &HD\,159330    & h &  20   &  10  & 240/4/8  &   63   & 20.3 & 0.050 & 1.18 & 17.3 &  30.7 & 6.66 & 6.78 & 0.21 \\
1342188840 &  233   &HD\,159330    & h &  20   &  10  & 240/4/8  &  117   & 23.4 & 0.049 & 1.16 & 20.1 &  35.3 & 7.23 & 6.33 & 0.15 \\
1342188839$+$40& 233&HD\,159330    & h &  20   &  20  & 240/4/8  &63$+$117& 21.7 & 0.034 & 0.81 & 26.9 &  32.8 & 6.90 & 6.60 & 0.14 \\
1342213583 &  628   &HD\,159330    & h &  20   &  15  & 180/4/10 &   70   & 20.9 & 0.036 & 0.86 & 24.4 &  31.6 & 6.69 & 6.67 & 0.16 \\
1342213584 &  628   &HD\,159330    & h &  20   &  15  & 180/4/10 &  110   & 19.3 & 0.036 & 0.86 & 22.4 &  29.1 & 6.46 & 5.60 & 0.13 \\
1342213583$+$84& 628&HD\,159330    & h &  20   &  30  & 180/4/10 &70$+$110& 20.3 & 0.024 & 0.58 & 35.2 &  30.7 & 6.57 & 6.15 & 0.12 \\
1342227973 &  843   &HD\,152222    & h &  20   &  33  & 180/4/10 &   70   & 13.7 & 0.025 & 0.60 & 23.0 &  20.7 & 6.59 & 6.40 & 0.13 \\
1342227974 &  843   &HD\,152222    & h &  20   &  33  & 180/4/10 &  110   & 13.9 & 0.024 & 0.56 & 24.9 &  20.9 & 7.01 & 6.55 & 0.13 \\
1342227973$+$74& 843&HD\,152222    & h &  20   &  66  & 180/4/10 &70$+$110& 13.7 & 0.017 & 0.40 & 34.3 &  20.7 & 6.76 & 6.56 & 0.09 \\
1342198537 &  400   &HD\,39608     & h &  20   &  35  & 180/4/10 &   70   & 11.9 & 0.024 & 0.57 & 20.9 &  17.9 & 7.57 & 7.21 & 0.17 \\
\hline                                   
\end{tabular}
\end{sidewaystable*}

\clearpage

\begin{sidewaystable*}[h!]
\caption{Scan map photometry measurements in the red (160\,$\mu$m filter).
Processing proceeded from SPG~v14.2.0 level 1 products with HIPE version 15 
build 1480. 
}             
\label{table:scanmapphotred10_1}      
\centering                          
\begin{tabular}{r r r r r r r c r r r r r r r r r}        
\hline\hline                 
            \noalign{\smallskip}
 OBSID     &   OD   &   Target     & G & Speed  & Rep. & Map params  & ScanAngles &f$_{\rm aper}$&$\sigma_{\rm pix}$&$\sigma_{\rm aper,corr}$& S/N &f$_{\rm star}$&  W$_{max}$ & W$_{min}$ & $\Delta$\,W\\
           &        &              &    &(\arcsec/s)&  &(\arcsec/\arcsec/no.)&(deg)& (mJy)     &      (mJy)     &       (mJy)        &     &   (mJy)    &  (\arcsec) & (\arcsec) &   (\arcsec)    \\
            \noalign{\smallskip}
\hline
            \noalign{\smallskip}
1342217348 &  684   &$\beta$\,Gem  & h &  20   &   3  & 180/4/10 &   70   & 345.7 & 0.123 & 4.59 & 75.3 & 502.0 & 11.37 &  9.95 & 0.16 \\
1342217349 &  684   &$\beta$\,Gem  & h &  20   &   3  & 180/4/10 &  110   & 351.8 & 0.124 & 4.62 & 76.2 & 510.9 & 11.67 &  9.78 & 0.18 \\
1342217351 &  684   &$\beta$\,Gem  & h &  20   &   3  & 180/4/10 &   70   & 346.8 & 0.145 & 5.39 & 64.4 & 503.6 & 11.35 &  9.83 & 0.17 \\
1342217352 &  684   &$\beta$\,Gem  & h &  20   &   3  & 180/4/10 &  110   & 341.4 & 0.127 & 4.72 & 72.2 & 495.8 & 11.51 &  9.67 & 0.16 \\
..48$+$49$+$51$+$52& 684&$\beta$\,Gem& h &20   &  12  & 180/4/10 &70$+$110& 345.2 & 0.066 & 2.44 &  141 & 501.2 & 11.40 &  9.80 & 0.28 \\
1342230120 &  872   &$\beta$\,Gem  & h &  20   &   3  & 180/4/10 &   70   & 336.7 & 0.121 & 4.51 & 74.7 & 489.0 & 10.97 &  9.78 & 0.18 \\
1342230121 &  872   &$\beta$\,Gem  & h &  20   &   3  & 180/4/10 &  110   & 345.7 & 0.127 & 4.74 & 72.9 & 502.0 & 11.08 &  9.76 & 0.18 \\
1342230122 &  872   &$\beta$\,Gem  & h &  20   &   3  & 180/4/10 &   70   & 350.2 & 0.144 & 5.36 & 65.4 & 508.5 & 10.87 &  9.86 & 0.19 \\
1342230123 &  872   &$\beta$\,Gem  & h &  20   &   3  & 180/4/10 &  110   & 343.9 & 0.112 & 4.17 & 82.6 & 499.3 & 11.05 &  9.74 & 0.18 \\
..20$+$21$+$22$+$23& 872&$\beta$\,Gem& h &20   &  12  & 180/4/10 &70$+$110& 343.1 & 0.066 & 2.45 &  141 & 498.2 & 11.06 &  9.78 & 0.31 \\
1342242770 & 1051   &$\beta$\,Gem  & h &  20   &   3  & 180/4/10 &   70   & 332.2 & 0.144 & 5.38 & 61.7 & 482.4 & 11.16 &  9.54 & 0.17 \\
1342242771 & 1051   &$\beta$\,Gem  & h &  20   &   3  & 180/4/10 &  110   & 340.9 & 0.122 & 4.53 & 75.2 & 495.0 & 11.46 &  9.77 & 0.20 \\
1342242772 & 1051   &$\beta$\,Gem  & h &  20   &   3  & 180/4/10 &   70   & 332.0 & 0.130 & 4.84 & 68.7 & 482.1 & 11.32 &  9.80 & 0.17 \\
1342242773 & 1051   &$\beta$\,Gem  & h &  20   &   3  & 180/4/10 &  110   & 344.6 & 0.131 & 4.88 & 70.6 & 500.5 & 11.45 &  9.55 & 0.21 \\
..70$+$71$+$72$+$73&1051&$\beta$\,Gem& h &20   &  12  & 180/4/10 &70$+$110& 339.0 & 0.069 & 2.58 &  131 & 492.3 & 11.37 &  9.68 & 0.26 \\
1342252879 & 1244   &$\beta$\,Gem  & h &  20   &   3  & 180/4/10 &   70   & 336.9 & 0.127 & 4.74 & 71.1 & 489.2 & 11.34 &  9.89 & 0.16 \\
1342252880 & 1244   &$\beta$\,Gem  & h &  20   &   3  & 180/4/10 &  110   & 343.3 & 0.128 & 4.76 & 72.1 & 498.5 & 10.93 &  9.62 & 0.16 \\
1342252881 & 1244   &$\beta$\,Gem  & h &  20   &   3  & 180/4/10 &   70   & 339.1 & 0.125 & 4.67 & 72.7 & 492.4 & 11.01 &  9.75 & 0.18 \\
1342252882 & 1244   &$\beta$\,Gem  & h &  20   &   3  & 180/4/10 &  110   & 343.5 & 0.114 & 4.23 & 81.2 & 498.7 & 11.11 &  9.66 & 0.17 \\
..79$+$80$+$81$+$82&1244&$\beta$\,Gem& h &20   &  12  & 180/4/10 &70$+$110& 338.9 & 0.059 & 2.20 &  153 & 492.1 & 11.30 & 10.02 & 0.32 \\
1342212862 &  614   &$\alpha$\,Ari & h &  20   &   3  & 180/4/10 &   70   & 230.1 & 0.133 & 4.95 & 46.4 & 334.1 & 11.42 & 10.11 & 0.21 \\
1342212863 &  614   &$\alpha$\,Ari & h &  20   &   3  & 180/4/10 &  110   & 220.1 & 0.121 & 4.51 & 48.8 & 319.6 & 10.89 &  9.88 & 0.22 \\
1342212864 &  614   &$\alpha$\,Ari & h &  20   &   3  & 180/4/10 &   70   & 228.9 & 0.121 & 4.50 & 50.9 & 332.4 & 10.81 &  9.66 & 0.21 \\
1342212865 &  614   &$\alpha$\,Ari & h &  30   &   3  & 180/4/10 &  110   & 227.0 & 0.120 & 4.48 & 50.6 & 329.6 & 11.27 &  9.80 & 0.22 \\
..62$+$63$+$64$+$65& 614&$\alpha$\,Ari&h& 20   &  12  & 180/4/10 &70$+$110& 225.2 & 0.063 & 2.34 & 96.0 & 327.1 & 11.01 &  9.87 & 0.43 \\
1342237396 &  974   &$\alpha$\,Ari & h &  20   &   3  & 180/4/10 &   70   & 221.1 & 0.118 & 4.41 & 50.2 & 321.0 & 11.27 & 10.04 & 0.23 \\
1342237397 &  974   &$\alpha$\,Ari & h &  20   &   3  & 180/4/10 &  110   & 227.7 & 0.121 & 4.49 & 50.7 & 330.7 & 11.01 & 10.01 & 0.21 \\
1342237398 &  974   &$\alpha$\,Ari & h &  20   &   3  & 180/4/10 &   70   & 223.8 & 0.140 & 5.21 & 42.9 & 325.0 & 10.85 &  9.69 & 0.20 \\
1342237399 &  974   &$\alpha$\,Ari & h &  20   &   3  & 180/4/10 &  110   & 225.4 & 0.141 & 5.26 & 42.9 & 327.3 & 10.82 &  9.66 & 0.22 \\
..96$+$97$+$98$+$99& 974&$\alpha$\,Ari&h& 20   &  12  & 180/4/10 &70$+$110& 223.7 & 0.071 & 2.65 & 84.5 & 324.8 & 11.10 &  9.81 & 0.35 \\
1342248027 & 1157   &$\alpha$\,Ari & h &  20   &   3  & 180/4/10 &   70   & 228.9 & 0.131 & 4.87 & 47.0 & 332.4 & 10.94 &  9.97 & 0.19 \\
1342248028 & 1157   &$\alpha$\,Ari & h &  20   &   3  & 180/4/10 &  110   & 230.4 & 0.122 & 4.55 & 50.6 & 334.6 & 10.69 & 10.13 & 0.18 \\
1342248029 & 1157   &$\alpha$\,Ari & h &  20   &   3  & 180/4/10 &   70   & 233.9 & 0.139 & 5.18 & 45.2 & 339.7 & 11.43 & 10.03 & 0.19 \\
1342248030 & 1157   &$\alpha$\,Ari & h &  20   &   3  & 180/4/10 &  110   & 232.3 & 0.127 & 4.72 & 49.1 & 337.3 & 10.91 &  9.99 & 0.22 \\
..27$+$28$+$29$+$30&1157&$\alpha$\,Ari&h& 20   &  12  & 180/4/10 &70$+$110& 230.4 & 0.078 & 2.91 & 79.2 & 334.5 & 11.03 &  9.98 & 0.34 \\
1342259799 & 1344   &$\alpha$\,Ari & h &  20   &   3  & 180/4/10 &   70   & 215.8 & 0.131 & 4.88 & 44.2 & 313.3 & 11.54 &  9.97 & 0.18 \\
1342259800 & 1344   &$\alpha$\,Ari & h &  20   &   3  & 180/4/10 &  110   & 221.0 & 0.116 & 4.32 & 51.2 & 320.9 & 10.78 &  9.97 & 0.19 \\
1342259801 & 1344   &$\alpha$\,Ari & h &  20   &   3  & 180/4/10 &   70   & 228.2 & 0.129 & 4.82 & 47.3 & 331.4 & 11.58 &  9.95 & 0.15 \\
1342259802 & 1344   &$\alpha$\,Ari & h &  20   &   3  & 180/4/10 &  110   & 221.3 & 0.127 & 4.71 & 47.0 & 321.3 & 10.79 &  9.95 & 0.22 \\
.799$+$800$+$01$+$02&1344&$\alpha$\,Ari&h&20   &  12  & 180/4/10 &70$+$110& 221.8 & 0.064 & 2.38 & 93.3 & 322.0 & 11.05 &  9.89 & 0.33 \\
            \noalign{\smallskip}
\hline                                   
\end{tabular}
\end{sidewaystable*}

\addtocounter{table}{-1}

\begin{sidewaystable*}[h!]
\caption{Scan map photometry measurements in the red (160\,$\mu$m filter) 
continued.
}             
\label{table:scanmapphotred10_2}      
\centering                          
\begin{tabular}{r r r r r r r c r r r r r r r r r}        
\hline\hline                 
            \noalign{\smallskip}
 OBSID     &   OD   &   Target     & G & Speed  & Rep. & Map params  & ScanAngles &f$_{\rm aper}$&$\sigma_{\rm pix}$&$\sigma_{\rm aper,corr}$& S/N &f$_{\rm star}$&  W$_{max}$ & W$_{min}$ & $\Delta$\,W\\
           &        &              &    &(\arcsec/s)&  &(\arcsec/\arcsec/no.)&(deg)& (mJy)     &      (mJy)     &       (mJy)        &     &   (mJy)    &  (\arcsec) & (\arcsec) &   (\arcsec)    \\
            \noalign{\smallskip}
\hline
            \noalign{\smallskip}
1342190969 &  284 &$\varepsilon$\,Lep& h & 20 &  1 & 150/4/10 &   70   & 160.3 & 0.265 &  9.85 & 16.3 & 232.8 & 10.60 &  9.84 & 0.27 \\
1342190970 &  284 &$\varepsilon$\,Lep& h & 20 &  1 & 150/4/10 &  110   & 164.2 & 0.236 &  8.77 & 18.7 & 238.5 & 11.64 & 10.58 & 0.29 \\
1342190969$+$70& 284&$\varepsilon$\,Lep&h& 20 &  2 & 150/4/10 &70$+$110& 161.6 & 0.188 &  7.01 & 23.1 & 234.6 & 10.78 &  9.83 & 0.41 \\
1342205202 &  502 &$\varepsilon$\,Lep& h & 20 &  1 & 180/4/10 &   70   & 182.0 & 0.212 &  7.88 & 23.1 & 264.3 & 10.59 & 10.48 & 0.34 \\
1342205203 &  502 &$\varepsilon$\,Lep& h & 20 &  1 & 180/4/10 &  110   & 163.1 & 0.213 &  7.92 & 20.6 & 236.8 & 10.79 &  9.87 & 0.31 \\
1342205204 &  502 &$\varepsilon$\,Lep& h & 20 &  3 & 180/4/10 &   70   & 148.9 & 0.141 &  5.26 & 28.3 & 216.2 & 10.96 &  9.80 & 0.19 \\
1342205205 &  502 &$\varepsilon$\,Lep& h & 20 &  3 & 180/4/10 &  110   & 159.6 & 0.128 &  4.77 & 33.4 & 231.7 & 11.30 &  9.94 & 0.19 \\
..02$+$03$+$04$+$05& 502&$\varepsilon$\,Lep&h&20&8 & 180/4/10 &70$+$110& 158.5 & 0.093 &  3.44 & 46.0 & 230.2 & 11.01 &  9.94 & 0.36 \\
1342214208 &  640 &$\varepsilon$\,Lep& h & 20 &  3 & 180/4/10 &   70   & 155.5 & 0.135 &  5.02 & 31.0 & 225.9 & 11.37 &  9.88 & 0.19 \\
1342214209 &  640 &$\varepsilon$\,Lep& h & 20 &  3 & 180/4/10 &  110   & 144.5 & 0.120 &  4.45 & 32.5 & 209.8 & 10.90 &  9.72 & 0.18 \\
1342214208$+$09& 640&$\varepsilon$\,Lep&h& 20 &  6 & 180/4/10 &70$+$110& 150.0 & 0.088 &  3.27 & 45.9 & 217.9 & 11.25 &  9.82 & 0.24 \\
1342227297 &  833 &$\varepsilon$\,Lep& h & 20 &  1 & 180/4/10 &   70   & 144.3 & 0.209 &  7.79 & 18.5 & 209.5 & 11.17 &  9.65 & 0.20 \\
1342227298 &  833 &$\varepsilon$\,Lep& h & 20 &  1 & 180/4/10 &  110   & 165.0 & 0.224 &  8.34 & 19.7 & 239.5 & 10.90 & 10.49 & 0.24 \\
1342227299 &  833 &$\varepsilon$\,Lep& h & 20 &  3 & 180/4/10 &   70   & 150.7 & 0.127 &  4.74 & 31.7 & 218.8 & 11.04 &  9.95 & 0.19 \\
1342227300 &  833 &$\varepsilon$\,Lep& h & 20 &  3 & 180/4/10 &  110   & 155.8 & 0.126 &  4.69 & 33.2 & 226.3 & 10.92 & 10.15 & 0.18 \\
297$+$98$+$99$+$300& 833&$\varepsilon$\,Lep&h&20&8 & 180/4/10 &70$+$110& 152.8 & 0.084 &  3.14 & 48.7 & 221.9 & 10.90 & 10.08 & 0.27 \\
1342241333 & 1034 &$\varepsilon$\,Lep& h & 20 &  1 & 180/4/10 &   70   & 156.0 & 0.238 &  8.87 & 17.6 & 226.5 & 11.79 &  9.75 & 0.23 \\
1342241334 & 1034 &$\varepsilon$\,Lep& h & 20 &  1 & 180/4/10 &  110   & 159.0 & 0.245 &  9.10 & 17.5 & 230.9 & 11.24 & 10.86 & 0.29 \\
1342241335 & 1034 &$\varepsilon$\,Lep& h & 20 &  3 & 180/4/10 &   70   & 162.9 & 0.124 &  4.62 & 35.3 & 236.6 & 11.84 &  9.93 & 0.26 \\
1342241336 & 1034 &$\varepsilon$\,Lep& h & 20 &  3 & 180/4/10 &  110   & 156.0 & 0.134 &  5.00 & 31.2 & 226.6 & 11.58 &  9.85 & 0.24 \\
..33$+$34$+$35$+$36&1034&$\varepsilon$\,Lep&h&20&8 & 180/4/10 &70$+$110& 158.8 & 0.093 &  3.44 & 46.2 & 230.7 & 11.78 & 10.13 & 0.28 \\
1342263902 & 1377 &$\varepsilon$\,Lep& h & 20 &  3 & 180/4/10 &   70   & 143.9 &0.157$^1$& 5.84$^1$&24.6$^1$& 209.0 &10.37&  9.60 & 0.20 \\
1342263903 & 1377 &$\varepsilon$\,Lep& h & 20 &  3 & 180/4/10 &  110   & 143.7 &0.157$^1$& 5.85$^1$&24.6$^1$& 208.6 &10.51&  9.56 & 0.22 \\
1342263904 & 1377 &$\varepsilon$\,Lep& h & 20 &  1 & 180/4/10 &   70   & 200.6 &0.255$^1$& 9.48$^1$&21.2$^1$& 291.3 &10.67&  9.47 & 0.36 \\
1342263905 & 1377 &$\varepsilon$\,Lep& h & 20 &  1 & 180/4/10 &  110   & 158.6 &0.292$^1$&10.85$^1$&14.6$^1$& 230.3 &11.58& 10.67 & 0.29 \\
..02$+$03$+$04$+$05&1377&$\varepsilon$\,Lep&h&20&8 & 180/4/10 &70$+$110& 151.8 &0.100$^1$& 3.71$^1$&40.9$^1$& 220.4 &10.28&  9.33 & 0.44 \\
            \noalign{\smallskip}
\hline                                   
\end{tabular} \\
$^1$ After OD\,1375 half of the red photometer array was lost, resulting in increased noise and reduced S/N wrt.\ pre-OD\,1375 observations  
\end{sidewaystable*}

\addtocounter{table}{-1}

\begin{sidewaystable*}[h!]
\caption{Scan map photometry measurements in the red (160\,$\mu$m filter) 
continued.
}             
\label{table:scanmapphotred10_3}      
\centering                          
\begin{tabular}{r r r r r r r c r r r r r r r r r}        
\hline\hline                 
            \noalign{\smallskip}
 OBSID     &   OD   &   Target     & G & Speed  & Rep. & Map params  & ScanAngles &f$_{\rm aper}$&$\sigma_{\rm pix}$&$\sigma_{\rm aper,corr}$& S/N &f$_{\rm star}$&  W$_{max}$ & W$_{min}$ & $\Delta$\,W\\
           &        &              &    &(\arcsec/s)&  &(\arcsec/\arcsec/no.)&(deg)& (mJy)     &      (mJy)     &       (mJy)        &     &   (mJy)    &  (\arcsec) & (\arcsec) &   (\arcsec)    \\
            \noalign{\smallskip}
\hline
            \noalign{\smallskip}
1342218544 &  697 &$\omega$\,Cap     & h & 20 &  3 & 180/4/10 &   70   & 106.1 & 0.131 &  4.89 & 21.7 & 154.0 & 11.65 & 10.17 & 0.16 \\
1342218545 &  697 &$\omega$\,Cap     & h & 20 &  3 & 180/4/10 &  110   & 115.0 & 0.122 &  4.55 & 25.3 & 166.9 & 10.62 &  9.20 & 0.16 \\
1342218546 &  697 &$\omega$\,Cap     & h & 20 &  3 & 180/4/10 &   70   & 104.4 & 0.129 &  4.79 & 21.8 & 151.5 & 10.81 &  9.45 & 0.17 \\
1342218547 &  697 &$\omega$\,Cap     & h & 20 &  3 & 180/4/10 &  110   & 107.9 & 0.126 &  4.70 & 23.0 & 156.7 & 11.13 & 10.09 & 0.19 \\
..44$+$45$+$46$+$47& 697&$\omega$\,Cap&h & 20 & 12 & 180/4/10 &70$+$110& 107.2 & 0.063 &  2.34 & 45.8 & 155.7 & 11.08 & 10.02 & 0.25 \\
1342231259 &  889 &$\omega$\,Cap     & h & 20 &  3 & 180/4/10 &   70   & 109.9 & 0.121 &  4.50 & 24.4 & 159.6 & 10.98 &  9.85 & 0.22 \\
1342231260 &  889 &$\omega$\,Cap     & h & 20 &  3 & 180/4/10 &  110   & 108.6 & 0.130 &  4.83 & 22.5 & 157.6 & 11.05 &  9.59 & 0.21 \\
1342231261 &  889 &$\omega$\,Cap     & h & 20 &  3 & 180/4/10 &   70   & 110.9 & 0.130 &  4.85 & 22.9 & 161.1 & 11.11 &  9.45 & 0.21 \\
1342231262 &  889 &$\omega$\,Cap     & h & 20 &  3 & 180/4/10 &  110   & 106.8 & 0.120 &  4.48 & 23.8 & 155.1 & 10.80 &  9.95 & 0.21 \\
..59$+$60$+$61$+$62& 889&$\omega$\,Cap&h & 20 & 12 & 180/4/10 &70$+$110& 108.6 & 0.065 &  2.43 & 44.7 & 157.7 & 10.62 &  9.54 & 0.38 \\
1342243778 & 1058 &$\omega$\,Cap     & h & 20 &  3 & 180/4/10 &   70   & 107.9 & 0.129 &  4.81 & 22.4 & 156.7 & 10.88 &  9.60 & 0.18 \\
1342243779 & 1058 &$\omega$\,Cap     & h & 20 &  3 & 180/4/10 &  110   & 116.4 & 0.126 &  4.70 & 24.8 & 169.0 & 10.24 &  9.71 & 0.21 \\
1342243780 & 1058 &$\omega$\,Cap     & h & 20 &  3 & 180/4/10 &   70   & 118.6 & 0.127 &  4.71 & 25.2 & 172.3 & 11.66 &  9.50 & 0.24 \\
1342243781 & 1058 &$\omega$\,Cap     & h & 20 &  3 & 180/4/10 &  110   & 117.9 & 0.118 &  4.37 & 27.0 & 171.2 & 11.10 &  9.99 & 0.20 \\
..78$+$79$+$80$+$81&1058&$\omega$\,Cap&h & 20 & 12 & 180/4/10 &70$+$110& 115.0 & 0.060 &  2.22 & 51.8 & 167.0 & 10.94 &  9.67 & 0.33 \\
1342254207 & 1266 &$\omega$\,Cap     & h & 20 &  3 & 180/4/10 &   70   & 116.6 & 0.132 &  4.89 & 23.8 & 169.2 & 11.77 &  9.55 & 0.22 \\
1342254208 & 1266 &$\omega$\,Cap     & h & 20 &  3 & 180/4/10 &  110   & 108.6 & 0.125 &  4.67 & 23.3 & 157.7 & 10.35 & 10.20 & 0.21 \\
1342254209 & 1266 &$\omega$\,Cap     & h & 20 &  3 & 180/4/10 &   70   & 108.1 & 0.136 &  5.05 & 21.4 & 157.0 & 12.12 &  9.76 & 0.23 \\
1342254210 & 1266 &$\omega$\,Cap     & h & 20 &  3 & 180/4/10 &  110   & 103.7 & 0.127 &  4.74 & 21.9 & 150.5 & 10.84 &  9.74 & 0.20 \\
..07$+$08$+$09$+$10&1266&$\omega$\,Cap&h & 20 & 12 & 180/4/10 &70$+$110& 108.3 & 0.070 &  2.59 & 41.8 & 157.3 & 11.25 &  9.82 & 0.33 \\
1342186142 &  160 &$\eta$\,Dra       & h & 10 &  1 & 120/5/9  &   80   &  50.5 & 0.305 & 11.37 &  4.4 &  73.4 &  9.86 &  8.88 & 0.57 \\
1342186143 &  160 &$\eta$\,Dra       & h & 10 &  1 & 120/5/9  &  100   &  58.8 & 0.285 & 10.60 &  5.5 &  85.4 & 10.46 &  9.31 & 0.46 \\
1342186151 &  160 &$\eta$\,Dra       & h & 10 &  1 & 120/5/9  &   80   &  67.9 & 0.281 & 10.45 &  6.5 &  98.5 & 11.03 &  9.87 & 0.52 \\
1342186152 &  160 &$\eta$\,Dra       & h & 10 &  1 & 120/5/9  &  100   &  68.3 & 0.303 & 11.28 &  6.1 &  99.2 & 12.95 & 10.70 & 0.68 \\
..42$+$43$+$51$+$52& 160&$\eta$\,Dra & h & 10 &  4 & 120/5/9  &80$+$100&  60.5 & 0.162 &  6.03 & 10.0 &  87.9 & 10.90 &  9.73 & 0.44 \\
1342186144 &  160 &$\eta$\,Dra       & h & 10 &  1 &  90/5/9  &   60   &  58.4 & 0.313 & 11.66 &  5.0 &  84.8 & 11.04 &  9.84 & 0.87 \\
1342186145 &  160 &$\eta$\,Dra       & h & 10 &  1 &  90/5/9  &  120   &  69.7 & 0.328 & 12.22 &  5.7 & 101.2 & 12.57 & 10.77 & 0.61 \\
1342186153 &  160 &$\eta$\,Dra       & h & 10 &  1 &  90/5/9  &   60   &  55.8 & 0.339 & 12.63 &  4.4 &  81.0 & 10.42 &  6.75 & 1.00 \\
1342186154 &  160 &$\eta$\,Dra       & h & 10 &  1 &  90/5/9  &  120   &  58.4 & 0.343 & 12.77 &  4.6 &  84.9 & 12.06 & 10.81 & 0.97 \\
..44$+$45$+$53$+$54& 160&$\eta$\,Dra & h & 10 &  4 &  90/5/9  &60$+$120&  58.6 & 0.172 &  6.41 &  9.1 &  85.0 & 10.94 &  9.61 & 0.65 \\
1342186146 &  160 &$\eta$\,Dra       & h & 20 &  1 &  90/5/9  &   70   &  93.8 & 0.367 & 13.67 &  6.9 & 136.2 & 15.47 & 11.05 & 0.70 \\
1342186147 &  160 &$\eta$\,Dra       & h & 20 &  1 &  90/5/9  &  110   &  80.4 & 0.319 & 11.89 &  6.8 & 116.7 & 10.92 & 10.14 & 0.54 \\
1342186155 &  160 &$\eta$\,Dra       & h & 20 &  1 &  90/5/9  &   70   &  69.5 & 0.329 & 12.22 &  5.7 & 100.9 & 13.48 & 10.65 & 0.58 \\
1342186156 &  160 &$\eta$\,Dra       & h & 20 &  1 &  90/5/9  &  110   &  80.1 & 0.294 & 10.92 &  7.3 & 116.4 &  9.66 &  8.67 & 0.77 \\
..46$+$47$+$55$+$56& 160&$\eta$\,Dra & h & 20 &  4 &  90/5/9  &70$+$110&  80.1 & 0.185 &  6.90 & 11.6 & 116.3 & 11.56 &  9.90 & 0.59 \\
1342186148 &  160 &$\eta$\,Dra       & h & 10 &  1 & 150/4/9  &   85   &  52.0 & 0.268 &  9.98 &  5.2 &  75.5 &  9.93 &  8.31 & 0.71 \\
1342186149 &  160 &$\eta$\,Dra       & h & 10 &  1 & 150/4/9  &   95   &  60.6 & 0.284 & 10.56 &  5.7 &  88.0 & 16.55 & 11.55 & 0.61 \\
1342186157 &  160 &$\eta$\,Dra       & h & 10 &  1 & 150/4/9  &   85   &  90.5 & 0.248 &  9.22 &  9.8 & 131.4 & 10.86 &  9.43 & 0.57 \\
1342186158 &  160 &$\eta$\,Dra       & h & 10 &  1 & 150/4/9  &   95   &  68.6 & 0.257 &  9.54 &  7.2 &  99.6 & 10.62 & 10.16 & 0.63 \\
..48$+$49$+$57$+$58& 160&$\eta$\,Dra & h & 10 &  4 & 150/4/9  & 85$+$95&  67.5 & 0.144 &  5.36 & 12.6 &  98.0 & 11.91 &  9.90 & 0.52 \\
            \noalign{\smallskip}
\hline                                   
\end{tabular}
\end{sidewaystable*}

\addtocounter{table}{-1}

\begin{sidewaystable*}[h!]
\caption{Scan map photometry measurements in the red (160\,$\mu$m filter) 
continued.
}             
\label{table:scanmapphotred10_4}      
\centering                          
\begin{tabular}{r r r r r r r c r r r r r r r r r}        
\hline\hline                 
            \noalign{\smallskip}
 OBSID     &   OD   &   Target     & G & Speed  & Rep. & Map params  & ScanAngles &f$_{\rm aper}$&$\sigma_{\rm pix}$&$\sigma_{\rm aper,corr}$& S/N &f$_{\rm star}$&  W$_{max}$ & W$_{min}$ & $\Delta$\,W\\
           &        &              &    &(\arcsec/s)&  &(\arcsec/\arcsec/no.)&(deg)& (mJy)     &      (mJy)     &       (mJy)        &     &   (mJy)    &  (\arcsec) & (\arcsec) &   (\arcsec)    \\
            \noalign{\smallskip}
\hline
            \noalign{\smallskip}
1342189191 &  244 &$\delta$\,Dra     & h & 20  &   1  & 240/4/8  &  117   &  36.7 & 0.266 & 9.91 &  3.7 &  53.2 & 10.01 &  9.43 & 0.48 \\
1342189192 &  244 &$\delta$\,Dra     & h & 20  &   1  & 240/4/8  &   63   &  53.9 & 0.253 & 9.41 &  5.7 &  78.3 & 11.57 &  9.81 & 0.51 \\
1342189191$+$92& 244&$\delta$\,Dra   & h & 20  &   2  & 240/4/8  &63$+$117&  47.3 & 0.188 & 7.01 &  6.7 &  68.7 &  9.76 &  9.18 & 0.65 \\
1342212497 &  607 &$\delta$\,Dra     & h & 20  &   3  & 180/4/10 &   70   &  47.4 & 0.120 & 4.45 & 10.7 &  68.8 & 10.20 &  9.89 & 0.39 \\
1342212498 &  607 &$\delta$\,Dra     & h & 20  &   3  & 180/4/10 &  110   &  65.0 & 0.126 & 4.69 & 13.9 &  94.4 & 10.64 & 10.35 & 0.34 \\
1342212499 &  607 &$\delta$\,Dra     & h & 20  &   3  & 180/4/10 &   70   &  62.1 & 0.127 & 4.71 & 13.2 &  90.2 &  9.87 &  9.83 & 0.29 \\
1342212500 &  607 &$\delta$\,Dra     & h & 20  &   3  & 180/4/10 &  110   &  59.0 & 0.121 & 4.49 & 13.1 &  85.7 & 10.73 &  9.99 & 0.30 \\
..497$+$98$+$99$+$500& 607&$\delta$\,Dra&h&20  &  12  & 180/4/10 &70$+$110&  58.9 & 0.058 & 2.15 & 27.4 &  85.5 & 10.44 & 10.12 & 0.41 \\
1342222147 &  751 &$\delta$\,Dra     & h & 20  &   3  & 180/4/10 &   70   &  56.5 & 0.131 & 4.88 & 11.6 &  82.0 & 10.38 &  9.72 & 0.29 \\
1342222148 &  751 &$\delta$\,Dra     & h & 20  &   3  & 180/4/10 &  110   &  64.0 & 0.134 & 4.99 & 12.8 &  92.9 & 11.63 &  9.74 & 0.40 \\
1342222149 &  751 &$\delta$\,Dra     & h & 20  &   3  & 180/4/10 &   70   &  58.1 & 0.124 & 4.60 & 12.6 &  84.3 &  9.90 &  9.22 & 0.30 \\
1342222150 &  751 &$\delta$\,Dra     & h & 20  &   3  & 180/4/10 &  110   &  62.9 & 0.119 & 4.41 & 14.3 &  91.4 & 10.92 & 10.66 & 0.29 \\
..47$+$48$+$49$+$50& 751&$\delta$\,Dra&h & 20  &  12  & 180/4/10 &70$+$110&  60.5 & 0.070 & 2.59 & 23.4 &  87.9 & 10.58 & 10.08 & 0.41 \\
1342233571 &  934 &$\delta$\,Dra     & h & 20  &   3  & 180/4/10 &   70   &  58.8 & 0.120 & 4.48 & 13.1 &  85.4 & 11.75 &  9.67 & 0.27 \\
1342233572 &  934 &$\delta$\,Dra     & h & 20  &   3  & 180/4/10 &  110   &  53.3 & 0.128 & 4.77 & 11.2 &  77.4 & 10.84 &  9.75 & 0.27 \\
1342233573 &  934 &$\delta$\,Dra     & h & 20  &   3  & 180/4/10 &   70   &  75.4 & 0.119 & 4.43 & 17.0 & 109.5 & 10.70 & 10.33 & 0.39 \\
1342233574 &  934 &$\delta$\,Dra     & h & 20  &   3  & 180/4/10 &  110   &  59.2 & 0.132 & 4.90 & 12.1 &  85.9 & 10.04 &  9.91 & 0.31 \\
..71$+$72$+$73$+$74& 934&$\delta$\,Dra&h & 20  &  12  & 180/4/10 &70$+$110&  61.5 & 0.061 & 2.29 & 26.9 &  89.3 & 10.66 &  9.87 & 0.39 \\
1342250093 & 1198 &$\delta$\,Dra     & h & 20  &   3  & 180/4/10 &   70   &  59.1 & 0.123 & 4.57 & 12.9 &  85.8 &  9.50 &  9.39 & 0.27 \\
1342250094 & 1198 &$\delta$\,Dra     & h & 20  &   3  & 180/4/10 &  110   &  47.7 & 0.135 & 5.03 &  9.5 &  69.2 & 10.49 &  9.57 & 0.42 \\
1342250095 & 1198 &$\delta$\,Dra     & h & 20  &   3  & 180/4/10 &   70   &  59.6 & 0.123 & 4.58 & 13.0 &  86.5 &  9.83 &  9.37 & 0.33 \\
1342250096 & 1198 &$\delta$\,Dra     & h & 20  &   3  & 180/4/10 &  110   &  55.4 & 0.122 & 4.53 & 12.2 &  80.4 & 11.12 & 10.10 & 0.37 \\
..93$+$94$+$95$+$96&1198&$\delta$\,Dra&h & 20  &  12  & 180/4/10 &70$+$110&  55.1 & 0.064 & 2.39 & 23.1 &  80.0 & 10.45 &  9.79 & 0.34 \\
1342257987 & 1328 &$\delta$\,Dra     & h & 20  &   3  & 180/4/10 &   70   &  62.1 & 0.134 & 4.98 & 12.5 &  90.2 & 10.82 & 10.17 & 0.44 \\
1342257988 & 1328 &$\delta$\,Dra     & h & 20  &   3  & 180/4/10 &  110   &  53.5 & 0.129 & 4.79 & 11.2 &  77.7 & 10.25 &  9.94 & 0.29 \\
1342257989 & 1328 &$\delta$\,Dra     & h & 20  &   3  & 180/4/10 &   70   &  61.6 & 0.122 & 4.55 & 13.5 &  89.4 & 11.16 & 10.74 & 0.56 \\
1342257990 & 1328 &$\delta$\,Dra     & h & 20  &   3  & 180/4/10 &  110   &  59.6 & 0.114 & 4.24 & 14.1 &  86.5 & 10.06 &  9.64 & 0.32 \\
..87$+$88$+$89$+$90&1328&$\delta$\,Dra&h & 20  &  12  & 180/4/10 &70$+$110&  58.6 & 0.074 & 2.74 & 21.4 &  85.1 & 10.69 & 10.18 & 0.38 \\
1342184574 &  139 &$\theta$\,Umi     & h & 10  &   1  & 210/15/4 &   90   &  37.3 & 0.412 &15.32 &  2.4 &  54.1 & 10.55 &  8.47 & 0.86 \\
1342184585 &  139 &$\theta$\,Umi     & h & 10  &   1  & 210/15/4 &   90   &  79.3 & 0.351 &13.07 &  6.1 & 115.1 & 11.45 &  8.32 & 1.05 \\
1342184574$+$85& 139&$\theta$\,Umi   & h & 10  &   2  & 210/15/4 &   90   &  56.1 & 0.311 &11.59 &  4.8 &  81.4 &  9.89 &  7.88 & 0.80 \\
1342184575 &  139 &$\theta$\,UMi     & h & 20  &   1  & 210/15/4 &   90   &  50.5 & 0.454 &16.89 &  3.0 &  73.3 &  9.69 &  6.51 & 0.51 \\
1342184586 &  139 &$\theta$\,Umi     & h & 20  &   1  & 210/15/4 &   90   &  47.2 & 0.412 &15.33 &  3.1 &  68.5 & 12.96 & 10.53 & 1.12 \\
1342184575$+$86& 139&$\theta$\,Umi   & h & 20  &   2  & 210/15/4 &   90   &  48.7 & 0.353 &13.13 &  3.7 &  70.8 & 10.42 &  8.37 & 0.49 \\
1342191982 &  300 &$\theta$\,Umi     & h & 20  &   6  & 150/4/10 &   70   &  41.4 & 0.091 & 3.38 & 12.2 &  60.1 & 10.95 &  9.94 & 0.38 \\
1342191983 &  300 &$\theta$\,Umi     & h & 20  &   6  & 150/4/10 &  110   &  44.6 & 0.096 & 3.58 & 12.5 &  64.8 & 10.65 &  9.61 & 0.36 \\
1342191982$+$83& 300&$\theta$\,Umi   & h & 20  &  12  & 150/4/10 &70$+$110&  42.6 & 0.067 & 2.50 & 17.0 &  61.9 & 10.72 &  9.77 & 0.40 \\
            \noalign{\smallskip}
\hline                                   
\end{tabular}
\end{sidewaystable*}

\addtocounter{table}{-1}

\begin{sidewaystable*}[h!]
\caption{Scan map photometry measurements in the red (160\,$\mu$m filter) 
continued.
}             
\label{table:scanmapphotred10_5}      
\centering                          
\begin{tabular}{r r r r r r r c r r r r r r r r r}        
\hline\hline                 
            \noalign{\smallskip}
 OBSID     &   OD   &   Target     & G & Speed  & Rep. & Map params  & ScanAngles &f$_{\rm aper}$&$\sigma_{\rm pix}$&$\sigma_{\rm aper,corr}$& S/N &f$_{\rm star}$&  W$_{max}$ & W$_{min}$ & $\Delta$\,W\\
           &        &              &    &(\arcsec/s)&  &(\arcsec/\arcsec/no.)&(deg)& (mJy)     &      (mJy)     &       (mJy)        &     &   (mJy)    &  (\arcsec) & (\arcsec) &   (\arcsec)    \\
            \noalign{\smallskip}
\hline
            \noalign{\smallskip}
1342190972 &  284 &HD\,41047         & h & 20  &   3  & 150/4/10 &   70   & 28.4 & 0.137 & 5.09 &  5.6 & 41.3 & 12.15 &  9.94 & 0.80 \\
1342190973 &  284 &HD\,41047         & h & 20  &   3  & 150/4/10 &  110   & 21.3 & 0.125 & 4.64 &  4.6 & 30.9 & 10.40 &  8.22 & 0.45 \\
1342190972$+$73& 284&HD\,41047       & h & 20  &   6  & 150/4/10 &70$+$110& 24.1 & 0.086 & 3.22 &  7.5 & 35.0 & 10.63 &  8.95 & 0.64 \\
1342203111 &  460 &42\,Dra           & h & 20  &  20  & 180/4/10 &   70   & 18.1 & 0.061 & 2.25 &  8.0 & 26.4 & 11.89 & 10.30 & 0.41 \\
1342203112 &  460 &42\,Dra           & h & 20  &  20  & 180/4/10 &  110   & 20.6 & 0.059 & 2.19 &  9.4 & 29.9 & 10.47 &  9.64 & 0.42 \\
1342203113 &  460 &42\,Dra           & h & 20  &  20  & 180/4/10 &   70   & 18.5 & 0.057 & 2.12 &  8.7 & 26.9 & 11.08 &  9.42 & 0.32 \\
1342203114 &  460 &42\,Dra           & h & 20  &  20  & 180/4/10 &  110   & 20.5 & 0.054 & 2.02 & 10.1 & 29.8 &  9.99 &  9.82 & 0.31 \\
..11$+$12$+$13$+$14& 460&42\,Dra     & h & 20  &  80  & 180/4/10 &70$+$110& 20.2 & 0.031 & 1.16 & 17.4 & 29.4 & 10.38 &  9.64 & 0.44 \\
1342243730 & 1057 &42\,Dra           & h & 20  &  20  & 180/4/10 &   70   & 22.6 & 0.056 & 2.10 & 10.8 & 32.8 & 11.08 &  9.93 & 0.34 \\
1342243731 & 1057 &42\,Dra           & h & 20  &  20  & 180/4/10 &  110   & 17.8 & 0.054 & 2.02 &  8.8 & 25.8 & 10.86 & 10.06 & 0.32 \\
1342243732 & 1057 &42\,Dra           & h & 20  &  20  & 180/4/10 &   70   & 19.7 & 0.057 & 2.13 &  9.2 & 28.6 & 11.34 & 10.23 & 0.24 \\
1342243733 & 1057 &42\,Dra           & h & 20  &  20  & 180/4/10 &  110   & 17.6 & 0.052 & 1.95 &  9.0 & 25.6 & 10.59 &  9.56 & 0.36 \\
..30$+$31$+$32$+$33&1057&42\,Dra     & h & 20  &  80  & 180/4/10 &70$+$110& 19.3 & 0.032 & 1.18 & 16.4 & 28.1 & 10.87 &  9.94 & 0.30 \\
1342184302 &  132 &HD\,138265        & h & 10  &   3  & 210/4/25 &   90   & 24.7 & 0.085 & 3.16 &  7.8 & 35.9 &  9.92 &  8.68 & 0.54 \\
1342184303 &  132 &HD\,138265        & h & 20  &   3  & 210/4/25 &   90   & 18.2 & 0.091 & 3.38 &  5.4 & 26.4 & 11.34 & 10.88 & 0.54 \\
1342188841 &  233 &HD\,138265        & h & 20  &  21  & 240/4/8  &   63   & 21.1 & 0.068 & 2.55 &  8.3 & 30.6 & 11.44 & 10.85 & 0.45 \\
1342188842 &  233 &HD\,138265        & h & 20  &  21  & 240/4/8  &  117   & 20.4 & 0.059 & 2.21 &  9.2 & 29.6 & 11.44 & 11.40 & 0.57 \\
1342188841$+$42& 233&HD\,138265      & h & 20  &  42  & 240/4/8  &63$+$117& 21.3 & 0.044 & 1.65 & 12.9 & 31.0 & 11.39 & 11.04 & 0.53 \\
1342191986 &  300 &HD\,138265        & h & 20  &   6  & 150/4/10 &   70   &  9.8 & 0.102 & 3.81 &  2.6 & 14.2 &  9.99 &  8.64 & 0.93 \\
1342191987 &  300 &HD\,138265        & h & 20  &   6  & 150/4/10 &  110   & 22.5 & 0.095 & 3.52 &  6.4 & 32.7 & 11.59 &  8.76 & 0.54 \\
1342191986$+$87& 300&HD\,138265      & h & 20  &  12  & 150/4/10 &70$+$110& 16.6 & 0.067 & 2.51 &  6.6 & 24.2 & 11.16 &  9.79 & 0.87 \\
1342227973 &  843 &HD\,152222        & h & 20  &  33  & 180/4/10 &   70   &  5.6 & 0.035 & 1.31 & 4.3  &  8.1 & 12.98 & 10.77 & 0.69 \\
1342227974 &  843 &HD\,152222        & h & 20  &  33  & 180/4/10 &  110   &  4.0 & 0.039 & 1.44 & 2.8  &  5.8 & 12.42 &  9.86 & 0.78 \\
1342227973$+$74& 843&HD\,152222      & h & 20  &  66  & 180/4/10 &70$+$110&  4.5 & 0.030 & 1.11 & 4.1  &  6.5 & 12.89 & 10.50 & 0.54 \\
1342240702 & 1028 &HD\,152222        & h & 20  &   6  & 180/4/10 &   70   &  6.5 & 0.079 & 2.93 & 2.2  &  9.5 & 10.20 &  8.42 & 1.69 \\
1342240703 & 1028 &HD\,152222        & h & 20  &   6  & 180/4/10 &  110   &  5.8 & 0.086 & 3.20 & 1.8  &  8.5 &  9.77 &  9.45 & 1.13 \\
1342240702$+$03&1028&HD\,152222      & h & 20  &  12  & 180/4/10 &70$+$110&  6.1 & 0.055 & 2.05 & 3.0  &  8.9 &  9.48 &  8.89 & 1.07 \\
1342198535 &  400 &HD\,39608         & h & 20  &  10  & 180/4/10 &   70   &  7.2 & 0.067 & 2.50 & 2.9  & 10.4 & 14.24 &  8.77 & 0.89 \\
1342198536 &  400 &HD\,39608         & h & 20  &  10  & 180/4/10 &  110   & 13.6 & 0.069 & 2.58 & 5.3  & 19.7 & 13.19 & 11.59 & 1.71 \\
1342198537 &  400 &HD\,39608         & h & 20  &  35  & 180/4/10 &   70   &  7.9 & 0.042 & 1.57 & 5.0  & 11.5 & 11.58 &  8.24 & 0.93 \\
1342198538 &  400 &HD\,39608         & h & 20  &  35  & 180/4/10 &  110   &  8.8 & 0.043 & 1.59 & 5.5  & 12.7 & 11.55 &  7.63 & 0.80 \\
..35$+$36$+$37$+$38& 400&HD\,39608   & h & 20  &  90  & 180/4/10 &70$+$110&  8.6 & 0.031 & 1.15 & 7.5  & 12.5 & 11.05 &  9.23 & 1.24 \\
            \noalign{\smallskip}
\hline                                   
\end{tabular}
\end{sidewaystable*}

\clearpage

\section{Chopped photometry}
\label{sect:appb}

\subsection{Photometry results of individual measurements}
\label{sect:individualchopphot}

Individual photometric results for the 70, 100, and 160\,$\mu$m filters are 
compiled in Tables~\ref{table:chopnodphotblue} to~\ref{table:chopnodphotred1}.
The applied radius for the photometric aperture was 5.6, 6.8 and 10.7\arcsec 
for the 70, 100 and 160\,$\mu$m filter, respectively. The number of output 
pixels (1\farcs1, 1\farcs4, and 2\farcs1 size, respectively) inside this 
photometric aperture is N$_{\rm aper}$= 81.42, 74.12, and 81.56, respectively. 
The corresponding correction factor for correlated noise are f$_{\rm corr}$ = 
6.33, 4.22, and 7.81, respectively. Aperture correction factors are 
c$_{\rm aper}$ = 1.61, 1.56 and 1.56 for the 70, 100 and 160\,$\mu$m filter, 
respectively. Proper motion correction was applied throughout.

The tables contain the following information: Col.~1: Unique observational 
identifier (OBSID) of the PACS observation; Col.~2: Herschel Observational 
Day (OD), including its phase; Col.~3: Target name; Col.~4: Applied gain (G) 
of the PACS bolometer electronics: h(igh)/l(ow);  Col.~5: Chopper dither 
pattern: y(es)/n(o); Col.~6: Number of repetitions (rep.) of the basic 
chop/nod cycle; Col.~7: Fitted peak flux intensity of the source; Col.~8: 
Measured flux inside the photometric aperture of this filter, f$_{\rm aper}$; 
Col.~9: Noise per pixel, $\sigma_{\rm pix}$; Col.~10: Noise corrected for 
correlated noise inside the measurement aperture, f$_{\rm aper}$, according to 
Eq.~\ref{eq:sigapercorr}. Col.~11: Achieved signal-to-noise ratio according to 
Eq.~\ref{eq:sn_meas}; Col.~12: Stellar flux f$_{\rm star}$ according to 
Eq.~\ref{eq:apercolcorr}; Cols.~13$+$14: Maximum and minimum Full Width (W) 
Half Maximum (in \arcsec) of the source PSF.

\begin{sidewaystable*}[h!]
\caption{Chop-nod photometry measurements in the blue (70\,$\mu$m filter).
Processing proceeded from SPG~v11.1.0 level 1 products with HIPE version 13 
build 2768. Gyro correction was applied for most of the cases.
}             
\label{table:chopnodphotblue}      
\centering                          
\begin{tabular}{c c c c c c c c c c c c c c}        
\hline\hline                 
            \noalign{\smallskip}
 OBSID     &       OD   &   Target     & G & Dith & Rep. & FitPeak &f$_{\rm aper}$&$\sigma_{\rm pix}$&$\sigma_{\rm aper,corr}$& S/N &f$_{\rm star}$& W$_{max}$ & W$_{min}$\\
           &            &              &   &      &      &  (mJy)  &  (mJy)  &  (mJy) &  (mJy) &     &   (mJy)  &(\arcsec)&(\arcsec)\\
            \noalign{\smallskip}
\hline
            \noalign{\smallskip}
1342217347 & 684.191528 & $\beta$\,Gem & h & y &  3 & 53.535 & 1608.2 & 0.193 &  11.0 & 146 & 2569.7 & 6.48 & 5.78 \\
1342184267 & 132.284763 &$\varepsilon$\,Lep&l&y&  1 & 27.686 &  734.0 & 0.335 &  19.1 &  38 & 1172.8 & 5.75 & 5.10 \\
1342184268 & 132.287321 &$\varepsilon$\,Lep&h&y&  1 & 28.453 &  742.0 & 0.242 &  13.8 &  54 & 1185.6 & 5.81 & 5.00 \\
1342186141 & 160.373009 & $\eta$\,Dra  & h & y &  2 & 10.717 &  318.8 & 0.161 &   9.2 &  35 &  509.4 & 5.86 & 5.72 \\
1342182975 & 108.553912 &$\delta$\,Dra & h & y &  2 &  9.897 &  266.8 & 0.156 &   8.9 &  30 &  426.3 & 5.74 & 5.17 \\
1342182976 & 108.558322 &$\delta$\,Dra & h & n &  2 &  9.963 &  277.2 & 0.159 &   9.1 &  31 &  443.0 & 5.89 & 5.06 \\
1342184293 & 132.472674 &$\delta$\,Dra & l & y &  2 & 10.015 &  277.2 & 0.193 &  11.0 &  25 &  443.0 & 6.12 & 5.17 \\
1342184496 & 138.176215 &$\delta$\,Dra & h & y &  2 & 10.630 &  276.2 & 0.167 &   9.5 &  29 &  441.4 & 5.72 & 4.94 \\
1342189190 & 244.175625 &$\delta$\,Dra & h & y &  4 &  9.234 &  275.8 & 0.110 &   6.3 &  44 &  440.6 & 5.86 & 5.74 \\
1342184576 & 139.165067 &$\theta$\,Umi & h & y &  1 &  6.432 &  181.9 & 0.213 &  12.2 &  15 &  290.6 & 5.74 & 5.57 \\
1342184577 & 139.167625 &$\theta$\,Umi & h & y &  1 &  6.331 &  179.8 & 0.217 &  12.4 &  15 &  287.4 & 5.88 & 5.32 \\
1342184578 & 139.170194 &$\theta$\,Umi & h & y &  1 &  6.572 &  180.0 & 0.227 &  13.0 &  14 &  287.5 & 5.72 & 5.15 \\
1342184579 & 139.173559 &$\theta$\,Umi & h & y &  1 &  5.950 &  169.0 & 0.230 &  13.1 &  13 &  270.1 & 5.68 & 5.36 \\
1342184580 & 139.175344 &$\theta$\,Umi & h & y &  1 &  6.210 &  185.0 & 0.217 &  12.4 &  15 &  295.6 & 6.11 & 5.46 \\
1342184581 & 139.177889 &$\theta$\,Umi & h & y &  1 &  6.496 &  183.3 & 0.230 &  13.1 &  14 &  292.9 & 5.99 & 5.32 \\
1342184582 & 139.180494 &$\theta$\,Umi & h & y &  1 &  6.295 &  174.9 & 0.231 &  13.2 &  13 &  279.5 & 5.83 & 5.16 \\
1342184583 & 139.183086 &$\theta$\,Umi & h & y &  1 &  6.193 &  181.7 & 0.239 &  13.7 &  13 &  290.3 & 6.07 & 5.50 \\
1342184584 & 139.185678 &$\theta$\,Umi & h & y &  1 &  6.477 &  180.8 & 0.250 &  14.3 &  13 &  288.8 & 5.74 & 5.54 \\
1342191981 & 300.807361 &$\theta$\,Umi & h & y & 20 &  5.745 &  175.0 & 0.052 &   3.0 &  59 &  279.6 & 6.18 & 5.65 \\
1342184296 & 132.725098 & HD\,138265   & h & y & 43 &  2.530 &   70.8 & 0.033 &   1.9 &  38 &  113.2 & 5.69 & 5.33 \\
1342184297 & 132.775573 & HD\,138265   & h & y & 10 &  2.457 &   71.0 & 0.067 &   3.8 &  19 &  113.5 & 5.66 & 5.61 \\
1342184298 & 132.797749 & HD\,138265   & h & y & 10 &  2.588 &   69.7 & 0.063 &   3.6 &  19 &  111.4 & 5.63 & 5.21 \\
1342185441 & 146.197794 & HD\,138265   & h & y &  1 &  2.424 &   72.0 & 0.122 &   7.0 &  10 &  115.0 & 5.77 & 5.40 \\
1342185442 & 146.211433 & HD\,138265   & h & y &  2 &  2.360 &   66.6 & 0.087 &   5.0 &  13 &  106.4 & 5.95 & 5.17 \\
1342185443 & 146.224583 & HD\,138265   & h & y &  4 &  2.504 &   72.4 & 0.077 &   4.4 &  16 &  115.6 & 5.95 & 5.40 \\
1342185444 & 146.238106 & HD\,138265   & h & y &  7 &  2.230 &   66.4 & 0.083 &   4.7 &  14 &  106.1 & 6.09 & 5.45 \\
1342185445 & 146.251663 & HD\,138265   & h & y & 12 &  2.272 &   70.9 & 0.085 &   4.9 &  15 &  113.2 & 6.11 & 5.73 \\
1342191984 & 300.874850 & HD\,138265   & h & y & 10 &  2.354 &   72.0 & 0.067 &   3.8 &  19 &  115.0 & 5.96 & 5.71 \\
1342186162 & 160.492905 & HD\,159330   & h & y & 10 &  1.486 &   37.7 & 0.067 &   3.8 & 9.9 &   60.3 & 5.45 & 5.04 \\
1342191963 & 300.492037 & HD\,152222   & h & y & 20 &  0.890 &   24.2 & 0.045 &   2.6 & 9.4 &   38.7 & 5.62 & 5.37 \\
1342182953 & 108.195255 & HD\,15008    & h & n & 10 &  0.537 &   14.3 & 0.060 &   3.4 & 4.2 &   22.8 & 6.30 & 4.81 \\
1342182956 & 108.254317 & HD\,15008    & h & y & 10 &  0.482 &   13.7 & 0.074 &   4.2 & 3.2 &   21.9 & 5.72 & 5.33 \\
1342183928 & 125.490660 & HD\,15008    & h & y & 16 &  0.483 &   14.0 & 0.057 &   3.3 & 4.3 &   22.4 & 6.11 & 5.44 \\
1342192766 & 316.230382 & HD\,15008    & h & y & 45 &  0.411 &   11.7 & 0.035 &   2.0 & 5.9 &   18.7 & 6.04 & 5.11 \\
            \noalign{\smallskip}
\hline                                   
\end{tabular}
\end{sidewaystable*}

\begin{sidewaystable*}
\caption{Chop-nod photometry measurements in the green (100\,$\mu$m filter).
Processing proceeded from SPG~v11.1.0 level 1 products with HIPE version 13 
build 2768. Gyro correction was applied for most of the cases.}             
\label{table:chopnodphotgreen}      
\centering                          
\begin{tabular}{c c c c c c c c c c c c c c}        
\hline\hline                 
            \noalign{\smallskip}
 OBSID     &       OD   &   Target     & G & Dith & Rep. & FitPeak &f$_{\rm aper}$&$\sigma_{\rm pix}$&$\sigma_{\rm aper,corr}$& S/N &f$_{\rm star}$& W$_{max}$ & W$_{min}$\\
           &            &              &   &      &      &  (mJy)  &  (mJy)  &  (mJy) &  (mJy) &     &   (mJy)  &(\arcsec)&(\arcsec)\\
            \noalign{\smallskip}
\hline
            \noalign{\smallskip}
1342217350 & 684.218345 & $\beta$\,Gem & h & y &  3 & 30.688 & 829.6 & 0.240 &  8.7 &  95 & 1266.7 & 7.47 & 7.01 \\
1342184266 & 132.282205&$\varepsilon$\,Lep&h&y &  1 & 15.193 & 379.8 & 0.352 & 12.8 &  30 &  579.9 & 7.08 & 6.39 \\
1342190968 & 284.817118&$\varepsilon$\,Lep&h&y &  2 & 13.141 & 360.1 & 0.250 &  9.1 &  40 &  549.8 & 7.82 & 6.63 \\
1342186150 & 160.410127 & $\eta$\,Dra  & h & y &  2 &  6.269 & 154.4 & 0.241 &  8.8 &  18 &  235.7 & 6.78 & 6.70 \\
1342182977 & 108.562731 &$\delta$\,Dra & h & y &  2 &  5.643 & 137.5 & 0.258 &  9.4 &  15 &  210.0 & 7.01 & 6.08 \\
1342182978 & 108.567141 &$\delta$\,Dra & h & n &  2 &  5.247 & 136.1 & 0.233 &  8.5 &  16 &  207.8 & 6.91 & 6.68 \\
1342184294 & 132.477141 &$\delta$\,Dra & l & y &  2 &  5.513 & 147.5 & 0.307 & 11.2 &  13 &  225.2 & 7.49 & 6.65 \\
1342184587 & 139.193292 &$\theta$\,Umi & h & y &  1 &  3.456 &  90.8 & 0.311 & 11.3 & 8.0 &  138.7 & 7.18 & 6.41 \\
1342184588 & 139.195850 &$\theta$\,Umi & h & y &  1 &  3.242 & 100.2 & 0.344 & 12.5 & 8.0 &  153.0 & 7.68 & 7.39 \\
1342184589 & 139.198419 &$\theta$\,Umi & h & y &  1 &  3.801 &  91.8 & 0.351 & 12.8 & 7.2 &  140.2 & 6.84 & 6.00 \\
1342184590 & 139.200953 &$\theta$\,Umi & h & y &  1 &  3.744 & 101.6 & 0.358 & 13.0 & 7.8 &  155.2 & 7.88 & 6.11 \\
1342184591 & 139.203546 &$\theta$\,Umi & h & y &  1 &  3.605 & 100.2 & 0.365 & 13.3 & 7.6 &  153.0 & 7.28 & 7.00 \\
1342184592 & 139.206138 &$\theta$\,Umi & h & y &  1 &  3.395 &  87.8 & 0.366 & 13.3 & 6.6 &  134.0 & 6.92 & 6.59 \\
1342184593 & 139.209497 &$\theta$\,Umi & h & y &  1 &  3.479 &  94.6 & 0.324 & 11.8 & 8.0 &  144.4 & 7.55 & 6.58 \\
1342184594 & 139.211300 &$\theta$\,Umi & h & y &  1 &  4.149 &  84.3 & 0.346 & 12.6 & 6.7 &  128.8 & 6.04 & 5.85 \\
1342184595 & 139.213881 &$\theta$\,Umi & h & y &  1 &  3.024 &  84.7 & 0.346 & 12.6 & 6.7 &  129.3 & 7.67 & 6.89 \\
1342190971 & 284.837708 & HD\,41047    & h & y & 10 &  2.244 &  57.5 & 0.107 &  3.9 &  15 &   87.8 & 7.31 & 6.37 \\
1342184299 & 132.819914 & HD\,138265   & h & y & 10 &  1.380 &  38.1 & 0.102 &  3.7 &  10 &   58.2 & 7.73 & 6.58 \\
1342184300 & 132.842102 & HD\,138265   & h & y & 10 &  1.456 &  39.5 & 0.082 &  3.0 &  13 &   60.4 & 7.16 & 6.68 \\
1342184301 & 132.874606 & HD\,138265   & h & y & 20 &  1.255 &  39.0 & 0.073 &  2.7 &  15 &   59.5 & 7.93 & 7.23 \\
1342191985 & 300.901319 & HD\,138265   & h & y & 20 &  1.339 &  38.4 & 0.075 &  2.7 &  14 &   58.6 & 7.69 & 7.07 \\
1342184503 & 138.314253 & HD\,159330   & h & y & 21 &  0.817 &  21.0 & 0.071 &  2.6 & 8.1 &   32.1 & 6.80 & 6.72 \\
1342186159 & 160.456273 & HD\,159330   & h & y & 10 &  0.804 &  22.5 & 0.090 &  3.3 & 6.9 &   34.3 & 7.99 & 6.54 \\
1342192782 & 316.508530 & HD\,159330   & h & y & 45 &  0.685 &  18.8 & 0.055 &  2.0 & 9.4 &   28.8 & 7.52 & 6.77 \\
            \noalign{\smallskip}
\hline                                   
\end{tabular}
\end{sidewaystable*}

\begin{sidewaystable*}[h!]
\caption{Chop-nod photometry measurements in the red (160\,$\mu$m filter).
Processing proceeded from SPG~v11.1.0 level 1 products with HIPE version 13 
build 2768. Gyro correction was applied for most of the cases.
}             
\label{table:chopnodphotred1}      
\centering                          
\begin{tabular}{c c c c c c c c c c c c c c}        
\hline\hline                 
            \noalign{\smallskip}
 OBSID     &       OD   &   Target     & G & Dith & Rep. & FitPeak &f$_{\rm aper}$&$\sigma_{\rm pix}$&$\sigma_{\rm aper,corr}$& S/N &f$_{\rm star}$& W$_{max}$ & W$_{min}$\\
           &            &              &   &      &      &  (mJy)  &  (mJy)  &  (mJy) &  (mJy) &     &   (mJy)  &(\arcsec)&(\arcsec)\\
            \noalign{\smallskip}
\hline
            \noalign{\smallskip}
1342217347 & 684.191528 & $\beta$\,Gem  & h & y &  3 & 10.781 & 343.7 & 0.337 & 23.8 &  14 & 499.1 & 13.10 & 10.38 \\
1342217350 & 684.218345 & $\beta$\,Gem  & h & y &  3 & 10.574 & 340.7 & 0.325 & 22.9 &  15 & 494.7 & 13.22 & 10.64 \\
1342184266 & 132.282205&$\varepsilon$\,Lep& h& y&  1 &  5.147 & 142.5 & 0.504 & 35.5 & 4.0 & 206.9 & 11.07 &  9.83 \\
1342184267 & 132.284763&$\varepsilon$\,Lep& l& y&  1 &  5.435 & 158.3 & 0.557 & 39.3 & 4.0 & 229.9 & 10.66 & 10.64 \\
1342184268 & 132.287321&$\varepsilon$\,Lep& h& y&  1 &  5.109 & 136.5 & 0.542 & 38.2 & 3.6 & 198.2 & 12.01 &  8.99 \\
1342190968 & 284.817118&$\varepsilon$\,Lep& h& y&  2 &  4.458 & 136.1 & 0.365 & 25.7 & 5.3 & 197.7 & 13.16 &  9.62 \\
1342186141 & 160.373009 & $\eta$\,Dra   & h & y &  2 &  2.442 &  64.2 & 0.304 & 21.4 & 3.0 &  93.2 & 12.53 &  8.87 \\
1342186150 & 160.410127 & $\eta$\,Dra   & h & y &  2 &  2.297 &  75.0 & 0.308 & 21.7 & 3.5 & 109.0 & 13.05 & 10.62 \\
1342182975 & 108.553912 &$\delta$\,Dra  & h & y &  2 &  1.652 &  46.6 & 0.315 & 22.2 & 2.1 &  67.7 & 14.38 &  8.78 \\
1342182976 & 108.558322 &$\delta$\,Dra  & h & n &  2 &  2.125 &  58.4 & 0.395 & 27.9 & 2.1 &  84.8 & 12.78 &  8.86 \\
1342182977 & 108.562731 &$\delta$\,Dra  & h & y &  2 &  1.935 &  52.3 & 0.341 & 24.1 & 2.2 &  75.9 & 11.93 &  9.24 \\
1342182978 & 108.567141 &$\delta$\,Dra  & h & n &  2 &  2.034 &  69.5 & 0.363 & 25.6 & 2.7 & 100.9 & 13.40 & 10.73 \\
1342184293 & 132.472674 &$\delta$\,Dra  & l & y &  2 &  1.791 &  57.5 & 0.344 & 24.3 & 2.4 &  83.4 & 13.26 &  8.97 \\
1342184294 & 132.477141 &$\delta$\,Dra  & l & y &  2 &  1.848 &  60.8 & 0.365 & 25.7 & 2.4 &  88.2 & 13.65 & 10.34 \\
1342184496 & 138.176215 &$\delta$\,Dra  & h & y &  2 &  1.950 &  59.8 & 0.367 & 25.9 & 2.3 &  86.8 & 13.38 &  9.68 \\
1342189190 & 244.175625 &$\delta$\,Dra  & h & y &  4 &  1.882 &  57.5 & 0.216 & 15.2 & 3.8 &  83.6 & 12.21 & 10.06 \\
1342184576 & 139.165067 &$\theta$\,Umi  & h & y &  1 &  1.727 &  55.3 & 0.387 & 27.3 & 2.0 &  80.4 & 12.84 &  8.89 \\
1342184577 & 139.167625 &$\theta$\,Umi  & h & y &  1 &  1.602 &  44.5 & 0.448 & 31.6 & 1.4 &  64.7 & 10.84 &  9.92 \\
1342184578 & 139.170194 &$\theta$\,Umi  & h & y &  1 &  1.711 &  33.8 & 0.460 & 32.4 & 1.0 &  49.1 & 11.09 &  6.68 \\
1342184579 & 139.173559 &$\theta$\,Umi  & h & y &  1 &  1.289 &  24.4 & 0.456 & 32.2 & 0.8 &  35.5 & 16.28 &  6.06 \\
1342184580 & 139.175344 &$\theta$\,Umi  & h & y &  1 &  2.036 &  39.0 & 0.490 & 34.6 & 1.1 &  56.6 & 10.66 &  6.32 \\
1342184581 & 139.177889 &$\theta$\,Umi  & h & y &  1 &  1.125 &  37.4 & 0.518 & 36.5 & 1.0 &  54.3 & 11.85 & 10.21 \\
1342184582 & 139.180494 &$\theta$\,Umi  & h & y &  1 &  1.284 &  46.6 & 0.521 & 36.7 & 1.3 &  67.6 & 19.99 &  8.28 \\
1342184587 & 139.193292 &$\theta$\,Umi  & h & y &  1 &  2.785 &  43.8 & 0.332 & 23.4 & 1.9 &  63.7 &  8.80 &  7.25 \\
1342184588 & 139.195850 &$\theta$\,Umi  & h & y &  1 &  1.454 &  34.2 & 0.465 & 32.8 & 1.0 &  49.6 &  9.90 &  9.37 \\
1342184589 & 139.198419 &$\theta$\,Umi  & h & y &  1 &  1.704 &  41.1 & 0.505 & 35.6 & 1.2 &  59.6 & 12.59 &  7.97 \\
1342184590 & 139.200953 &$\theta$\,Umi  & h & y &  1 &  1.516 &  45.3 & 0.483 & 34.1 & 1.3 &  65.8 & 16.06 &  8.04 \\
1342184591 & 139.203546 &$\theta$\,Umi  & h & y &  1 &  1.499 &  42.2 & 0.520 & 36.7 & 1.2 &  61.3 & 12.55 &  8.36 \\
1342184592 & 139.206138 &$\theta$\,Umi  & h & y &  1 &  2.357 &  63.1 & 0.549 & 38.7 & 1.6 &  91.6 & 12.02 &  9.29 \\
1342184593 & 139.209497 &$\theta$\,Umi  & h & y &  1 &  1.547 &  37.3 & 0.409 & 28.8 & 1.3 &  54.2 & 14.40 &  6.77 \\
1342184594 & 139.211300 &$\theta$\,Umi  & h & y &  1 &  2.035 &  30.5 & 0.527 & 37.2 & 0.8 &  44.2 &  9.37 &  7.05 \\
1342191981 & 300.807361 &$\theta$\,Umi  & h & y & 20 &  1.139 &  36.3 & 0.108 &  7.6 & 4.8 &  52.8 & 13.36 & 10.17 \\
           \noalign{\smallskip}
\hline                                   
\end{tabular}
\end{sidewaystable*}

\addtocounter{table}{-1}
  
\begin{sidewaystable*}[h!]
\caption{continued. Chop-nod photometry measurements in the red (160\,$\mu$m 
filter).
} 
\label{table:chopnodphotred2}      
\centering                          
\begin{tabular}{c c c c c c c c c c c c c c c c}        
\hline\hline                 
            \noalign{\smallskip}
 OBSID     &       OD   &   Target     & G & Dith & Rep. & FitPeak &f$_{\rm aper}$&$\sigma_{\rm pix}$&$\sigma_{\rm aper,corr}$& S/N &f$_{\rm star}$& W$_{max}$ & W$_{min}$\\
           &            &              &   &      &      &  (mJy)  &  (mJy)  &  (mJy) &  (mJy) &     &   (mJy)  &(\arcsec)&(\arcsec)\\
            \noalign{\smallskip}
\hline
            \noalign{\smallskip}
1342190971 & 284.837708 & HD\,41047     & h & y & 10 &  0.684 &  19.7 & 0.164 & 11.6 & 1.7 &  28.6 & 11.03 &  9.64 \\
1342184296 & 132.725098 & HD\,138265    & h & y & 43 &  0.604 &  20.2 & 0.104 &  7.3 & 2.8 &  29.4 & 12.31 & 11.22 \\
1342184297 & 132.775573 & HD\,138265    & h & y & 10 &  0.672 &  25.2 & 0.150 & 10.6 & 2.4 &  36.5 & 13.82 & 11.12 \\
1342184298 & 132.797749 & HD\,138265    & h & y & 10 &  0.657 &  26.5 & 0.142 & 10.0 & 2.6 &  38.4 & 14.26 & 11.84 \\
1342184299 & 132.819914 & HD\,138265    & h & y & 10 &  0.705 &  21.7 & 0.144 & 10.2 & 2.1 &  31.5 & 13.72 &  9.04 \\
1342184300 & 132.842102 & HD\,138265    & h & y & 10 &  0.755 &  22.1 & 0.139 &  9.8 & 2.3 &  32.1 & 12.87 &  8.97 \\
1342184301 & 132.874606 & HD\,138265    & h & y & 20 &  0.618 &  20.9 & 0.129 &  9.1 & 2.3 &  30.3 & 13.21 & 10.56 \\
1342185443 & 146.224583 & HD\,138265    & h & y &  4 &  0.724 &  17.9 & 0.168 & 11.8 & 1.5 &  26.0 & 11.20 &  8.61 \\
1342185444 & 146.238106 & HD\,138265    & h & y &  7 &  0.708 &  18.4 & 0.167 & 11.8 & 1.6 &  26.7 & 14.15 &  7.47 \\
1342185445 & 146.251663 & HD\,138265    & h & y & 12 &  0.755 &  18.7 & 0.185 & 13.0 & 1.4 &  27.2 & 12.88 &  7.93 \\
1342191984 & 300.874850 & HD\,138265    & h & y & 10 &  0.653 &  23.1 & 0.151 & 10.7 & 2.2 &  33.5 & 14.90 & 10.40 \\
1342191985 & 300.901319 & HD\,138265    & h & y & 20 &  0.616 &  19.5 & 0.124 &  8.8 & 2.2 &  28.4 & 13.16 & 10.10 \\
           \noalign{\smallskip}
\hline                                   
\end{tabular}
\end{sidewaystable*}

\clearpage

\section{Comparison scan map with chop/nod photometry}
\label{sect:appc}

In Table~\ref{table:compscanchop} we list the flux ratios of scan map 
photometry and chop/nod photometry for ten sources, which were observed in 
both modes. The comparison between the two photometry modes gives the 
following result:

For 70\,$\mu$m photometry, the consistency of the fluxes is better than 3\% 
for seven out of nine sources. The two excursions, HD\,159330 and HD\,15008, 
are consistent within the larger error margin which is caused by a larger 
uncertainty because of only one chop/nod measurement (HD\,159330) or the 
faintness of the source (HD\,15008).

For 100\,$\mu$m photometry, the consistency of the fluxes is better than 2\% 
for five out of eight sources. For HD\,138265 the flux consistency is 
$\approx$4\%, but the derived error margin is smaller. For $\eta$\,Dra and 
HD\,41047 there is only one chop/nod measurement, which introduces a high 
uncertainty, but fluxes are consistent within the error margin. 

For 160\,$\mu$m photometry, the consistency of the fluxes is better than 3\% 
for four out of seven sources. For $\varepsilon$\,Lep the scan map flux is 9\% 
higher than the chop/nod one. There is only a small (4) number of chop/nod 
measurements versus a large (18) number of scan map measurements. We therefore 
consider the scan map mode result as the more reliable one. The opposite is 
the case for the number of photometric measurements of $\theta$\,Umi, with 2 
scan map measurements versus 16 chop/nod measurements. Here the scan map flux 
is 13\% higher than the chop/nod one. However, the 2 scan map measurements, 
each with 6 repetitions, have the best S/N of all measurements and are 
therefore quite reliable. Fifteen out of 16 chop/nod measurements have a 
repetition factor of only 1. They still allow a reasonable detection of the 
source at the expected location but show considerable scatter in the resulting 
(colour-corrected) fluxes between 35.5 and 91.6\,mJy (expected flux according 
to the model: 53.9\,mJy). Only one chop/nod measurement has 20 repetitions 
with a S/N comparable to the two scan maps. Its resulting flux of 52.8\,mJy is 
13\% lower than the average 60.9\,mJy from the two scan maps. Here we should 
note that the annulus used for background determination is closer to the 
source and narrower for chop/nod aperture photometry (radius 
24 -- 28\arcsec,~\citet{nielbock13}) than for scan map photometry (radius 
35 -- 45\arcsec,~\citet{balog14}). As we discuss in 
Sect.~\ref{sect:starsfirexcess}, the scan map measurements prove contamination 
of the source flux by FIR cirrus emission in the order of 10\% explaining the 
excess over the model flux. The maps also show that there is additional 
emission around the source which is much more picked up by the background 
annulus of the chop/nod photometry, resulting in a higher subtracted 
background value. This leads to the result that the chop/nod photometry is 
close to the expected model flux, because the underlying cirrus emission is by 
chance properly compensated for by the background subtraction, while the scan 
map photometry reveals the extra emission inside the aperture. The photometric 
result must therefore be associated by an additional uncertainty of 10\%, 
because the background subtraction strongly depends on the selected background 
area geometry (c.f.\ Table~\ref{table:photminiscan}). For HD\,41047 there is 
only one chop/nod measurement with a very high assigned flux uncertainty, so 
that also the flux ratio of scan map to chop/nod photometry is highly uncertain.

%
%
\begin{table}[ht!]
\caption{Ratios of fluxes obtained in scan map mode photometry 
(Table~\ref{table:photminiscan}) versus chop/nod mode photometry 
(Table~\ref{table:photchopnod}) as a measure of consistency between the two 
photometry modes. Values in italics have a high uncertainty. 
}             
\label{table:compscanchop}      
\centering                          
\begin{tabular}{r c c c c}        
\hline\hline                 
            \noalign{\smallskip}
   HD   &    Name           &   R$_{70}^{S/C}$  &   R$_{100}^{S/C}$  &   R$_{160}^{S/C}$  \\    
            \noalign{\smallskip}
\hline                        
            \noalign{\smallskip}
 62509  &  $\beta$\,Gem     & 1.031$\pm$0.005 & 1.014$\pm$0.011  & 1.000$\pm$0.006  \\
 32887  &$\varepsilon$\,Lep & 0.987$\pm$0.006 & 1.019$\pm$0.025  & 1.093$\pm$0.038  \\
148387  &   $\eta$\,Dra     & 0.993$\pm$0.034 & 1.059$\pm$0.068  & 0.969$\pm$0.095  \\
180711  &   $\delta$\,Dra   & 0.992$\pm$0.007 & 1.002$\pm$0.009  & 1.026$\pm$0.043  \\
139669  &   $\theta$\,Umi   & 1.009$\pm$0.008 & 1.018$\pm$0.043  & 1.131$\pm$0.070  \\
 41047  &   HR\,2131        &                 & 1.101$\pm$0.063  &{\it 1.248$\pm$0.652} \\  
138265  &   HR\,5755        & 1.001$\pm$0.011 & 0.958$\pm$0.011  & 1.000$\pm$0.050  \\
159330  &   HR\,6540        & 1.076$\pm$0.118 & 1.016$\pm$0.116  &        --        \\
152222  &                   & 1.008$\pm$0.099 &        --        &        --        \\
 15008  &   $\delta$\,Hyi   & 1.110$\pm$0.064 &        --        &        --        \\
           \noalign{\smallskip}
\hline                                   
\end{tabular}
\end{table}

\clearpage

\section{ISOPHOT Highly Processed Data Product (HPDP) photometry}
\label{sect:appd}
%
%
\begin{table}[ht!]
\caption{ISOPHOT~\citep{lemke96} Highly Processed Data Product (HPDP)
         photometry of P22 mini-maps of normal stars 
        (https://www.cosmos.esa.int/web/iso/highly-processed-data-products: 
         Mo\'or et al., 2003, "Far-infrared observations of normal stars 
         measured with ISOPHOT in mini-map mode"). The values in column
         f$_{\nu}$ are the original HPDP fluxes (for SED $\propto \nu^{-1}$).
         They have to be divided by the colour-correction factor cc,
         which is for a 5000\,K BB SED.
}             
\label{table:isophothpdpphot}      
\begin{center}
    \begin{tabular}{l c c l c c}
   \hline\hline
            \noalign{\smallskip}
Star               & Filter & $\lambda_{\rm c}$ & ISO TDT no. &  f$_{\nu}$   & cc \\
                   &        &     ($\mu$m)     &             &    (mJy)    &    \\   
            \noalign{\smallskip}
\hline
            \noalign{\smallskip}
$\alpha$\,Ari      & C\_180 &        180       &  79001902   &  314$\pm$19 &  1.10 \\

            \noalign{\smallskip}
\hline
            \noalign{\smallskip}
$\varepsilon$\,Lep & C\_60  &         60       &  65701315   & 1779$\pm$71 &  1.06 \\
                   & C\_50  &         65       &  65701312   & 1835$\pm$65 &  1.29 \\
                   & C\_70  &         80       &  65701309   & 1113$\pm$46 &  1.23 \\
                   & C\_90  &         90       &  65701318   &  918$\pm$64 &  1.17 \\
                   & C\_100 &        100       &  65701306   &  692$\pm$36 &  1.10 \\
                   & C\_105 &        105       &  65701303   &  601$\pm$44 &  1.05 \\
                   & C\_120 &        120       &  65002709   &  507$\pm$43 &  1.21 \\
                   & C\_135 &        150       &  65002103   &  292$\pm$21 &  1.10 \\
                   & C\_160 &        170       &  65002406   &  225$\pm$21 &  1.20 \\
            \noalign{\smallskip}
\hline
            \noalign{\smallskip}
$\omega$\,Cap      & C\_60  &         60       &  72701415   & 1232$\pm$65 &  1.06 \\
                   & C\_50  &         65       &  72701412   & 1258$\pm$45 &  1.29 \\
                   & C\_70  &         80       &  72701409   &  779$\pm$32 &  1.23 \\
                   & C\_90  &         90       &  72701418   &  543$\pm$38 &  1.17 \\
                   & C\_100 &        100       &  72701406   &  490$\pm$25 &  1.10 \\
                   & C\_105 &        105       &  72701403   &  441$\pm$32 &  1.05 \\
                   & C\_120 &        120       &  73401709   &  337$\pm$28 &  1.21 \\
                   & C\_135 &        150       &  73401603   &  221$\pm$16 &  1.10 \\
                   & C\_160 &        170       &  73401706   &  219$\pm$26\tablefootmark{1} &  1.20 \\
            \noalign{\smallskip}
\hline
            \noalign{\smallskip}
$\eta$\,Dra        & C\_90  &         90       &  78300677   &  365$\pm$26 &  1.17 \\
                   & C\_160 &        170       &  35800501   &  123$\pm$16 &  1.20 \\
            \noalign{\smallskip}
\hline
            \noalign{\smallskip}
     \end{tabular}
\end{center}
$\tablefoottext{1}$~Measurement not used in Fig.~\ref{fiducial_star_scale}
\end{table}

\end{appendix}


\end{document}